\documentclass[a4paper,fleqn,usenatbib]{mnras}

\usepackage{newtxtext,newtxmath}

\usepackage[T1]{fontenc}
\usepackage{ae,aecompl}

\usepackage{graphicx}	
\usepackage{amsmath}	
\usepackage{amssymb}	
\usepackage{mathrsfs}
\usepackage{color}

\newcommand{\fracd}[2]{\frac{\partial #1}{\partial #2}}
\newcommand{\del}{\nabla}

\title[Radiative Feedback and Nuclear Gas Disc Formation]{Gas Inflow and Star Formation near Supermassive Black Holes: The Role of Nuclear Activity}

\author[C. Frazer, F. Heitsch]{
Christopher C. Frazer,$^{1}$\thanks{E-mail: cfrazer@live.unc.edu}
Fabian Heitsch$^{1}$
\\
$^{1}$Department of Physics \& Astronomy, University of North Carolina 
at Chapel Hill, Chapel Hill, NC 27599-3255, USA}

\date{Accepted 2019 July 23. Received 2019 June 18; in original form 2019 June 18}

\pubyear{2019}

\begin{document}
\label{firstpage}
\pagerange{\pageref{firstpage}--\pageref{lastpage}}
\maketitle

\begin{abstract}
 Numerical models of gas inflow towards a supermassive black hole (SMBH) show that star formation may occur in such an environment through the growth of a gravitationally unstable gas disc. We consider the effect of nuclear activity on such a scenario. We present the first three-dimensional grid-based radiative hydrodynamic simulations of direct collisions between infalling gas streams and a $4 \times 10^6~\text{M}_\odot$ SMBH, using ray-tracing to incorporate radiation consistent with an active galactic nucleus (AGN). We assume inflow masses of $ \approx 10^5~\text{M}_\odot$ and explore radiation fields of 10\% and 100\% of the Eddington luminosity ($L_\text{edd}$). We follow our models to the point of central gas disc formation preceding star formation and use the Toomre Q parameter ($Q_T$) to test for gravitational instability. We find that radiation pressure from UV photons inhibits inflow. Yet, for weak radiation fields, a central disc forms on timescales similar to that of models without feedback. Average densities of $> 10^{8}~\text{cm}^{-3}$ limit photo-heating to the disc surface allowing for $Q_T\approx1$. For strong radiation fields, the disc forms more gradually resulting in lower surface densities and larger $Q_T$ values. Mass accretion rates in our models are consistent with 1\%--60\% of the Eddington limit, thus we conclude that it is unlikely that radiative feedback from AGN activity would inhibit circumnuclear star formation arising from a massive inflow event. 
\end{abstract}

\begin{keywords}
hydrodynamics -- radiative transfer -- Galaxy:centre -- stars:formation 
\end{keywords}



\section{Introduction}

\subsection{Star Formation in the Galactic Centre}
Stellar orbits in the Milky Way's Galactic Centre (GC) serve as direct evidence for the existence of a $4\times10^6~\text{M}_\odot$ supermassive black hole (SMBH), coinciding with the radio source, Sgr A* \citep{2002Natur.419..694S,2003ApJ...586L.127G,2009ApJ...707L.114G,2017ApJ...837...30G}. Several hundred massive stars orbit within the central parsec of Sgr A*, many of which belong to a clockwise orbiting disc extending from $0.05$ to $0.5$~pc \citep{2003ApJ...594..812G,2006ApJ...643.1011P,2009ApJ...697.1741B,2010ApJ...708..834B,2009ApJ...690.1463L,2013ApJ...764..155L,2014ApJ...783..131Y} with an eccentricity of $-eps-converted-to.pdfilon \approx 0.3$ \citep{2009ApJ...697.1741B,2014ApJ...783..131Y}. The stellar disc age of $<$ 6 Myr \citep{2006ApJ...643.1011P,2013ApJ...764..155L} strongly suggests that these stars formed in situ. However, the immense tidal field of the SMBH is expected to inhibit star formation within the central few parsecs of the GC (see discussion in \citet{2016LNP...905..205M}).

Possible explanations for the observed stellar population include the disruption of a newly formed massive star cluster migrating inwards via dynamical friction \citep{2001ApJ...546L..39G}. Such a process could potentially be expedited by the gravitational influence of an intermediate mass black hole \citep{2003ApJ...593L..77H,2004ApJ...607L.123K,2005ApJ...635..341L} or an overabundance of massive stars \citep{2005ApJ...6218..236G}, though observational and timescale constraints do not strongly support such scenarios \citep{2008ApJ...675.1278S,2010RvMP...82.3121G,2006ApJ...643.1011P,2005MNRAS.364L..23N}. The alternative, in-situ star formation, remains favored.                          

Theory suggests that star formation in the immediate vicinity of a SMBH can occur via rapid cooling and fragmentation of an accretion disc \citep{2003ApJ...590L..33L,2007MNRAS.379...21N,2005A&A...437..437N,2006ApJ...643.1011P}. The formation of a sufficiently dense accretion disc is a natural consequence of the tidal disruption of a $\approx 10^5~\text{M}_\odot$ molecular gas stream on a low angular momentum orbit about the GC \citep{2008ApJ...683L..37W}. Hydrodynamic models of this process reproduce stellar discs in agreement with observed stellar orbits \citep{1998MNRAS.294...35S,2013MNRAS.433..353L,2008Sci...321.1060B, 2012ApJ...749..168M,2011MNRAS.412..469A}. The origin of such gas inflow remains uncertain, though models suggest that gas clump collisions at $\approx 1~\text{pc}$ could supply sufficient inflow to incite a star formation episode \citep{2009MNRAS.394..191H, 2013ApJ...771..119A}. Furthermore, observational estimates for an inflow rate of $0.1$-$1~\text{M}_\odot~\text{yr}^{-1}$ in the GC \citep{1996ARA&A..34..645M} are consistent with the infall of $\approx  10^5~\text{M}_\odot$ gas streams on a timescale of a few $\times$ Myr.
\subsection{Nuclear Activity in the Galactic Centre}
Models which explore the process of stellar disc formation resulting from gas stream capture also show evidence of accretion rates onto the SMBH at considerable fractions of the Eddington limit \citep{2008Sci...321.1060B,2009MNRAS.394..191H,2011MNRAS.412..469A}: 
\begin{equation}
    \label{eq:eddM}
    \dot M_\text{edd} = 2 \times 10^{-8} \left(\frac{M_\text{BH}}{\text{M}_\odot}\right) \ \text{M}_\odot \  \text{yr}^{-1} \ .
\end{equation}
During such an accretion episode, an active galactic nucleus (AGN) can radiate at large fractions of the Eddington luminosity:
\begin{equation}
    \label{eq:eddL}
    L_\text{edd} = \dot M_\text{edd} c^2 -eps-converted-to.pdfilon_r = 3\times 10^4 \left(\frac{M_\text{BH}}{\text{M}_\odot}\right) \left( \frac{-eps-converted-to.pdfilon_r}{0.1} \right) L_\odot \ ,
\end{equation}
where $-eps-converted-to.pdfilon_r$ is the radiative efficiency and $c$ is the speed of light. Despite estimates for large scale mass inflow in the GC, Sgr~A* shows no evidence of current accretion activity (see \citet{1996ARA&A..34..645M} and references therein). Furthermore, observations of the GC limit the bolometric luminosity of Sgr A* to $\approx 10^{-10}-10^{-9}~L_\text{edd}$ over the past few hundred years \citep{1993ApJ...407..606S,2003ApJ...591..891B}. Yet, there are at least two pieces of evidence that point to past AGN activity in the GC. First, the existence of two extended gamma-ray sources referred to as the Fermi bubbles \citep{2010ApJ...724.1044S} may be the result of either a Galactic outflow triggered by AGN activity 6 Myr ago \citep{2012MNRAS.424..666Z} or an AGN jet that existed 1-3 Myr ago \citep{2012ApJ...756..181G}. Second, as previously noted, the population of several hundred massive stars within the central parsec of Sgr A* is difficult to explain in the absence of rapid gas inflow towards Sgr A*. 

So far, no models exploring tidal disruption of inflowing gas and resulting star formation have considered the effect of radiative feedback from an accretion episode onto the SMBH. Yet, radiative-hydrodynamic (RHD) simulations of gas clouds subject to AGN radiation have been presented in several works. Using two-dimensional models, \citet{2011MNRAS.415..741S} demonstrated that the fate of infalling gas clouds largely depends on the column density of the gas. Only in cases with sufficiently strong shielding can gas withstand the impinging radiation and complete its approach towards the SMBH. This requirement for sufficiently dense gas columns is also noted in studies of AGN winds \citep{2013ApJ...763L..18W,2014MNRAS.441.3055B} and relativistic jets \citep{2012ApJ...757..136W}. 

Using three-dimensional RHD simulations of gas complexes at distances of $\approx 10 \ \text{pc}$ from a radiating SMBH, \citet{2011A&A...536A..41H,2010A&A...522A..24H} explored the effect of X-ray feedback on the initial mass function (IMF). Assuming that UV radiation is obscured by interior gas and dust, these models showed that X-ray radiation alone leads to gas compression and heating, the latter of which promotes the formation of higher mass protostars. Most recently, \citet{2014MNRAS.443.2018N} explored the effect of both UV and X-ray radiation on infalling gas clouds with galactocentric distances of 5 pc and 50 pc using three-dimensional hydrodynamic simulations, including a detailed chemical network. They parameterize the radiation field by the ionization parameter $U_\text{ion}$, which measures the ratio of the photon density to the number density of the irradiated gas. These models show that photo-evaporation dominates when $U_\text{ion}$ is low, whereas radiation pressure becomes more important for large values of $U_\text{ion}$. Yet, models which follow the evolution of such clouds through a direct collision with the central SMBH, as is required for the birth of a nuclear stellar disc, have yet to be considered.

\subsection{Motivation and Outline}

We explore the role of radiative feedback from accretion onto the central SMBH in the formation and evolution of a circum-nuclear gas disc. Specifically, we consider the gas inflow scenario which is known to result in both the formation of a stellar disc and accretion rates at large fractions of the Eddington limit.  

Numerical methods, including relevant physics for radiative transfer, are outlined in \S~\ref{sec:methods}. In \S~\ref{sec:simsetup}, we describe the initial conditions for our stream inflow models. We present our simulation results in \S~\ref{sec:simresults} and discuss the implications in the context of nuclear star formation in \S~\ref{sec:results}. We provide a summary of key results from this study in \S~\ref{sec:conclusion}. In addition, we provide an overview of the radiative transfer routine used for this work as well as standard tests of its accuracy in Appendix~\ref{ap:radtrans}.

\section{Methods}
\label{sec:methods}
We use a modified version of \textsc{athena} 4.2 \citep{2008ApJS..178..137S} to solve the following set of equations (see Appendix~\ref{app:athenamod} for specifics on our modifications):
\begin{align} 
    \label{eq:dcons}
    \fracd{\rho}{t} + \del \cdot \left(\rho {\bf v}\right) &= 0  \\
    \label{eq:mcons}
    \fracd{\rho {\bf v}}{t} + \del \cdot \left(\rho {\bf v} {\bf v} + {\bf I} P \right) &= - \rho \del \phi + \rho~{\bf a}_\gamma  \\
    \label{eq:econs}
    \fracd{E}{t} + \del \cdot \left({\bf v} (E + P) \right) &= -\rho {\bf v} \cdot \del \phi + \rho {\bf v}\cdot \bf{a}_\gamma + G - L  \\  
    \label{eq:scalcons}
    \fracd{C \rho}{t} + \del \cdot \left(C \rho {\bf v}\right)  &= 0 \\    
    \label{eq:ioncons}
    \fracd{\rho_{\text{HII}}}{t} + \del \cdot \left(\rho_{\text{HII}} {\bf v}\right) &=  m_H \left(I - R\right);
\end{align}
with the gas density $\rho$, the fluid velocity vector ${\bf v}$, the gas pressure $P$, the unit dyad ${\bf I}$, the energy density 
\begin{equation} \label{eq:energy}
E = \frac{1}{2} \rho {\bf v} \cdot {\bf v}  + \frac{P}{\gamma -1}  \ ,
\end{equation}
and a static gravitational potential
\begin{equation} \label{eq:pot}
    \phi = - \frac{G M_{\text{BH}}}{r}   \ 
\end{equation}
where $M_\text{BH}$ is the black hole mass which is situated at the origin. We include two source terms, ${\bf G}$ and ${\bf L}$, which account for energy gains, and for losses due to radiative processes (see section~\ref{sec:thermphys}).  A colour field, $C = \frac{m_\text{in}}{m_\text{cell}}$, is advected to trace inflowing gas.

We track the ionization of hydrogen gas via Eq.~\ref{eq:ioncons}. The mass density of ionized hydrogen, $\rho_\text{HII}$, depends on two source terms, $I$ and $R$, which are the ionization and recombination rates per volume (see \S~\ref{sec:radiation}). In addition, we include an acceleration, ${\bf a}_\gamma$, to account for radiation pressure.

For all models in this work we use the directionally un-split Van-Leer (VL) integrator \citep{2009NewA...14..139S} with second order reconstruction in the primitive variables \citep{1984JCoPH..54..174C} and the HLLC Riemann solver \citep{toro2009riemann}. 
For all models we assume a pure and neutral hydrogen gas with a constant mean molecular weight of $\mu = 1$.
We modify the adiabatic equation of state with $\gamma=5/3$ via the thermal physics described in Sec.~\ref{sec:thermphys},
and we use a Cartesian geometry for our computational mesh.
                               
\subsection{Radiation}
\label{sec:radiation}
To include the effect of radiation, we have outfitted \textsc{athena} with an adaptive ray-tracing routine that follows both a previous implementation into the code \citep{2007ApJ...671..518K} and the radiation module from \textsc{enzo-moray} \citep{2011MNRAS.414.3458W}. In short, we solve the equation of radiative transfer iteratively at the beginning of every hydrodynamic step. Because the radiation timescale is often much shorter than the dynamical timescale, we sub-cycle multiple radiation times-eps-converted-to.pdf per hydrodynamic timestep. At each radiation cycle, an adaptive ray tree is traced from the radiation source outwards throughout the computational mesh. As rays traverse through cells, attenuation of incident radiation leads to photon deposition, gas heating, ionization, and radiation pressure. A full description of our radiative transfer module as well as standard tests of its accuracy are provided in Appendix~\ref{ap:radtrans}. Here, we detail only the components concerning radiative-hydrodynamic coupling. We discuss the exact radiation model used for our simulations in section~\ref{sec:radfield}.

\subsubsection{Ionization Physics} 
 To conserve photon number, each ionizing photon ($E_\gamma > 13.6$ eV) must be exchanged exactly for one ionization of a hydrogen atom, thus the ionization rate is   
\label{sec:ionphys}
\begin{equation}               
    \label{eq:ionrate}
    I_i = \frac{\delta N_{\gamma,i}}{\Delta t_{\gamma} \Delta x^3} \ ,
\end{equation}
where $\delta N_{\gamma,i}$ is the number of photons deposited, $\Delta t_{\gamma}$ is the radiation timestep, and $\Delta x^3$ is the cell volume assuming a uniform aspect ratio for grid cells. The subscript $i$ indicates the photon species. In addition to photo-ionization, we also include collisional ionization such that  
\begin{equation}
    \label{eq:collion}
    I_\text{coll} = k_\text{coll} n_\text{HI} n_\text{HII} \ ,
\end{equation}
where the collisional ionization rate coefficient is \citep{1986MNRAS.221..635T}:
\begin{equation}
    \label{eq:collionrate}
    k_\text{coll} = 5.84 \times 10^{-11} \sqrt{\frac{T}{K}} e^{-E_H/k_B T} \text{cm}^{3} \text{s}^{-1} \ .
\end{equation}
$E_\text{H} = 13.6~\text{eV}$ is the ionization potential of hydrogen, $n_{\text{HI}}$ is the neutral hydrogen number density, $n_\text{HII}$ is the ionized hydrogen number density, $T$ is the temperature of the gas, and $k_B$ is the Boltzmann constant. The net ionization rate which enters into Eq.~\ref{eq:ioncons} is the sum of collisional and radiative ionization terms: 
\begin{equation}
    \label{eq:ionsum}
    I = \sum_i I_i + I_\text{coll} \ ,
\end{equation}
where the sum is over contributions from each photon species. For photon energies in excess of the the binding energy of hydrogen, photo-heating also occurs so that
\begin{equation}
\label{eq:ionheat}   
    \Gamma_{\text{ion},i}  = \frac{\delta N_{\gamma,i} \left(E_{\gamma,i}-E_\text{H}\right)}{n_\text{H} \Delta x^3 \Delta t_\gamma}   
\end{equation}
is the volumetric heating rate due to photo-absorption.

\subsubsection{Recombination}   
\label{sec:recomb}
For recombination of ionized hydrogen, we assume the ``on-the-spot" approximation in which recombinations of hydrogen atoms to the ground state emit ionizing photons that are re-absorbed by the intervening medium \citep{1989agna.book.....O}. Recombinations to excited states of hydrogen are assumed to emit photons to which the surrounding gas is optically thin. Under this assumption, the recombination rate is 
\begin{equation}
    \label{eq:recomb}
R = \alpha_B(T) n_e n_p    \ ,
\end{equation}
where $\alpha_B(T)$ is the recombination coefficient for case B recombination,
\begin{equation}
    \label{eq:alphaB}
    \alpha_B(T) = 2.59 \times 10^{-13} \left(\frac{T}{10^4 K}\right)^{-0.7} \text{cm}^3 \ \text{s}^{-1} \ ,
\end{equation}  
and $n_e$ and $n_p$ are the electron number density and free proton number density of the gas. Under our assumption of a pure hydrogen gas, the number density of electrons is equal to the number density of free protons. Therefore, the recombination rate in our case simplifies to $R= n_\text{HII}^2 \alpha_B(T)$.  

\subsubsection{Compton Heating}
In the X-ray regime, Compton heating from the scattering with free electrons leads to photon energy loss rather than absorption. Because our radiative transfer scheme uses monochromatic photon bins, we cannot change the energy of photons. Instead, we follow the approach of \citet{2011ApJ...738...54K} by proportionally decreasing the photon number flux to account for energy loss from Compton scattering such that
\begin{equation}
    \label{eq:comptonabs}
    \delta N_{\gamma,\text{comp}} = N_\gamma (1- e^{-\tau_e}) \Delta E(T_e)/E_\gamma   \  ,
\end{equation}
where $N_\gamma$ is the incident number of photons. $\tau_e \approx n_e \sigma_{KN} \Delta s$ is the optical depth where $n_e$ is the electron number density ($n_e = n_\text{HII}$), $\sigma_{KN}$ the Klein-Nishina cross section \citep{1979rpa..book.....R}, and  $\Delta s$ is the path length of a ray segment though a grid cell. For the non-relativistic energies considered in this work, we take $\sigma_{KN} \approx \sigma_T$, where $\sigma_T$ is the Thomson scattering cross section. The energy lost in a Compton scattering event is
\begin{equation}
\Delta E(T_e) = 4 k_B T_e \frac{E_\gamma}{m_e c^2}   \ ,
\end{equation}
where $m_e$ is the electron mass and $T_e$ is the electron temperature. The Compton heating term is written in the same manner as heating from photo-absorption:
\begin{equation}
    \label{eq:comptonheat}
    \Gamma_\text{comp} = \frac{\delta N_{\gamma,\text{comp}} E_\gamma}{n_e \Delta x^3 \Delta t_\gamma}  \ .
\end{equation}

\subsubsection{Secondary Ionization}
\label{sec:secondion}
Photons with energies much greater than the ionization potential of hydrogen (E$_\gamma$ > 100eV) may result in the ionization of multiple hydrogen atoms. The fractional amount of energy allotted to the heating and ionization of hydrogen for these photons is \citep{1985ApJ...298..268S}:  
\begin{align}
    Y_\Gamma &= 0.9971*\left(1-\left(1-x^{0.2663}\right)^{1.3163}\right)       \\       
    Y_\text{H,ion} &= 0.3908*\left(1-x^{0.4092}\right)^{1.7592} \ ,
\end{align}
where $x = n_\text{HII}/n_\text{H}$ is the ionization fraction of the gas. It should be noted that we do not explicitly include helium in our models, though the energy fraction which contributes to ionization of helium is:
\begin{equation}
    Y_\text{He,ion} = 0.0554*\left(1-x^{0.4614}\right)^{1.6660} \ .
\end{equation}
We include ionization of helium atoms implicitly such that the total fractional energy input for ionization is $Y_{\text{ion}} = Y_{\text{H,ion}} + Y_{\text{He,ion}}$. The net ionization and heating rates are then given as:
\begin{align}
    I_{\text{i,secondary}} &= \frac{\delta N_{\gamma,i} E_{\gamma,i}}{\Delta t_\gamma \Delta x^3} \frac{Y_{\text{ion}}}{E_\text{H}}  \\
    \Gamma_{\text{i,secondary}} &= \frac{\delta N_{\gamma,i} E_{\gamma,i}}{n_\text{H}\Delta t_\gamma \Delta x^3} Y_\Gamma
\end{align}
For X-rays, we substitute the ionization and heating terms above into equations \ref{eq:ionsum} and \ref{eq:ionheat}. 

\subsubsection{Radiation Pressure}
\label{sec:radpres}
As photons are absorbed by the intervening medium, they transfer momentum ($p_\gamma = E_\gamma/c$) to the gas. The force due to this process is
\begin{equation}
 F_{\gamma,i} = \frac{\delta N_{\gamma,i} E_{\gamma,i}}{c \Delta t_\gamma} \  \hat r  \ ,
\end{equation}
where the index $i$ is again used to distinguish monochromatic photon bins. Momentum injection is aligned with the photon propagation direction, denoted by $\hat r$. The resulting acceleration that enters into equations $\ref{eq:mcons}$ and \ref{eq:econs} is determined by dividing by the cell mass:
\begin{equation}
    {\bf a}_\gamma =   \frac{\delta N_{\gamma,i} E_{\gamma,i}}{c \Delta t_\gamma \rho \Delta x^3}\ \hat r   \ .
\end{equation}
We apply the source terms for radiation pressure in tandem with the ionization and thermal updates in the radiation routine. An alternative approach would be to include this term in the hydrodynamic integration step with other force terms \citep{2011MNRAS.414.3458W}. Despite this simplification, our implementation shows good agreement with theory as demonstrated in Appendix~\ref{sec:radprestest}.

\subsection{Thermal Physics} 
\label{sec:thermphys}
The heating term, $\textbf{G}$, in Eq.~\ref{eq:econs} is the sum of contributions from ionization, heating of neutral gas from a constant background radiation field, and heating from Compton scattering:
\begin{equation}
    G = n_\text{HI}  \Gamma_\text{amb} + \sum n_\text{H} \Gamma_{ion,i} + n_e \Gamma_\text{comp} 
\end{equation}
We take the background heating term to be \citep{2002ApJ...564L..97K}
\begin{equation}
    \Gamma_{\text{amb}} = G_0  \ (2 \times 10^{-26})  \ \text{erg} \ \text{s}^{-1}    \ ,
\end{equation}
We set $G_0 = 1000$ to account for the strong interstellar radiation field in the GC \citep{2013ApJ...768L..34C}. To avoid overheating of low density gas, we use a hyperbolic tangent function to smoothly drive the ambient heating term to 0 for temperatures above $10^4$ K. For neutral gas, we assume a modified version of the cooling function from \citet{2002ApJ...564L..97K}:
\begin{align}
    \Lambda_\text{n} &= 2\times 10^{-26} \Bigg( 10^7 \exp \frac{-118400}{T+1000} \\ \nonumber
                &  \ \ \  + 0.014 \sqrt{T} \exp \frac{-92 \ \beta_{\text{T}}}{T} \Bigg) \   \text{erg}  \ \text{s}^{-1}  \ \text{cm}^3 \ .
\end{align}
The parameter $\beta_{\text{T}}$ is introduced to approximate the effect of cosmic rays which penetrate deep into dense gas structures \citep{1978ApJ...222..881G}. The cosmic ray ionization rate in the GC is roughly a thousand times greater than the solar neighborhood \citep{2013ApJ...768L..34C} which results in a minimum gas temperature of $\approx$ 100 K \citep{1995ApJ...443..152W,2011MNRAS.414.1705P}. To approach this minimum temperature smoothly, we set $\beta_\text{T}$ = 10.  

For ionized gas, we include recombination cooling and free-free cooling \citep{1989agna.book.....O}, 
\begin{align}
    \Lambda_\text{rec} &= 8.418 \times 10^{-26} T^{0.11} \   \text{erg} \  \text{s}^{-1} \ \text{cm}^{3}  \\
    \Lambda_\text{ff} &= 1.427 \times 10^{-27} 1.3 \sqrt{T}     \    \text{erg} \  \text{s}^{-1} \ \text{cm}^{3}   
\end{align}
We also follow the treatment for collisionally excited radiation in \citet{1989agna.book.....O}, but reduce the resulting cooling rate to  a piecewise approximation which incorporates trace amounts of NII, NIII, OII, OIII, NeII, and NeIII (see Appendix~\ref{app:CLE}):
\begin{align}
    \label{eq:CLE}
    \Lambda_\text{CLE}(T) &= 
\begin{cases}
    3.47 \times 10^{-29} \ T^{1.915}  & 0 < T  < 10^2  \\
    2.34 \times 10^{-26} \ T^{0.500}  & 10^2 < T  < 10^{2.8}  \\
    1.11 \times 10^{-24} \ T^{-0.099}  & 10^{2.8} < T  < 10^{3.6}  \\
    1.08 \times 10^{-32} \ T^{2.127}  &  10^{3.6} < T  < 10^4  \\
    2.67 \times 10^{-30} \ T^{1.529}  & 10^4 < T  < 10^{4.5}  \\
    1.74 \times 10^{-24} \ T^{0.237}   &  10{^4.5} <  T  < 10^5  \\
    1.10 \times 10^{-21} \ T^{-0.323}  & 10^5 < T  < 10^6  \\
    7.49 \times 10^{-21} \ T^{-0.462}  &   10^6  < T \\
\end{cases} \\    
    & \text{ergs} \  \text{s}^{-1} \  \text{cm}^{3}      \nonumber
\end{align}
The net cooling rate is a combination of all cooling terms:
\begin{equation}
    L = n_\text{HI}^2 \Lambda_\text{n}(T)  + n_\text{HII}^2 \left(\Lambda_\text{ff}(T) + \Lambda_\text{rec}(T) + \Lambda_\text{CLE}(T)\right)
\end{equation}
Unlike \citet{2007ApJ...671..518K}, we did not restrict cooling in mixed-ionization cells, as this led to incorrect propagation speeds of ionization fronts in our HII region expansion test (see Appendix~\ref{sec:dtypeion}).

\section{Simulation Set-up}
\label{sec:simsetup}
We use a $(4~\text{pc})^3$ computational box centered on the origin. Boundary conditions on the box allow outflow but prohibit inflow. A SMBH of $4\times 10^6~\text{M}_\odot$ \citep{2009ApJ...707L.114G} sits at the origin, implemented via a static gravitational potential. To resolve fluid flow around the SMBH while also minimizing computational cost we use three levels of static mesh refinement (SMR), resulting in an effective resolution of $512^3$ or $7.8$~mpc at the finest level. The geometry of the computational mesh, including refinement zones, is shown in Table~\ref{tab:SMR}.

\begin{table}
    \centering
    \caption{Mesh and Refinement Geometry}
    \begin{tabular}{ccccccc}
        \hline
        Level & Dimensions  & x$_0$  & y$_0$ & z$_0$ & Res.  & $\Delta$x    \\
              & [pc]        & [pc]   & [pc]  & [pc]  &             & [mpc]\\
        \hline
        1     & 4$\times$4$\times$4       & -2     & -2    & -2    & 64$^3$      & 62.5 \\
        2     & 2$\times$2$\times$2       & -1     & -1    & -1    & 64$^3$      & 31.3 \\
        3     & 1$\times$1$\times$1       & -0.5   & -0.5  & -0.5  & 64$^3$      & 15.6 \\
        4     & 0.5$\times$0.5$\times$0.5 & -0.25  & -0.25  & -0.25 & 64$^3$      & 7.8 \\
        \hline
    \end{tabular}
    \label{tab:SMR}
\end{table}

We initialize a uniform ambient medium with a number density of $n_0 = 1~\text{cm}^{-3}$ and a gas temperature of $T_0 = 5803 \ \text{K}$, corresponding to the equilibrium temperature of gas ionized by UV radiation in our thermal model. In models including radiation, the ambient gas is assumed to be fully ionized. To ensure that the ambient medium does not collapse under the influence of the SMBH, we set the colour field to zero. At every hydrodynamic timestep, we reset cells with $C<10^{-10}$ to the ambient initial condition. This approach is similar to \citet{2012ApJ...750...58B}, though we have lowered the threshold colour field as the ambient gas profile is not convectively unstable.

The densities and temperatures in our models range over several orders of magnitude, requiring density and temperature floors to avoid occasional failures in the integration scheme. We choose a minimum number density of $n = 1~\text{cm}^{-3}$, consistent with the ambient background. Similarly, we assume a temperature minimum of $T=100$~K, which is consistent with the minimum equilibrium temperature in our thermal model. We enforce these floors at the end of the hydrodynamic update. Imposing a density floor is equivalent to adding mass. Yet, over the duration of the simulation, the mass accumulated in this way is negligible.

\subsection{Accretion Boundary}
\label{sec:accretionboundary}
An accretion boundary with radius $R_\text{acc} \approx 40~\text{mpc}$, or 5 cells on the highest refinement level, encloses the SMBH at the origin. This accretion boundary is orders of magnitude larger than the inner-most stable orbit ($R_\text{ISCO}\approx 1\mu\text{pc}$) as computational limitations prohibit our simulations from resolving gas flow in this regime. Conversely, the accretion boundary is much less than the Bondi-Hoyle radius of the inflowing gas ($R_\text{Bondi-Hoyle} = \frac{2 G M}{c_s^2 + v_\text{in}^2} \approx 3~\text{pc}$), ensuring that inflow of mass to the SMBH is resolved. Outflow is not explicitly imposed on the accretion boundary. Instead, we smoothly remove momentum and mass from gas which enters this region using the radial smoothing profile, 
\begin{equation}
    s = \left(1 - \frac{r}{R_\text{acc}} \right)^2  \text{min}\left(\frac{\Delta t}{t_\text{s}},1\right)      \ ,
\end{equation}
with the hydrodynamic timestep $\Delta t$, and the smoothing timescale $t_\text{s} = 0.1 \ \text{yr}$. The density, velocity, and temperature within the accretion boundary are then rescaled as
\begin{align}                                                 
    n' &= n \left(1-s\right) + n_0 s \\          
    T' &= T \left(1-s\right) + T_0 s \\          
    {\bf{v}}' &= {\bf{v}} (1-s)   \ ,         
\end{align}
where updated values are primed. We scale the colour field proportionally with the density, and we calculate the change of mass in this region ($\rho C \Delta x^3$) throughout this process to track accretion. This implementation does not consider the ratio of the kinetic energy of gas parcels to the binding energy with respect to the SMBH, thus gas which orbits on a near-radial trajectory will always be consumed irrespective of the incident velocity. It should also be noted that the mass lost to the SMBH through this process is not added to the static potential used in Eq.~\ref{eq:pot}. This choice is made for two reasons. First, the viscous timescales on which accretion onto the SMBH occurs are much longer than the timescales considered in our models (see \S~\ref{subsec:viscous}). Second, we adopt a constant radiation field to represent feedback due to accretion as prescribed by Eq.~\ref{eq:eddL}. By maintaining a constant SMBH mass, we preserve the ratio of the preset luminosity to the Eddington limit.

\subsection{Inflow Conditions}
\begin{figure}
    \includegraphics[width=\linewidth]{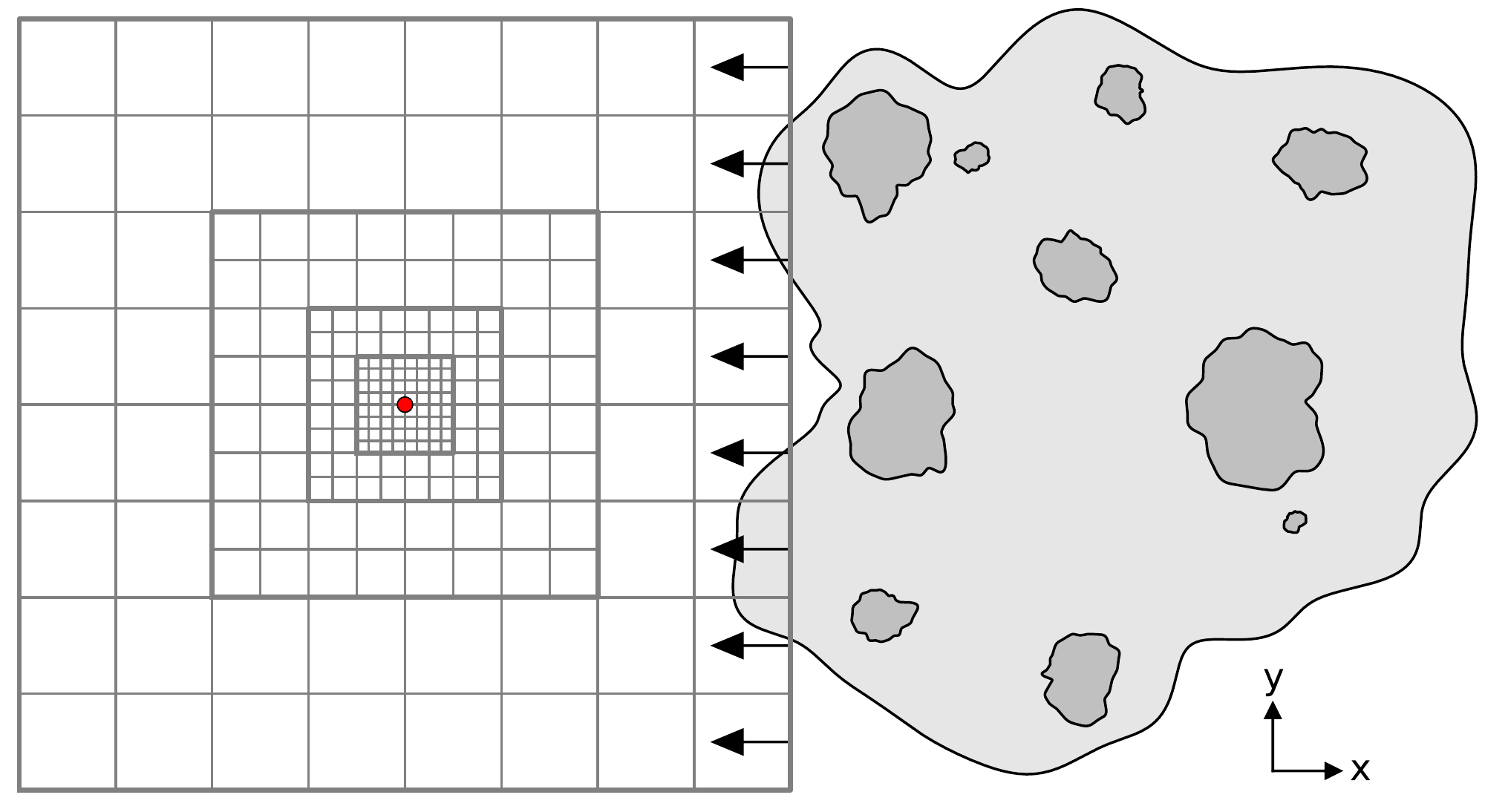} 
    \caption{Two-dimensional representation of the computational mesh and inflow conditions used for our models. Four refinement levels are nested and centered on an SMBH which rests at the origin (red circle). A gas stream with density perturbations flows into the mesh via a time-dependent inflow condition denoted by black arrows.     }
    \label{fig:inflowcartoon}
\end{figure}
\label{sec:inflow}
We model infalling gas streams via an inflow condition on the $+\hat x$ face of the computational box. We show a two-dimensional representation of this set-up in Figure~\ref{fig:inflowcartoon}. We add density perturbations to the inflowing gas for two reasons. First, inhomogeneity in the inflow mimics the substructure observed in interstellar gas. Second, in the process of gravitational focusing during the cloud's infall, streams of gas passing the SMBH in opposite directions collide, leading to specific angular momentum cancellation. Uniform inflow results in the well known Bondi-Hoyle-Lyttleton accretion process \citep{1944MNRAS.104..273B}, whereas inhomogeneity in the gas leads to the retention of angular momentum that is essential for the formation of a dense gas disc and stars \citep{2008ApJ...683L.147Y}. Following \citet{2008ApJ...683L.147Y}, we exclude velocity fluctuations because of the highly supersonic bulk motion of the flow. Velocity dispersions characteristic of molecular gas complexes in the GC are typically 15--50 km s$^{-1}$ \citep{1988ApJ...324..223B}. In comparison, the orbital velocities in the inner 0.1~pc of the GC $\gtrsim 400~\text{km  s}^{-1} $, nearly an order of magnitude greater than the expected peak velocity fluctuations.

Density perturbations are calculated as a sum of incoherent sine waves. We first determine the perturbation exponent as   
\begin{align}
    \label{eq:pertexp}
    \delta(\vec{x}) &=  \sum_{k_k,k_j,k_i=1}^{k_\text{max}} |k|^{-\alpha}   \sin \left( \frac{2 \pi k_i}{L_\text{max}} (x-v_\text{in} t) + \phi_x(\vec{k})\right)   \nonumber \\
     &\times \  \sin\left(\frac{2\pi k_j y}{L_\text{max}} +\phi_y(\vec{k})\right)  \sin\left( \frac{2\pi k_k z }{L_\text{max}} + \phi_z (\vec{k})\right)  \  ,  
\end{align}
with random phases $\phi_x$, $\phi_y$, and $\phi_z$ between 0 and $2 \pi$ for all values of $\vec{k} = (k_i,k_j,k_k)$. $v_\text{in}$ is the inflow velocity of the gas. $L_\text{max}$ is the longest length scale included, which is set to 5~pc. The maximum wavenumber is set by the ratio of this inflow length and the ``clump scale", $L_c$, which represents the size of the smallest structures in the inflow. We set $L_c = 0.1~\text{pc}$, corresponding to a maximum normalized wavenumber $k_\text{max} = \text{ceil} (L_\text{max}/(2 L_\text{c})) = 26 $. This is consistent with lower limits of observed clump sizes in both the circumnuclear disc (CND) ($r\approx0.125$~pc; \citet{2005ApJ...622..346C}) and the 50 km s$^{-1}$ cloud ($ r > 0.15$~pc; \citet{2012PASJ...64..111T}).  Somewhat motivated by turbulent cloud structure, we choose a power law index of $\alpha=3$ \citep{1981MNRAS.194..809L}. We then scale the perturbation exponent to the inflow density as
\begin{align}
    \delta_{\text{norm}}(\vec{x}) &= \frac{\delta(\vec{x}) - \min(\delta)}{\max(\delta)-\min(\delta)}  \ \log_{10} \left(\frac{\rho_\text{in,max}}{\rho_\text{in,min}}\right) \nonumber \\  
    &+   \log_{10}(\rho_\text{in,min}) \ ,
\end{align}
where $\rho_\text{in,min}$ and $\rho_\text{in,max}$ are the minimum and maximum mass densities of the inflow. The inflow mass density at each position is then calculated as:
\begin{equation}
    \rho_{in}(\vec{x}) = 10^{\ \delta_\text{norm}(\vec{x})}
\end{equation}   

Scaling the density perturbations logarithmically is necessary to generate a gas stream with multiple isolated clumps similar to interstellar gas observed in the central 100 pc of the GC \citep{1990A&A...234..133Z}. Lastly, to give the cloud finite extent in the direction perpendicular to the inflow, we use a hyperbolic tangent function to smoothly bring the inflow density to the ambient gas density for $y^2 + z^2$ > 2~pc, effectively removing high angular momentum clumps from the inflow. The formulation outlined above generates a sufficient number of ``cloud fragments" distributed in a manner that allows the inflow structure to retain a net angular momentum perpendicular to the inflow direction. While angular momentum cancellation occurs via collisions of gas clumps passing the SMBH in opposite directions, residual angular momentum promotes the formation of a gaseous disc.

We store the inflow structure in an array with the cell spacing of the coarsest grid. Inflowing densities are then linearly interpolated from this grid, and fed into the boundary cells of the computational box. We assume a neutral inflowing gas, thus we set the gas temperature to the density-dependent equilibrium temperature for neutral gas ($\Gamma_\text{n}$ = $n_\text{HI} \Lambda_\text{n}$). We also set the colour field $C=1$ for all inflowing mass in order to distinguish it from the ambient background.

\subsection{Radiation Field}
\label{sec:radfield}
In our treatment of radiation we use a binned monochromatic spectrum with photon energies of 16~eV for UV radiation and 1~keV for X-ray radiation. We assume a piecewise spectral energy distribution for an AGN \citep{2005A&A...437..861S}:
\begin{equation}
    \label{eq:AGNSED}
    L_\lambda \propto 
\begin{cases}
    \lambda^{-1}    &  \lambda  < 500 \text{\AA}  \\
    \lambda^{-0.2}  & 500 \text{\AA}  < \lambda  < 121.5 \text{nm}  \\
    \lambda^{-1.54} & 121.5 \text{nm} < \lambda < 10 \mu \text{m} \\
    \lambda^{-4}    & 10 \mu \text{m}  < \lambda 
\end{cases}
\end{equation}      
\begin{figure}
    \includegraphics[width=\linewidth]{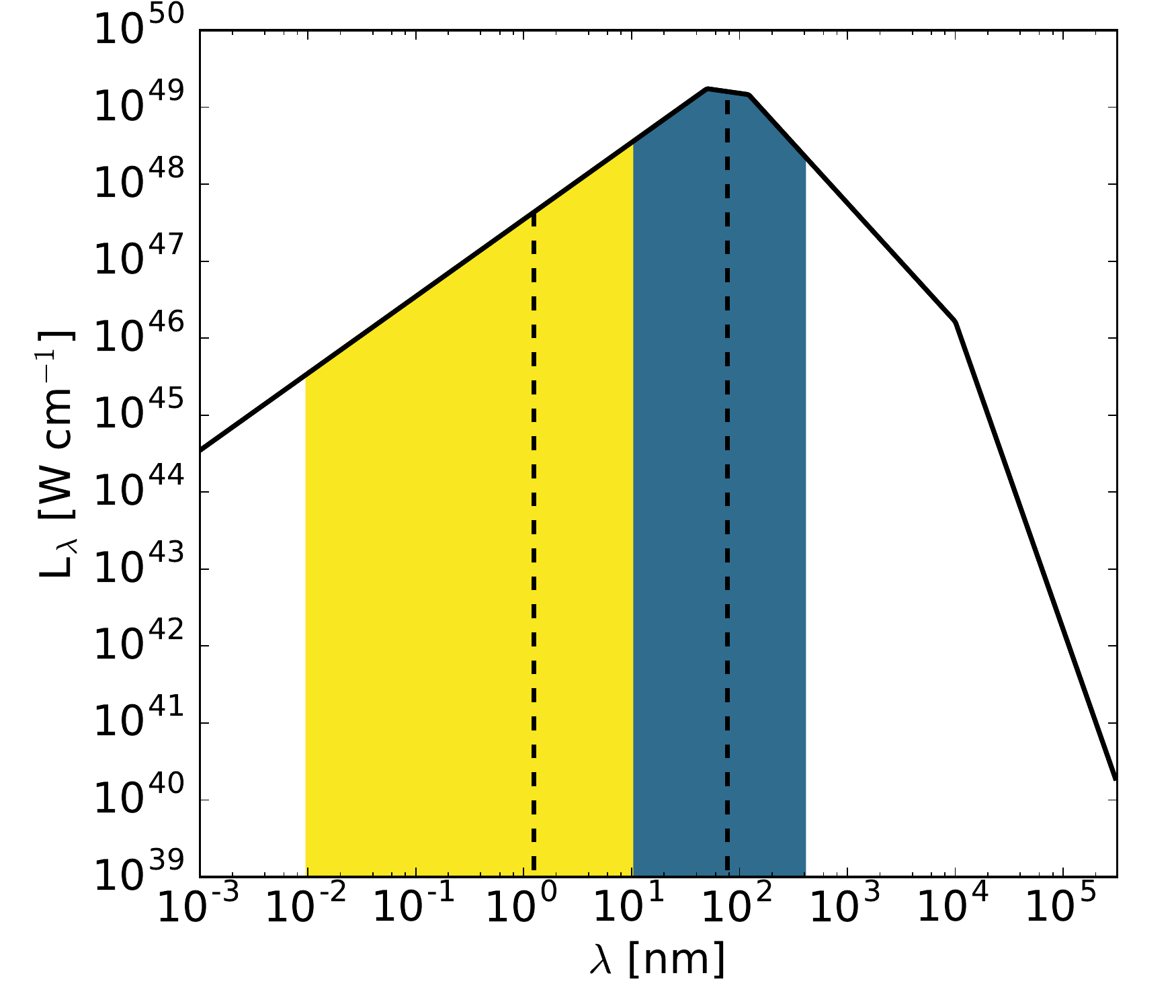} 
    \caption{Model spectral energy distribution for a $4\times10^6~\text{M}_\odot$ SMBH radiating at the Eddington luminosity. The coloured regions represent the wavelength ranges for X-ray (left, yellow) and UV (right, blue) photons. The vertical lines at 1.24~nm and 77.5~nm correspond to the photon energies of 1~keV and 16~eV used in this work.}
    \label{fig:AGNSED}
\end{figure}   

In Figure~\ref{fig:AGNSED} we show the normalized spectral energy distribution for a $4\times10^6 \ \text{M}_\odot$ SMBH radiating at the Eddington luminosity (Eq.~\ref{eq:eddL}). Assuming an input bolometric luminosity, we integrate the spectral energy distribution and calculate the normalization factor $L_\text{norm} = L_\text{bol} / \int_0^\infty L_\lambda d\lambda$. For each photon species, we calculate the net luminosity within the appropriate wavelength range and multiply by this normalization factor. For UV photons, we integrate in the range of 10--400~nm, and for X-rays we set this range to 0.01--10~nm. The photon emission rate is then determined by dividing by the photon energy. Using this model, the photon emission rates at the Eddington luminosity of a $4\times 10^6 \text{M}_\odot$ SMBH are approximately $Q_\text{UV} = 10^{55} \ \text{s}^{-1}$ and $Q_\text{X} =10^{51} \ \text{s}^{-1}$.

\subsection{Models}
We present two sets of inflow models. The naming convention of our models uses a prefix to denote the inflow structure of the gas and a suffix to indicate the radiation field. The inflow is initialized with a minimum number density of $10^4 \ \text{cm}^{-3}$ for both cases. The peak number density for the C1 inflow condition is set to $5\times10^6 \ \text{cm}^{-3}$ which yields an average number density of $9.7\times 10^4 \ \text{cm}^{-3}$ and total inflow mass of $ 10^5 \ \text{M}_\odot$. For the C2 model inflow, the peak number density is set to $10^7 \ \text{cm}^{-3}$ which results in an average number density of $1.6\times10^5~\text{cm}^{-3}$ and a total mass of $1.6\times10^{5} \  \text{M}_\odot$. We use unique random seeds to calculate perturbations in C1 and C2. Following the analytical model of \citet{2008ApJ...683L..37W}, we impose a uniform inflow velocity of 100~km~s$^{-1}$ for all models. This velocity is comparable to the Keplerian orbital velocity of the circumnuclear disk (CND), which is the innermost molecular gas reservoir in the GC \citep{2005ApJ...622..346C} and may have formed through an infall event similar to that modeled here \citep{2016A&A...585A.161M,2018ApJ...864...17T}. Velocity perturbations are excluded as they are expected to be unimportant due to the highly supersonic bulk flow \citep{2008ApJ...683L.147Y}. We follow the evolution of this system for 100 kyr, during which gas is injected into the $+\hat x$ boundary on the coarsest level for the first $25 \ \text{kyr}$. We use a hyperbolic tangent function to smoothly transition the gas density and momentum between inflow and ambient conditions over a $\approx$ 1 kyr period, carefully avoiding the introduction of steep gradients to the boundary of the simulation. Parameters for the two inflow conditions are shown in Table~\ref{tab:inflow}.      

\begin{table}
    \centering
    \caption{Inflow Parameters}
    \begin{tabular}{ccccccc}
        \hline
        prefix  & n$_\text{min}$ & n$_\text{max}$ & $\bar{\text{n}}$  & M$_\text{in}$  & seed  \\
                & cm$^{-3}$      & cm$^{-3}$      & cm$^{-3}$         & M$_\odot$      &    \\
        \hline
        C1      & $10^4$         & $5\times10^6$  & $9.7\times 10^4$  & $1.0\times10^5$  & 1\\
        C2      & $10^4$         & $10^7$         & $1.6\times 10^5$  & $1.6\times10^5$  & 2\\
        \hline
    \label{tab:inflow}
    \end{tabular}
\end{table}
\begin{table}
    \centering
    \caption{Radiation Parameters}
    \begin{tabular}{lcccc}
        \hline
        suffix  & $Q_\text{UV}$ & $Q_\text{X}$   & notes \\
              & s$^{-1}$      & s$^{-1}$          &   \\
        \hline
         C        & 0        & 0    &  no radiation \\
         C-128$^*$ & 0 & 0 &  resolution increase $\times$2 \\
         C-256$^*$ & 0 & 0 &  resolution increase $\times$4 \\
        RL   & 1e54     & 1e50 &               \\
        RH      & 1e55     & 1e51 &  L = L$_\text{edd}$\\
        X$^\dagger$     & 0    & 1e51 & X-rays only \\
        UV$^\dagger$    & 1e55 & 0    & UV photons only  \\
        NORP$^\dagger$  & 1e55 & 1e51 & no radiation pressure \\
        \hline 
        \multicolumn{4}{l}{$*$ used for testing convergence, only for C2 (see sec.~\ref{subsec:tidal_limit})}\\   
        \multicolumn{4}{l}{$\dagger$ used for testing radiation field components (see secs.~\ref{subsec:uv}--\ref{subsec:rp})}   
    \end{tabular}
    \label{tab:radiation}
\end{table}

For each inflow condition, we run simulations without radiation as a control case. These simulations are denoted with the suffix C. We include our full radiative transfer scheme at 10\% and 100\% of the Eddington luminosity. These simulations are labelled with the suffixes RL (``low") and RH (``high"), respectively. We repeat runs C1-RH and C2-RH three times each to explore the influence of various components of the radiation field. In these models, we scale the radiation field to the Eddington limit and (i) only include X-rays  (ii) only include UV photons (iii) exclude radiation pressure. These simulations are labelled X, UV, and NORP respectively. A list of the radiation parameters used in our models is given in Table~\ref{tab:radiation}. We also repeat model C2-C twice at two and four times the resolution listed in Table~\ref{tab:SMR}. These models are named C2-128 and C2-256, respectively, and are used in the discussion of convergence and computational limits of our RHD simulations (see \S\ref{subsec:discsubstructure}).

\subsection{Disc Finding and Stability Measure}
\label{subsec:discfindstable}

We do not include self-gravity in our models, thus we cannot follow the evolution of formed gas discs to the point of star formation. This is partly due to the fact that our simulations are resolution limited, and we do not sufficiently resolve the Jeans length ($\lambda_J = (\pi c_s^2/ G\rho)^{1/2}$,  \citealp{1902RSPTA.199....1J}) of dense gas. Peak densities in our models reach $10^9 \  \text{cm}^{-3}$, thus, assuming an equilibrium temperature of $100$~K, the Jeans length of $6 \ \text{mpc}$ is roughly equal to the cell size on the highest refinement level. When considering additional constraints on the spatial resolution required for monitoring gravitational collapse in grid-based simulations (i.e. \citet{1997ApJ...489L.179T}) or threshold densities motivated by the tidal limit \citep{2008Sci...321.1060B,2013MNRAS.433..353L,2009MNRAS.394..191H} the required resolution is not computationally feasible with our radiation framework. 

As an alternative to directly monitoring disc fragmentation, we test for gravitational instability of formed discs in our models via an approximate measurement of the Toomre Q parameter \citep{1964ApJ...139.1217T}:
\begin{equation}
    \label{eq:toomreq}
    Q_\text{T} \approx \frac{c_s \Omega}{ \pi G \Sigma} \ ,
\end{equation}
where $c_s$ is the sound speed of the gas, $\Omega$ is the orbital frequency, G is the gravitational constant, and $\Sigma$ is the surface density of the disc. A Keplerian disc is subject to gravitational instability for values of $Q_\text{T} < 1$.

Because of the random structure of the inflow, we cannot perfectly predict the orientation or extent of formed gas discs. Therefore, we have implemented a parallelized disc finding routine that is detailed in Appendix~\ref{app:discfind}. In short, the routine uses search volumes with radii of 0.25 pc, 0.5 pc, 1 pc, and 2 pc centered on the SMBH to calculate the net angular momentum of enclosed gas. A disc is found if a significant fraction of the mass within the search volume has an angular momentum vector roughly parallel to the net angular momentum vector. The routine then interpolates the computational mesh to a frame aligned with the net angular momentum vector, preserving the mesh refinement when possible. We exclude gas that does not have a parallel angular momentum vector from this interpolation step in order to isolate the disc. To extract only the dense central component of the disc, we also exclude gas with a mass density $\rho < 10^{-3} \rho_\text{peak}$, where $\rho_\text{peak}$ is the peak mass density within the search volume. We run this routine at 1 kyr intervals throughout the simulation. We then use the interpolated disc frame mesh to calculate the mass-weighted average sound speed ($\bar{c_s} = \sum (c_s \rho \Delta x^3)/\sum (\rho \Delta x^3)$), average surface density, average orbital frequency, and the total mass of each disc. These values are then inserted into Eq.~\ref{eq:toomreq}. To avoid redundant measurements of the disc at each time interval, we only retain values for the smallest search volume that contains at least 99\% of the total disc mass. Our values of $Q_\text{T}$ should be treated as conservative estimates as they do not account for the radial dependence of any of the quantities which enter into the Toomre Q calculation. 

\section{Simulation Overview} \label{sec:simresults}
\subsection{Dynamics}

\begin{figure*}
    \includegraphics[width=\linewidth]{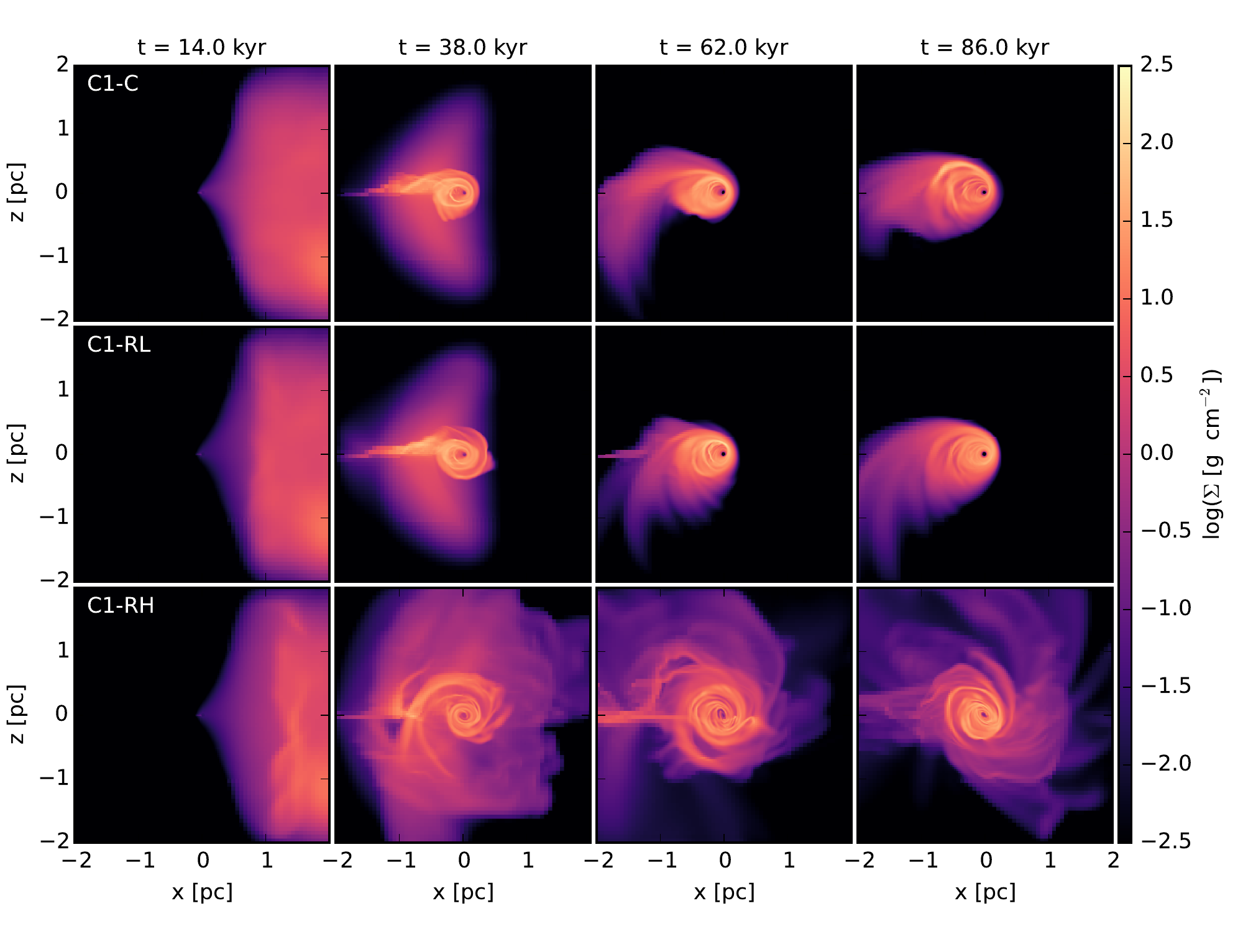} 
    \caption{ Gas column density perpendicular to the direction of inflow in a time sequence of models C1-C (top), C1-RL (middle) and C1-RH (bottom). A SMBH of $\text{M}=4\times10^6~\text{M}_\odot$ sits at the origin. The models are characterized by radiation fields that increase in strength from top to bottom (L = 0, 0.1 L$_\text{edd}$, L$_\text{edd}$). The leftmost panel demonstrates the effect of radiative compression occurring during inflow, which increases in proportion to the radiation field strength. The progression of disc formation and settling is shown from left to right. While there is similarity between the low radiation field case (C1-RL) and the control model (C1-C), radiation inhibits inflow and results in a more extended gas disc at the Eddington luminosity (C1-RH).} 
    \label{fig:dynamics1}
\end{figure*}

Figure~\ref{fig:dynamics1} shows the time evolution for models C1-C, C1-RL, and C1-RH. In the control model, C1-C, inflowing gas is gravitationally focused by the SMBH, leading to stream collisions and partial angular momentum loss. From $\approx 10-30$~kyr, residual angular momentum from collisions leads to the onset of disc formation. From $\approx 30-70~\text{kyr}$ the disc continues to accumulate mass from a post-collision stream. Angular momentum accretion only mildly alters the orientation of the disc over time. An eccentric disc forms within $\approx 1$~pc of the origin (see Sec.~\ref{subsubsec:discecc} for details on eccentricity). High surface density streams form in the central 0.5 pc, but these features are transient due to strong shearing. 

The time evolution of C1-RL ($L = 0.1 L_\text{edd}$) is nearly identical to C1-C with only minor morphological differences. During inflow, competing forces from radiation and the gravitational pull of the SMBH compress the irradiated face of the inflowing gas, leading to a build-up of mass at x~$\approx$~1~pc. Stream collisions provide sufficient angular momentum loss for disc formation from $\approx \text{10--30 kyr}$. The sub-structure of the formed disc differs from the control model because of additional angular momentum loss that is supplied via photo-compressions during inflow. The density of the disc is slightly elevated and is more uniform than in C1-C.

For C1-RH ($L=L_\text{edd}$), the influence of radiation is more pronounced. Evidence of photo-compression during initial inflow at $t = 14$~kyr extends beyond $x=1$~pc. The build-up of inflowing gas competing with the impinging radiation leads to higher density sub-structure. A period of disc building occurs from $t \approx \text{30--70 kyr}$, though radiation partially inhibits inflow. As a result, the forming disc is surrounded by a low density envelope of gas. Stream collisions continue to supply mass to the inner 1 pc, supporting a more gradual period of disc growth. This process continues through $\approx 62~\text{kyr}$. The central component of the disc eventually forms with similar extent to those seen in C1-C and C1-RL, albeit with a different orientation. Peak densities seen in C1-RH are lower than in the previous cases, likely due to the prolonged period of disc building. 

                                                                                                      Models with the C2 (more massive) inflow structure follow this general sequence for both the dynamical evolution and radiation field strength. Yet, disc mass is increased due to both the more massive inflow and elevated substructure densities.

\subsection{Thermal Evolution}
\label{sec:thermalevol}

\begin{figure*}
    \includegraphics[width=\linewidth]{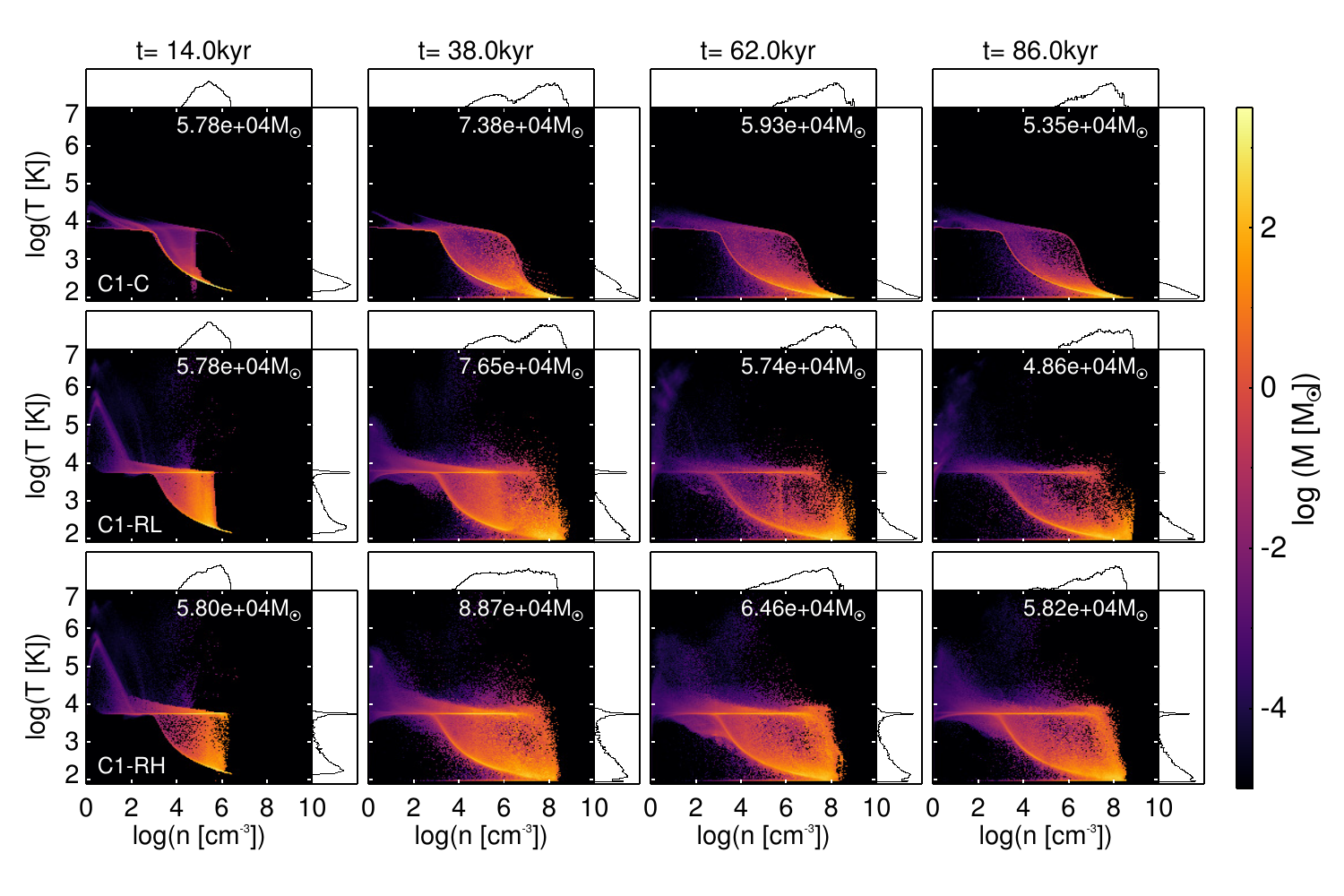} 
    \caption{Time evolution of the temperature vs. number density mass distribution for C1-C (top), C1-RL (middle), and C1-RH (bottom). Time increases from left to right with times identical to those shown in Figure~\ref{fig:dynamics1}. The one dimensional histograms along each axis show the total mass at each temperature or density, and are plotted logarithmically with a range extending from $10^2 \text{M}_\odot$  to $10^{4.2} \text{M}_\odot$. The total mass in each histogram is plotted in the top right of each image.}
    \label{fig:TN2D_cloud1}
\end{figure*}

We show the mass distribution in density-temperature space over time for C1-C, C1-RL and C1-RH in Figure~\ref{fig:TN2D_cloud1}. The times used match those in Figure~\ref{fig:dynamics1}. Note the presence of mass distributions in both density and temperature affixed to the appropriate axes on each panel. The histograms are logarithmically scaled and have bounds from $10^2~\text{M}_\odot$ to $10^{4.2}~\text{M}_\odot$. The top row shows the thermal evolution of C1-C. At $t=14$~kyr, inflowing gas has reached the central SMBH. A majority of the mass congregates at density-dependent equilibrium temperatures for neutral gas (i.e. $\Lambda_n = \Gamma_n/n_\text{HI}$) so that the equilibrium curve appears as a thin line which asymptotes to $5\times 10^3 \ \text{K}$ at low density and to $100$~K for high densities. A transition between temperature extremes occurs between  $n = 10^3 \ \text{cm}^{-3}$ and  $n=10^8 \ \text{cm}^{-3}$. Adiabatic pressure response ($P \propto \rho^\gamma$) to gravitational compression or shocks drives a fraction of the mass away from thermal equilibrium. 

As inflowing mass forms into a disc ($t\approx10-70~\text{kyr}$), a second peak emerges along the density axis, consistent with the high density central disc shown in the dynamical evolution of C1-C (Figure~\ref{fig:dynamics1}). At $t=86 \ \text{kyr}$, the mass distribution peaks at $\approx 10^8 \ \text{cm}^{-3}$. A transition from $\approx 5800 \text{K}$ to 100 K begins at $n = 10^6 \ \text{cm}^{-3}$. Above this density cooling rates are sufficiently high to maintain the neutral equilibrium temperature. In the absence of radiation, the onset of disc formation is evident in the depletion of the low density peak at $t=14$~ kyr and the growth of a high density peak. Adiabatic pressure response to stream collisions is reflected in the temperature histogram at t=38~kyr. Efficient cooling permits the formation of a gas disc near thermal equilibrium (T=few$\times$ 100~K), and the distribution of temperatures narrows as the gas settles.

The left column of Figure~\ref{fig:TN2D_cloud1} demonstrates how radiation affects the initial inflow. The irradiated face of the inflowing gas is ionized and photo-heated, thus both C1-RL and C1-RH show an increased amount of mass in the region between the equilibrium curve for neutral gas and the equilibrium temperature of 5803 K for a primarily UV-ionized gas. The prominence of the 5803~K peak in the temperature histogram (projected along the y axis) increases proportionally to the radiation field strength. Low density ($n<10^2 \ \text{cm}^{-3}$) gas in both C1-RL and C1-RH reaches temperatures in excess of $10^6$~K. C1-C lacks this feature due to the assumption of neutral gas which leads to overly-efficient cooling in low density gas. This effect is not likely of dynamical significance as the pressure of this gas falls orders of magnitude below that of both the disc and the surrounding high-density gas streams. Similarly, following the density histograms along the x-axis through the first column shows evidence of photo-compression as the distribution skews toward higher densities.

During the simulation, the mass fraction of ionized gas drops. This occurs for two reasons. First, the forming disc is extremely thin - only a small portion of the ionizing photons are intercepted by the intervening gas. Second, the high density of the disc results in strong shielding that restricts the incoming photons to the disc surface. The drop in ionized mass is most evident by comparing the first row of Figure~\ref{fig:TN2D_cloud1} (C1-C) to the second row (C1-RL). C1-C shows no evidence of ionization. This is by construction as we have assumed a neutral gas. In C1-RL, the 5803~K peak is clearly present during inflow due to ionization. Following the temperature histogram over time, one can see this peak diminish as the disc forms indicating a lower mass fraction of ionized gas. The same is true for the bottom row of the figure (C1-RH), though a larger mass fraction remains ionized post-disc-formation. 

More mass migrates to higher densities over time, eventually extending to $\approx 10^9~\text{cm}^{-3}$. Both models with radiation show a preference to higher densities than are seen in C1-C. This is likely a consequence of photo-compression occurring during the initial inflow which lowers the angular momentum of the gas and increases the density of inflowing clumps. The density distribution in C1-RH differs from C1-C and C1-RL at $t=\text{38 kyr}$ with a suppression of the high density peak indicative of the onset of dense disc formation. Although the formation of the disc is delayed in the presence of the strong radiation field, densities in C1-RH still manage to reach to $> 10^8\text{cm}^{-3}$. A peak at T~=~5803~K indicates that a fraction of ionized gas survives even after disc formation. In both radiation models, a turn off is seen above $10^8~\text{cm}^{-3}$, marking the density above which recombination rates are too high, and shielding too effective, for gas to remain fully ionized.

\section{Results} \label{sec:results}
In the previous section, we qualitatively demonstrated that gas inflow during AGN activity (here implemented as a static UV and X-ray radiation field) may still result in the formation of a central gas disc. The dynamical and thermal evolution of these discs suggests that radiation has two noteworthy effects (i) photo-heating and ionization drives a small fraction of the disc mass away from neutral thermal equilibrium. (ii) Radiation both inhibits the initial inflow of material resulting in an increase in disc density for weak radiation fields and a delay in disc formation for strong radiation fields.

Here, we consider the relative impact of X-rays (\S~\ref{subsec:xray}) and UV photons (\S\ref{subsec:uv}), and the dynamical influence of radiation pressure in disc formation (\S\ref{subsec:rp}). We discuss the effects of radiation on disc geometry and estimate the gravitational stability of formed gas discs in our models via conservative measurements of the Toomre Q parameter (\ref{subsec:discform}). We summarize the effects of radiation on inflow kinematics, particularly with regard to the distribution of radial inflow velocities (\S~\ref{subsec:kinematics_feedback}). We discuss the mass accretion rates measured in our models and estimate the expected radiative feedback for such gas inflow assuming approximate viscous transport timescales for the unresolved accretion flow (\S\ref{subsec:massacc}). We consider the formation of gravitationally unstable structures within the disc by comparing densities within the disc to the tidal stability limit (\S\ref{subsec:discsubstructure}). We lastly speculate about star formation rates and efficiencies for gas discs in our models (\S\ref{subsec:SFRs}).

\subsection{Effects of X-Rays} \label{subsec:xray}
\begin{figure}
    \includegraphics[width=\linewidth]{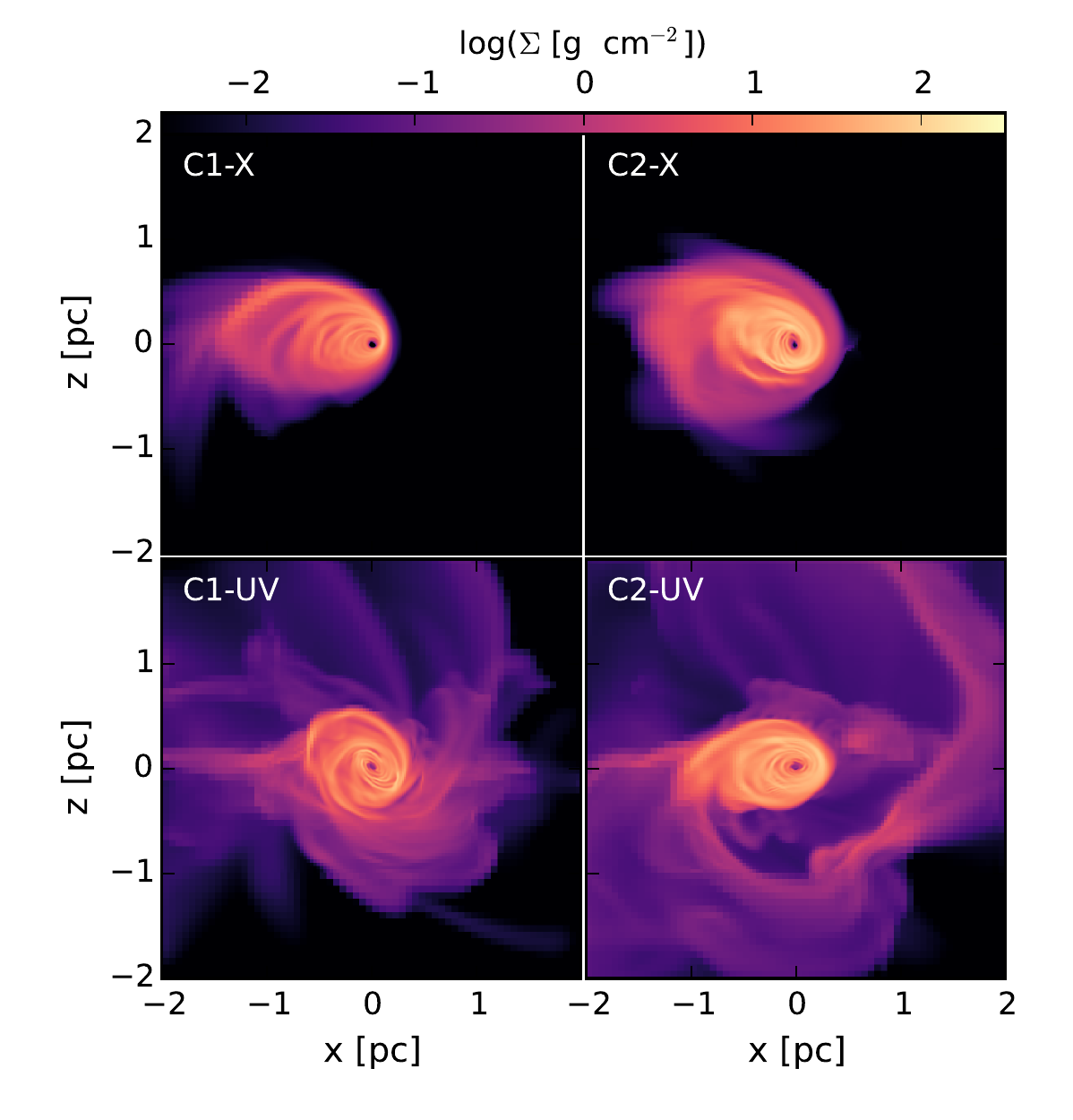} 
    \caption{Column density perpendicular to the inflow direction for models C1-X, C2-X, C1-UV, and C2-UV. All snapshots are taken at t = 86 kyr.}
    \label{fig:UVXRAYsnap}
\end{figure}

\begin{figure}
    \includegraphics[width=\linewidth]{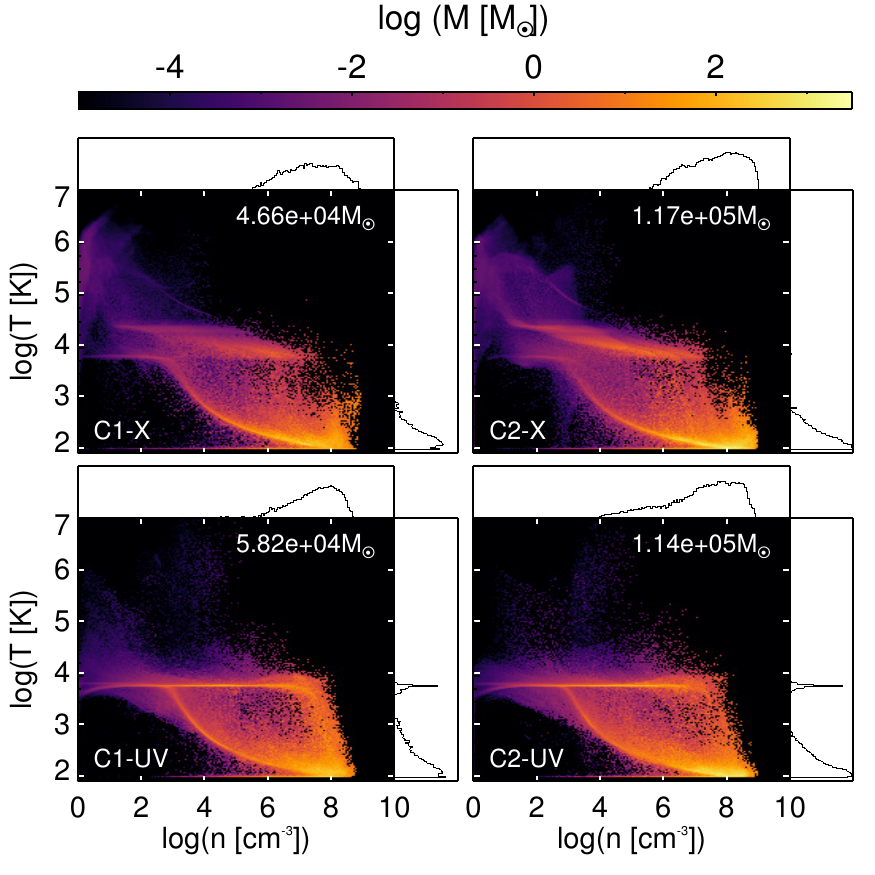} 
    \caption{Temperature vs. number density mass distribution for C1-UV, C2-UV, C1-X, C2-X at $t=86$~kyr. The one dimensional histograms along each axis show the total mass at each temperature or density and are plotted logarithmically with a range extending from $10^2 \text{M}_\odot$  to $10^{4.2} \text{M}_\odot$. The total mass included is plotted in the top right of each figure.}
    \label{fig:UVXRAY_TN_C1}
\end{figure}

In the top row of Figure~\ref{fig:UVXRAYsnap}, we show the column density of the disc in models C1-X and C2-X for which the radiation field is set to the Eddington limit and UV photons are removed. The discs are morphologically similar to models without radiation (Figure~\ref{fig:dynamics1}), suggesting that X-rays do not dramatically affect the formation of the central gas disc. The top row of Figure~\ref{fig:UVXRAY_TN_C1} shows the temperature-density distribution for these models at the same time. The mass distribution along the density axis is most similar to the control models, though peak densities are higher. Between $n=10^4~\text{cm}^{-3}$ and $n=10^7~\text{cm}^{-3}$ mass congregates at $T=10^4$~K. At low densities, temperatures exceed $10^6$~K. As the equilibrium temperature of an X-ray ionized gas is $\approx 1.267 \times 10^6~\text{K}$, the lack of mass in this temperature regime indicates that X-rays only partially ionize the gas. 

To understand the lack of dynamical influence by high energy X-rays, we can characterize the radiation field in terms of the ionization parameter, $ U_\text{ion} = Q_\text{ion}/(4\pi r^2 c n_\text{H} $), which serves as a measure of the radiation field strength. First, we can consider the effect of X-rays during the initial inflow. Taking the photon emission rate to be $Q_\text{ion} = Q_\text{X} = 10^{51}~\text{s}^{-1}$, and assuming a distance of $\approx 1~\text{pc}$ and an average gas density of $10^5~\text{cm}^{-3}$ for the inflow, the ionization parameter is $3\times 10^{-3}$. As stated in \citet{2014MNRAS.443.2018N}, an ionization parameter of $10^{-2}$ is considered a ``low" radiation field in which the evolution of the irradiated gas is dominated by photo-evaporation. Therefore, the X-ray flux is not sufficient to significantly influence the inflowing gas via radiation driven outflow. Second, we can consider the impact X-rays have on the formed disc. For the average disc density of $10^8~\text{cm}^{-3}$, and a minimum distance of $r = 40~\text{mpc}$, which marks the disc's inner edge, the ionization parameter is equally low at $U_\text{ion} = 2\times 10^{-3}$. Yet, it is unclear to what extent the conclusions of \citet{2014MNRAS.443.2018N} generalize to nuclear disc structures. It is clear though, that the X-ray photon density is too low to strongly influence the evolution and structure of the disc. 

Our results differ from previous models that highlight the importance of X-ray driven compression in gas clouds at distances of $\approx 10$~pc from an AGN \citep{2010A&A...522A..24H,2011A&A...536A..41H}. This discrepancy may be caused by several factors. First, the circumnuclear distances considered in this work are over an order of magnitude lower than in previous models, thus the gas suffers from the effect of strong tides. Second, our inflow includes high density substructure for which the optical depth is sufficiently high to rapidly absorb X-rays. Given the photo-ionization cross-section for 1 keV photons of $\sigma_\text{pi} = 10^{-23}~\text{cm}^2$, the mean free path of a photon in gas clumps with densities of $n>10^6~\text{cm}^{-3}$ is $\approx$~30~mpc which is less than a cell size on the coarsest resolution level. Lastly, the density of the disc rapidly surges to values much greater than those considered in prior models. With an average density of $10^8~\text{cm}^{-3}$ the mean free path of X-ray photons drops to 0.3 \text{mpc}, or a fraction of the cell size on the highest refinement level. Alternatively, as a consequence of the low ionization parameter for X-rays, the St\"{o}mgren length ($l_s \approx \frac{Q}{4 \pi r^2} \frac{1}{\alpha_B n_\text{H}^2}$) within the disc is $\approx 5\times10^{-9}~\text{pc}$, also indicating that the flux is insufficient to penetrate into the disc. This effect is compounded by the fact that the disc is geometrically thin, limiting photon absorption.

\subsection{Effects of UV Photons}  \label{subsec:uv}
In the bottom row of Figure~\ref{fig:UVXRAYsnap}, we show the column density of the disc in models C1-UV and C2-UV for which the radiation field is set to the Eddington limit and X-rays are removed. These snapshots are taken at $t=86$~kyr. Both of these models show a dense central disc surrounded by low density streams extending to the edge of the computational domain. In the bottom row of Figure~\ref{fig:UVXRAY_TN_C1} we show the temperature-density distribution of theses models at $t = 86~\text{kyr}$. For both C1-UV and C2-UV, the mass distribution along the density axis is most similar to the full radiation models. For C1-UV, mass congregates at the 5803 K equilibrium temperature for a UV-ionized gas. Gas temperatures also extend upwards to 1000~K due to photo-heating on the surface of the disc. It should be noted that the high-temperature, low-density gas seen in the C1-X and C2-X is not present, confirming that this parameter space is only accessed via X-ray photo-heating. 
\begin{figure*}                                                                                       
    \includegraphics[width=0.9\linewidth]{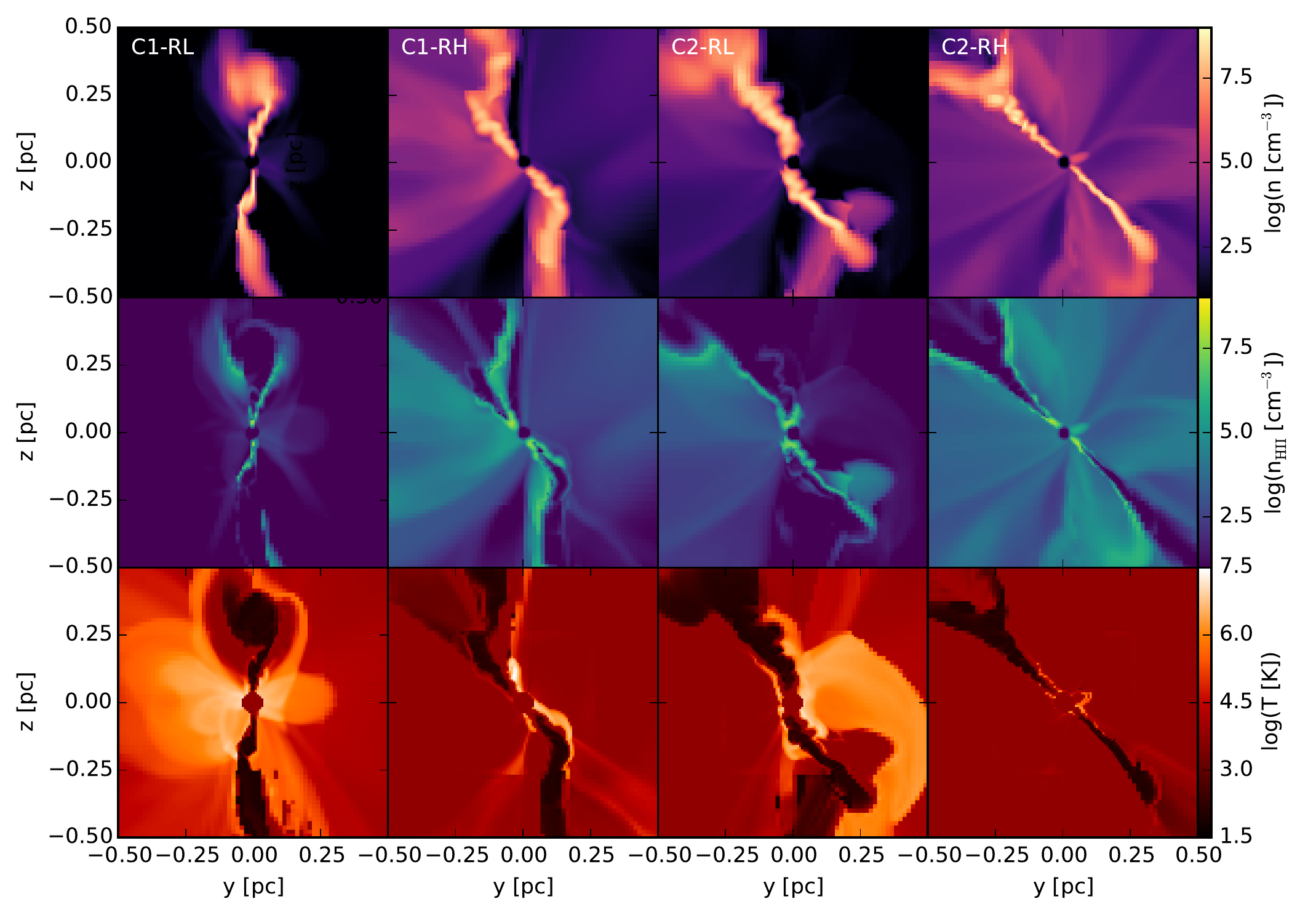} 
    \caption{Midplane slice in the y-z plane of the total density (top), the ionized hydrogen density (center) and the temperature (bottom) for the two inflow clouds C1 and C2, and the two radiation field strengths RL and RH. Snapshots are taken at $t=86~\text{kyr}$ and are zoomed into the second refinement zone.}
    \label{fig:shielding}
\end{figure*}

To quantify limits to the effects of UV photons, we follow the same arguments used for X-rays. First, considering the period of initial inflow, we take the photon emission rate to be $Q=Q_\text{UV} = 10^{55}~\text{s}^{-1}$, assume a density of $10^5~\text{cm}^{-3}$, and set the distance to $\approx 1$~pc. For these values, the ionization parameter is $\approx 30$. As stated in \citet{2014MNRAS.443.2018N}, this constitutes a ``high" radiation field where the role of radiation pressure becomes dominant, suppressing photo-evaporation. The structure of the initial inflow of C1-RH shown in Figure~\ref{fig:dynamics1} shows evidence of strong compression driven by radiation pressure. Given the considerably low ionization parameter of X-rays, it is clear that UV photons alone dominate the inflow structure. Second, assuming the disc density of $10^8~\text{cm}^{-3}$ and inner edge of 40~mpc, the ionization parameter for UV photons is $\approx 18$, suggesting that ionization and heating still occur on the inner portion of the disc. 

The Str\"{o}mgren length of UV photons within the disc is $\approx 6 \times 10^{-5}$~pc, thus strong shielding allows obscured portions of the disc to remain neutral despite the considerable radiation field. We demonstrate this effect in Figure~\ref{fig:shielding} where we show the gas number density, ionized gas number density, and temperature through the disc in C2-RL and C2-RH. In both cases, photo-ionization and photo-heating of the disc are restricted to the surface. Because the discs are partially warped, unobscured edges are also photo-ionized. Fully ionized, high-density gas is seen at the inner edge of the disc around $r\lesssim0.1$~pc. Beyond this ionized region, the midplane of the gas disc remains almost completely neutral and cool. Low density gas surrounding the disc is nearly fully ionized with the exception of regions shielded by the central disc.

\subsection{The Role of Radiation Pressure}  
\label{subsec:rp}
\begin{figure}
    \includegraphics[width=\linewidth]{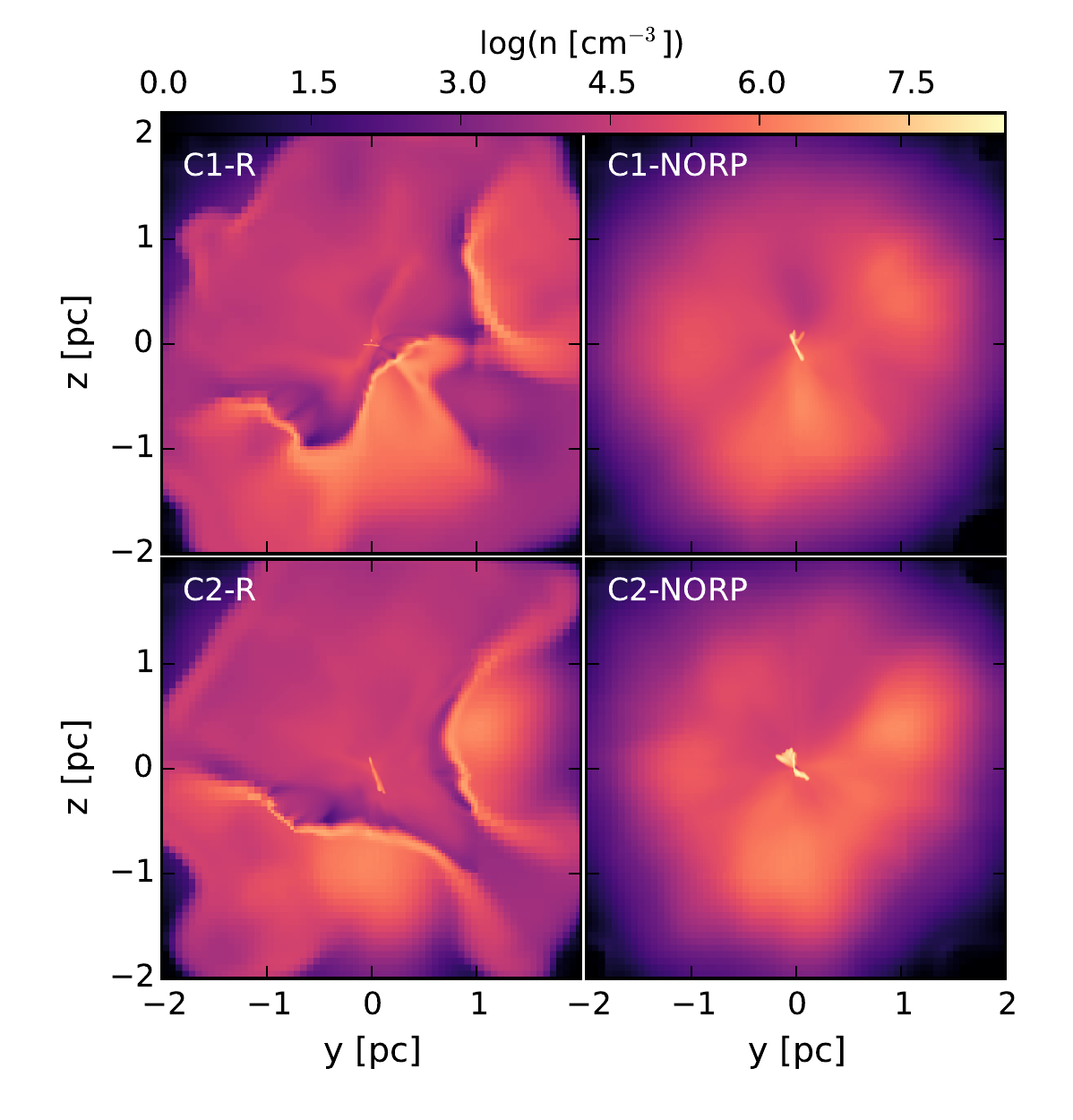} 
    \caption{Midplane slice of the number density anti-parallel to the inflow direction for C1-RH, C2-RH, C1-NORP, and C2-NORP. Snapshots are taken at $t = 28~\text{kyr}$. }
    \label{fig:radpressnap}
\end{figure}
In Figure~\ref{fig:radpressnap}, we show a midplane slice perpendicular to the inflow direction of the number density for models C1-RH, C2-RH, C1-NORP, and C2-NORP. In the absence of radiation pressure, disc formation is consistent with lower radiation fields. The forming disc can be seen in C1-NORP and C2-NORP within the central 0.5 pc. In contrast, C1-RH and C2-RH show evidence of a low density central disc. Streams of gas which approach the origin are photo-compressed. Only gas with sufficiently low angular momentum and sufficiently high column density with respect to the origin is able to continue the approach towards the central SMBH. As shown in the bottom row of Figure~\ref{fig:dynamics1}, streams of gas are forced to larger radii, delaying collisions, angular momentum loss, and the formation of a central disc.  
           
In contrast to both \citet{2014MNRAS.443.2018N} and \citet{2011MNRAS.415..741S}, radiation pressure does not completely disrupt the inflowing gas in our models. This is firstly because the inflow models considered here do not begin at rest, thus the gas is exposed to the radiation field for only a fraction of the time. Second, the average gas density in our inflow is roughly an order of magnitude higher than in previous models, thus shielding diminishes both the mean free path of photons and the Str\"{o}mgren length. We note that we do not include self-gravity in our models, thus we are unable to track the formation of filamentary structures in the photo-compressed gas as is seen in gas cloud models at larger radii. It is clear, though, that radiation pressure inhibits both the initial stages of gas disc formation and continued disc growth.

\subsection{Disc Formation}   
\label{subsec:discform}
\begin{figure}
    \includegraphics[width=\linewidth]{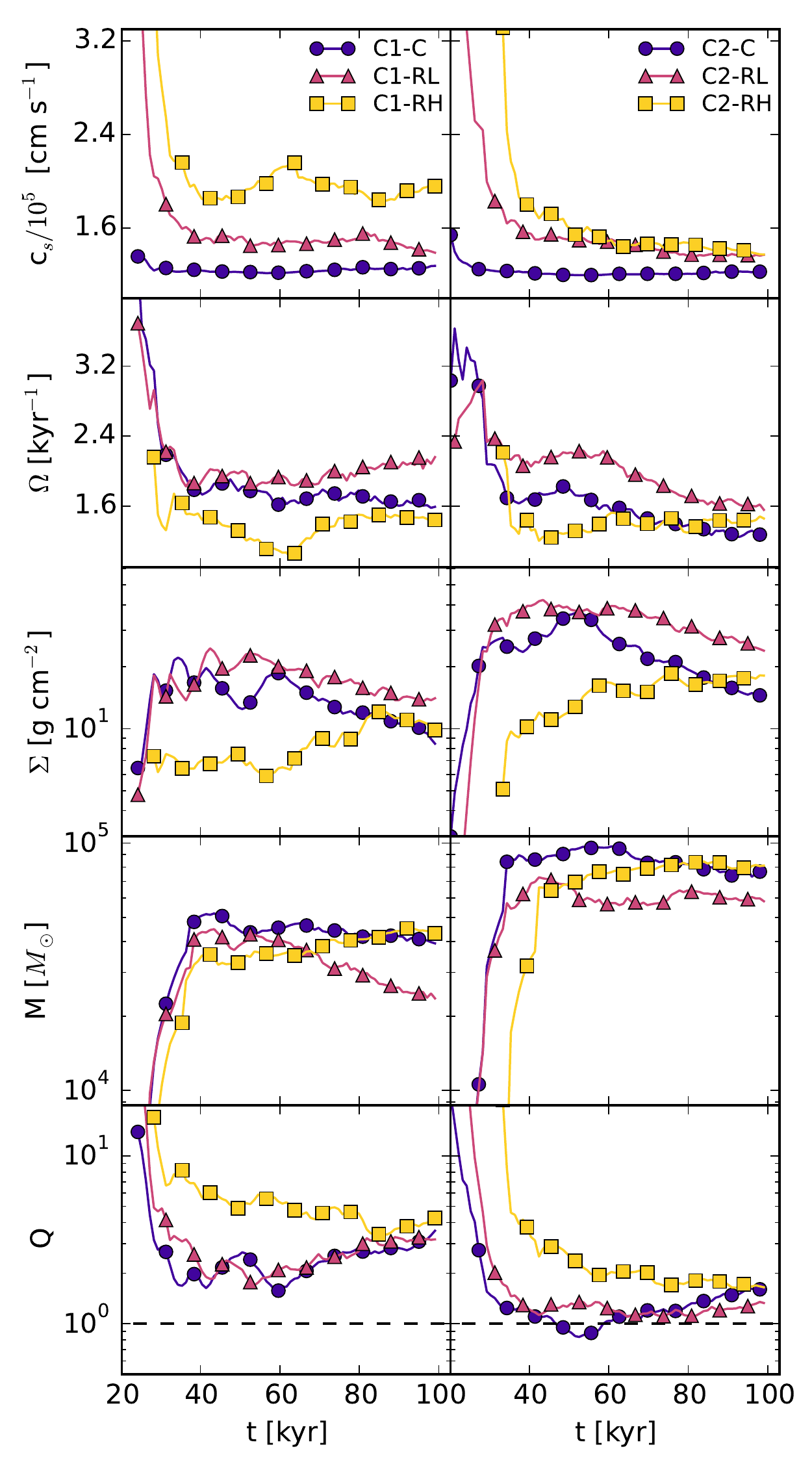} 
    \caption{Disc parameters and Toomre Q parameter ($Q_\text{T}$) for models with the C1 (left) and C2 (right) inflow conditions. The gas sound speed, orbital frequency, surface density, disc mass, and $Q_T$ are shown from top to bottom for models without radiation (purple circles), with a radiation field 10\% of the Eddington luminosity (pink triangles), and with a radiation at 100\% of the Eddington luminosity (yellow squares). }  
    \label{fig:discfindall}
\end{figure}

In Figure~\ref{fig:discfindall} we show the disc parameters and Toomre $Q$ parameter for all models. In C1-C, a disc is first found at $t \approx 20$~kyr. Over time, the sound speed of the disc remains roughly constant, indicating that a bulk of the disc material is in thermal equilibrium.  The surface density of the disc rapidly increases from 20--30~kyr which is reflected in a increase of total disc mass. Peak surface densities occur between 40--60~kyr, which results in $Q_T \lesssim 2$ during this period. The value of $Q_T$ does not drop below unity during the disc building phase, though our measurements of $Q_T$ are conservative averages and do not account for local density enhancements that are clearly present around the period of minimum $Q_T$ in Figure~\ref{fig:dynamics1}. 

For C1-RL, a disc is formed at nearly the same time as C1-C. The sound speed rapidly drops to values slightly above the equilibrium value for neutral gas, indicating that the gas disc is only partially ionized. The orbital velocity rises in time, which is reasonable considering the fact that the central disc is less extended at late times than in C1-C. The surface density is elevated with respect to C1-C likely due to the fact that angular momentum losses also occurs during the stream's initial approach. The elevated sound speed of the disc is compensated by the increased surface density, thus the values of $Q_T$ are comparable to the model without radiation. The disc evolution of C1-RH differs from the previous case. The surface density of the disc rises gradually over time, eventually approaching the values seen in C1-C at late times. The disc is characterized by $Q_T>3$ for the duration of the simulation. The sound speed also remains consistently higher because of photo-heating.

The disc parameters for the C2 (higher mass) inflow models are shown in the right panel of Figure~\ref{fig:discfindall}. In C2-C, the sound speed of the gas is roughly constant. The surface density increases rapidly from $20-60$~kyr, with a maximum occurring at $\approx 55~\text{kyr}$. For $t>60$~kyr, the surface density steadily decreases. At $\approx 55~\text{kyr}$, $Q_T <1$, indicating that such conditions may be gravitationally unstable. The resulting disc mass is roughly twice that seen in C1-C, as is expected for the higher mass inflow.

C2-RL shows both higher temperatures and surface densities with respect to the control model. As a result, the values of $Q_T$ follow the trend seen in C2-C. A period of $Q_T < 1$ is not seen for C1-RL, though $Q_T \approx 1$ from $t=40~\text{kyr}$ through the end of the simulation. The disc mass in C2-RL is suppressed with respect to C2-C. The sound speed in C2-RH evolves similarly to C2-RL, and is only slightly larger. This is likely due to the fact that the disc which forms in this model is extremely thin (Figure~\ref{fig:shielding}), very little disc mass is exposed to the radiation source. The surface density of the disc steadily increases over time, giving rise to a central disc with comparable mass to C2-C. The lack of a rapid initial disc building phase results in $Q_T >1$ for the entirety of the simulation. The downward trend in $Q_T$ may indicate a delayed period of $Q_T < 1$ beyond the simulation time.

\subsubsection{Disc Eccentricity}\label{subsubsec:discecc} 
Our models do not include passive particles which would be ideal for reconstructing the time-resolved orbits of individual gas parcels within the formed accretion disk. As an alternative approach to probing the geometry of the resulting gas disk, we use the following procedure to estimate the instantaneous disc eccentricity: (i) We randomly select one thousand points within the computational domain. (ii) Streamlines are calculated from these initial points using a forward Euler integration with trilinearly interpolated velocities at each iterated position. We restrict the timestep so that the distance travelled at each iteration does not exceed 25\% of the local cell size. (iii) We remove streamlines that do not at least partially trace the disc structure. To do this, we first exclude streamlines that do not trace a full 2$\pi$ in the orbit plane. We also remove streamlines that fail to intercept regions with densities in excess of $10^7 \text{cm}^{-3}$. If either of these exclusion criteria are met, a new random starting point is selected.

(iv) Each streamline is rotated into a frame in which the net angular momentum vector of all iteration points along the streamline is perpendicular to the x-y plane. (this process is similar to the rotation used in our disc finding method outlined in Appendix~\ref{app:discfind}). Although the streamlines may be warped and will have vertical extent in this rotated frame, we ignore the z dimension for the purpose of calculating the eccentricity. We use a least squares estimate following \citet{Fitzgibbon:1999:DLS:302943.302950} to determine the properties of the fit ellipse. Streamlines which result in a least squares fit error $>10^{-3}$ or result in a eccentricity $>$ 1 (hyperbolic) are excluded and a new streamline is calculated. (v) The streamlines are organized into bins according to the corresponding semi-major axis of the fit ellipse. (vi) Eccentricities are averaged in each bin. We repeat this process for every output timestep for each of our six simulations. 

\begin{figure}         
    \includegraphics[width=\linewidth]{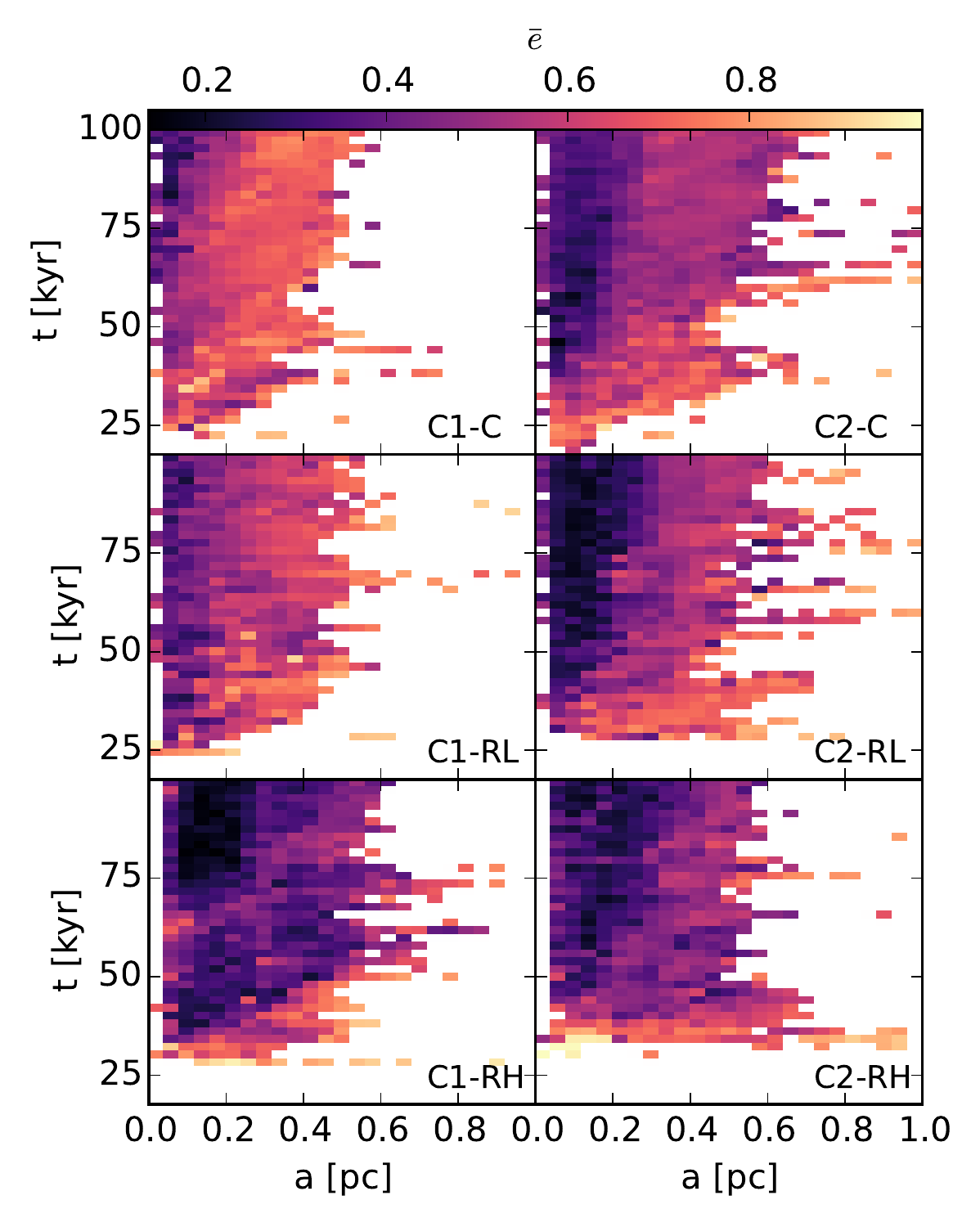} 
    \caption{Average disc eccentricity histogram vs. semi-major axis. vs. time. Each panel shows the eccentricity history of the forming accretion disc over time (model indicated in bottom right of each panel).  We note that the eccentricities shown are not time-resolved (orbits of individual gas parcels) but are instantaneous (flow lines within the disc). In general, the average eccentricity decreases with increasing radiation field. Furthermore, orbits tend to be more circular for lower semi-major axes.}
    \label{fig:eccstream_all}
\end{figure}

The results of this analysis are shown in Figure~\ref{fig:eccstream_all}. In both control models, disc-like orbits are captured beginning at t~$\approx 52\text{kyr}$. Initially, the orbits are highly eccentric. As stream collisions take place and the disc builds in mass, circularization takes place. At late times, two distinct orbit populations manifest. For C1-C the average eccentricity at r~$\lesssim 0.3$ is $< 0.5$. At larger radii the eccentricity tends to increase to $\approx$ 0.8. This behaviour is also exhibited in C2-C, though the peak eccentricity values are lower, presumably due to the increased efficiency of angular momentum loss through collisions for the higher mass inflow.

An increase in the radiation field strength reduces the angular momentum of both the inflowing gas and the resulting accretion disc. For our models with radiation, the eccentricity values are typically lower than in the control case. Again, the presence of two structures is clear at late times with a low eccentricity component enclosed by a high eccentricity gas stream. Note that in C1-RH and the C2 models, the average eccentricities at 0.2--0.3~pc are consistent with the low eccentricity ($\bar{\text{e}} \approx 0.3$) orbits of existing stars in the GC \citep{2009ApJ...697.1741B,2014ApJ...783..131Y}.

\subsubsection{Angular Momentum and Disc Orientation}
\label{subsec:disc_orientation}
\begin{figure}
    \includegraphics[width=\linewidth]{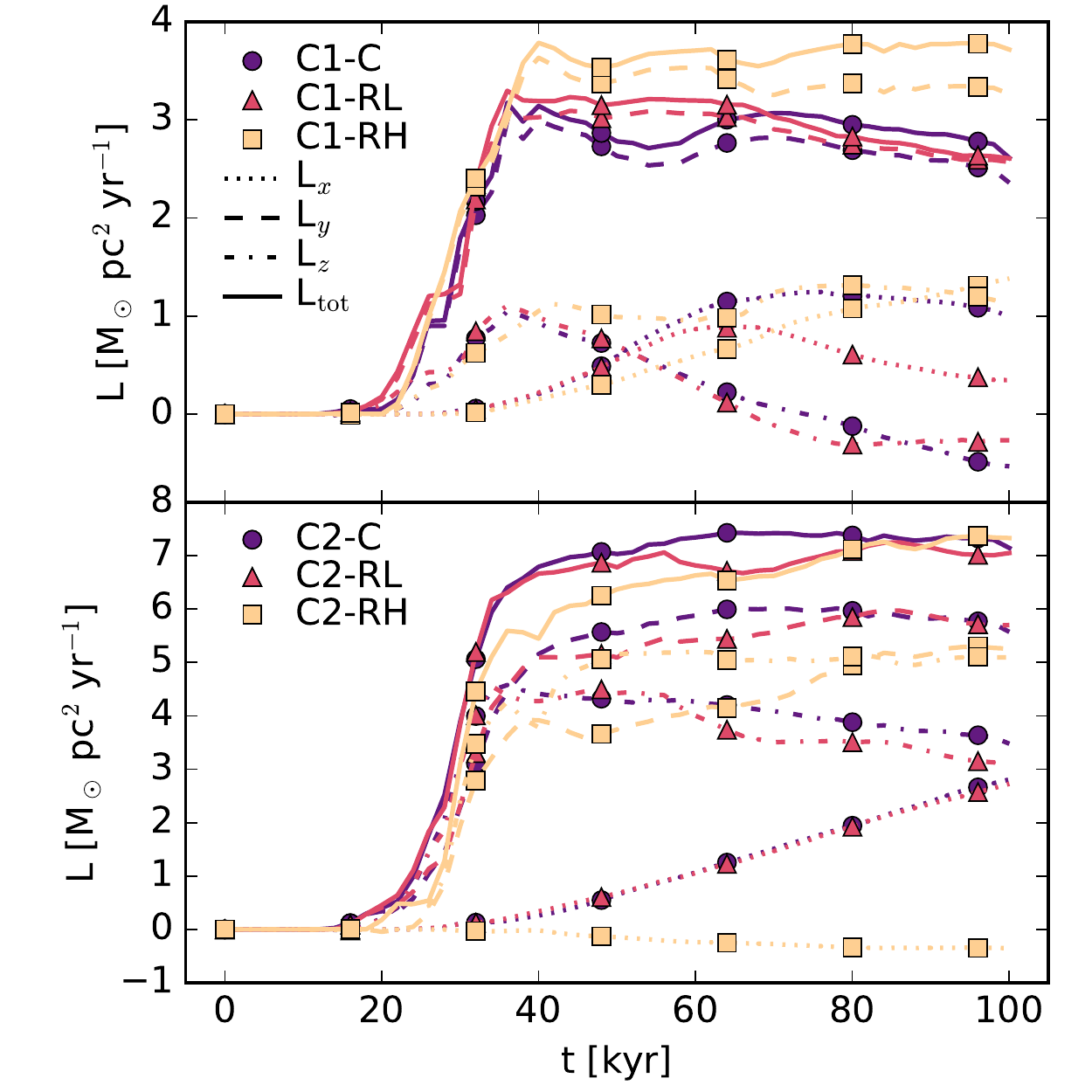} 
    \caption{Disc angular momentum over time for all models. The top panel shows angular momentum evolution of the disc for all C1 models, and the bottom shows the same for the C2 models. Solid lines show the net angular momentum of the formed disc. The components of the net angular momentum are indicated with dashed and dotted lines according to the legend in the top panel. }
    \label{fig:disc_angmom_t}
\end{figure}

We use the discs extracted from our disc finding routine (see \S~\ref{subsec:discfindstable}) to calculate angular momentum and disc orientation over time (Figure~\ref{fig:disc_angmom_t}). By comparing the total angular momentum of all models, we show that (1) the angular momenta of discs formed from the higher mass inflow are roughly twice that of the lower mass counterparts, and (2) the net angular momentum rapidly saturates after disc formation. The downward trend in the net angular momentum for both C1-C and C1-RL is consistent with the decrease in disc mass seen Figure~\ref{fig:discfindall}. C1-RH produces a disc with the highest angular momentum of the three models. This suggests that, although angular momentum is lost during inflow, the efficiency of angular momentum loss through stream collisions that follow also decreases. 

As discussed in \S~\ref{subsec:uv}, the effects of ionizing radiation are negligible once an accretion disc has formed. Strong shielding at the surface of the disc limits photo-ionization and photo-heating of the gas. Yet, it is clear that the dynamical evolution of the inflow is subject to change when the gas is exposed to a sufficiently intense radiation field. Due to the inhomogeneous structure of the inflow, compression and angular momentum loss are also non-uniform. As a consequence, the changes in the inflow alter the structure and orientation of the accretion disc that later forms. We show this variation in the component angular momentum between models in Figure~\ref{fig:disc_angmom_t}. The component angular momentum differ across radiation field strength. This can also be seen in the midplane slices of the radiation models in Figure~\ref{fig:radpressnap}. It is worth noting that neither the component nor the net angular momentum suffer rapid change after disc formation. Although this is not an indication that star formation will occur, the formation and fragmentation of dense structures within the disc would only benefit from such stability.

\label{subsubsec:DiscGrowth}

\subsection{Inflow Kinematics and Feedback}
\label{subsec:kinematics_feedback}
Numerical studies of gas dynamics under the influence of an SMBH detail the intimate relationship between various feedback mechanisms and the structure of inflowing material \citep{2014MNRAS.441.3055B,2011MNRAS.412..469A,2017MNRAS.468.4956Z,2015MNRAS.451.3627Z}. In particular, \citet{2014MNRAS.441.3055B} show via a deceleration argument that ram pressure $P_\text{ram}$ in an active nucleus can accelerate gas, driving material from the black hole. Yet, along lines of sight with column densities in excess of a critical value $\Sigma_\text{cr} = P_\text{ram}t/v_0$, gas infalling at $v_0$ overcomes this repulsive force to restore inflow. Such a phenomenon cannot be expected for radiation fields explored in this work due to the fact that the Eddington limit is derived from the balance of radiation pressure and gravitational acceleration. In the absence of strong inflow conditions, one could expect to see this balance. Even more so, for radiation fields in excess of the Eddington limit, a direct analog to the critical column density achieved from overcoming ram pressure could be found. Yet since neither condition is met in our models, we defer modelling of this behaviour to future studies and continue into a comparable analysis of the inflow structure exhibited in our models.

\begin{figure}
    \includegraphics[width=\linewidth]{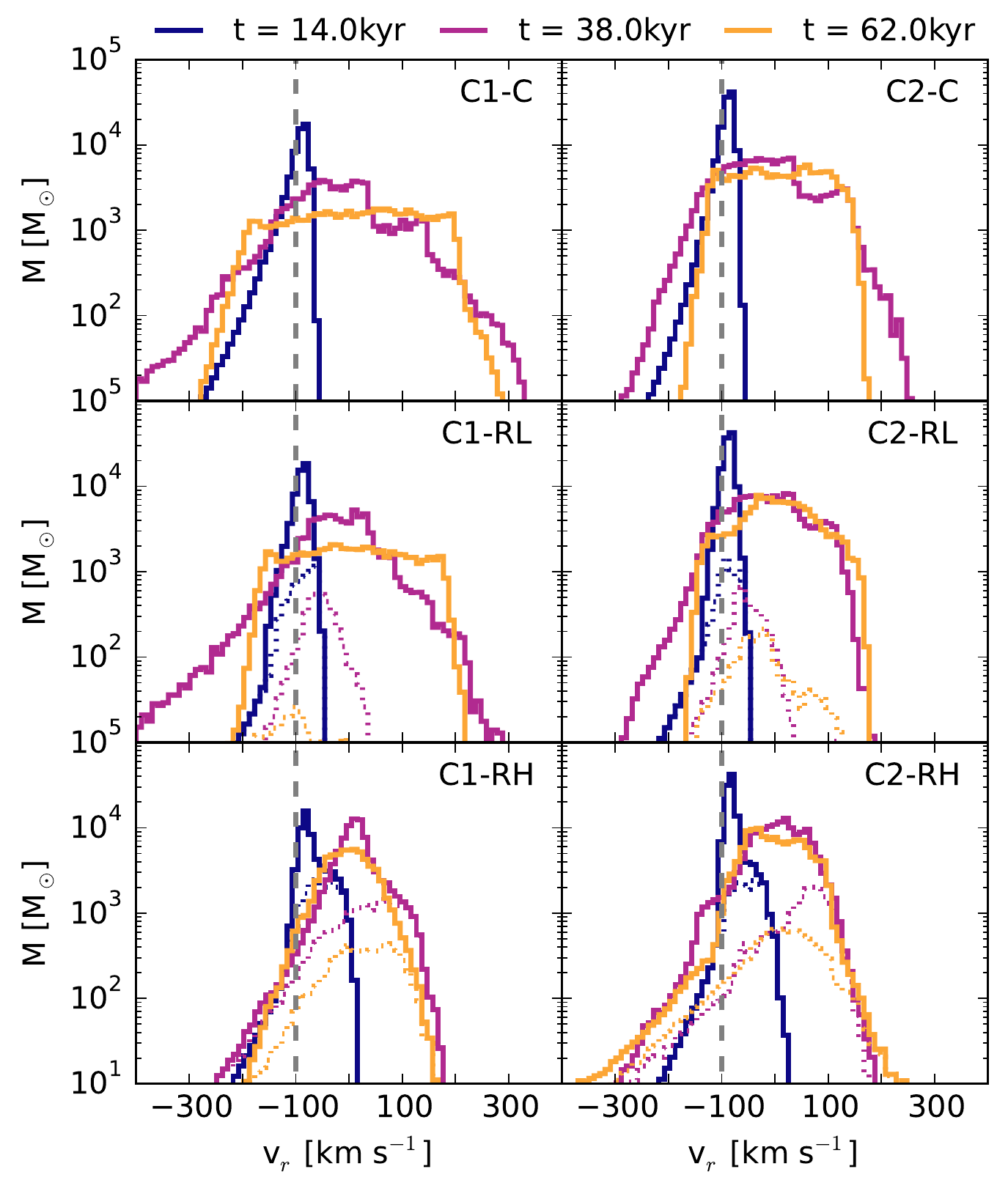} 
    \caption{Mass distribution of radial velocities over time for all models. The vertical dashed line indicates the magnitude of the initial inflow velocity of the gas. Solid lines are used to indicate the total mass in each bin. Dashed lines show the distribution of ionized mass. Radiation strength increases along each column,and the simulation name is indicated in the top right of each panel. Note that the mass fraction of ionized gas increases with radiation field strength. In general the total mass distribution does not vary strongly along the columns. The distribution is notably narrower at late times for both strong radiation field models (C1-RH and C2-RH), consistent with angular momentum loss occurring during inflow. Note that ionized mass tends to higher, positive radial velocities with increased radiation field.}
    \label{fig:vradhist_all}
\end{figure}   

\begin{figure*}
    \includegraphics[width=\linewidth]{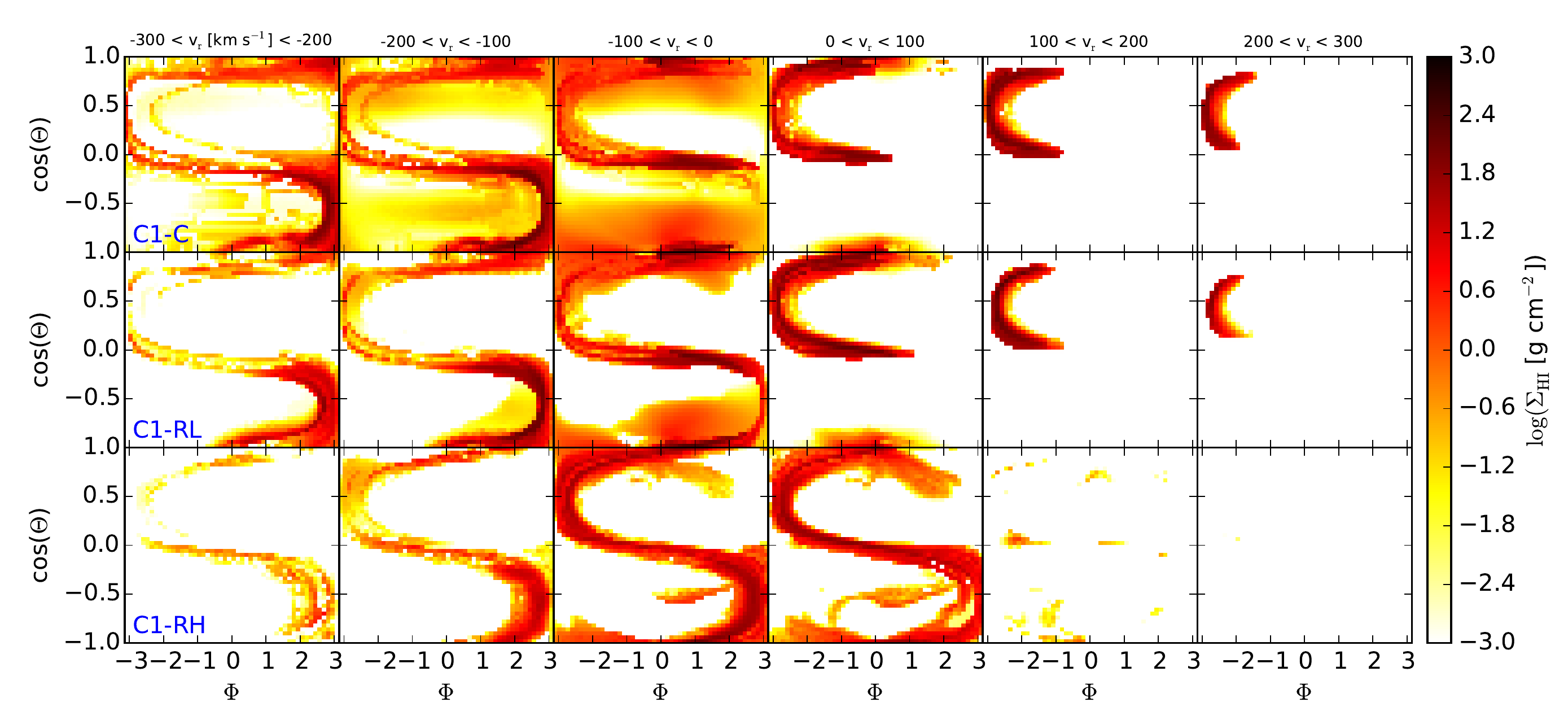} 
    \caption{Neutral gas column density from the vantage of the SMBH binned in radial velocity channels for all C1 models. Column densities are computed at t = 38~kyr, well after an accretion disc has formed. Velocity channels are listed at the top of each column with inward motion to the left and outward motion to the right. Bin limits are reported in km s$^{-1}$ as indicated above the first column. Radiation field strength increases along each column from top to bottom. The accretion disc appears as an ``S-shaped" high column density feature. The top row shows results from a model without radiation (ionization), thus low neutral column densities are present.}
    \label{fig:sigma_vrad_frames}
\end{figure*}

In Figure~\ref{fig:vradhist_all} we show the distribution of radial velocities for all models, using three times to showcase evolution throughout the simulations. A vertical line in each panel shows the initial inflow velocity as reference. As the gas stream approaches the SMBH it is accelerated, giving rise to a negative tail that is seen in C1-C and C2-C. As the radiation field increases in strength, this tail is suppressed. This is most evident in C1-RH and C2-RH where a few thousands solar masses of gas approach zero velocity at early times. This is consistent with the dynamical evolution seen in Figure~\ref{fig:dynamics1} where the leading edge of the inflow is swept up by the radiation field. 

As time progresses and the collision with the SMBH occurs, a distribution of radial velocities manifests between -300 km s$^{-1}$ and 300 km s$^{-1}$. At 10\% of the Eddington limit, the gas dynamics are largely governed by the dominant gravitational field of the SMBH, thus C1-RL and C2-RL roughly follow the control models. In the case of the lower mass inflow, there is evidence of angular momentum loss during inflow that manifests itself in a slightly lower mass fraction with positive radial velocity at t = 38 kyr. At late times, when the disc has formed, the distribution of radial velocities narrows with increasing radiation field which can be attributed once more to angular momentum loss occurring during inflow. This behaviour is only amplified for C1-RH and C2-RH in which the acceleration of the inflow is heavily suppressed at early times. The increased symmetry about $v_\text{r} \approx 0 \text{ \ km \ s}^{-1}$ for C1-RH and C2-RH  agrees with the circularization of the accretion discs shown in Figure~\ref{fig:eccstream_all}.

Figure~\ref{fig:vradhist_all} also shows the distribution or radial velocities for ionized gas. During inflow for both C1-RL and C2-RL, the ionized mass accounts for a fraction of gas with larger (less negative) radial velocities due to radiative effects. The combination of forces from differential pressure from photo-heating and from radiation pressure serve to slow the gas. At later times, the ionized gas is mostly at negative radial velocities, indicating that only gas infalling to the SMBH is being affected by the radiation field. It is worth noting that the ionized mass fraction dramatically decreases by t = 62~kyr as columns of gas within the forming disc are sufficiently high to maintain a high neutral fraction. In contrast, at the Eddington limit, the ionized gas fraction tends towards positive radial velocities and does not diminish considerably over time. From Figure~\ref{fig:dynamics1} it is clear that the radiation field is pervasive in these cases, sustaining an low density envelope around the central disc. Figure~\ref{fig:shielding} also shows that the highly dynamic environment surrounding the disk is fully ionized.

In Figure~\ref{fig:sigma_vrad_frames} we show a radial velocity binning of neutral gas columns surrounding the SMBH for the C1 models at t = 38~kyr, shortly after disc formation. Column densities are computed via ray-trace along each line of sight from the SMBH. Both the radial velocities and neutral gas densities are tri-linearly interpolated from the computational grid at the midpoint of each ray segment along the ray trace. The top row of the figure shows the results for C1-C. The disc is clearly seen as an ``S-shape" that appears throughout the velocity channels. A low column density envelope of gas surrounds the disc in negative velocity bins. As the radiation field increases, these low column densities vanish due to ionization. This is consistent with the fact that the low density gas envelope surrounding the disc in C1-RH is nearly completely ionized. The radial velocity distribution for C1-RL echoes C1-C, consistent with the mass histograms shown in Figure~\ref{fig:vradhist_all}. It is worth noting that the column densities reported at both extremes are slightly lower for C1-RL. Yet, as the radiation field increases, the trend towards lower radial velocity magnitudes continues. For C1-RH, inflow along the disc structure still persists, though less so than in the lower radiation case. This is due to the fact that the formed disc in this model is more circular. 


\subsection{Mass Accretion}  
\label{subsec:massacc}
\begin{figure}
    \includegraphics[width=\linewidth]{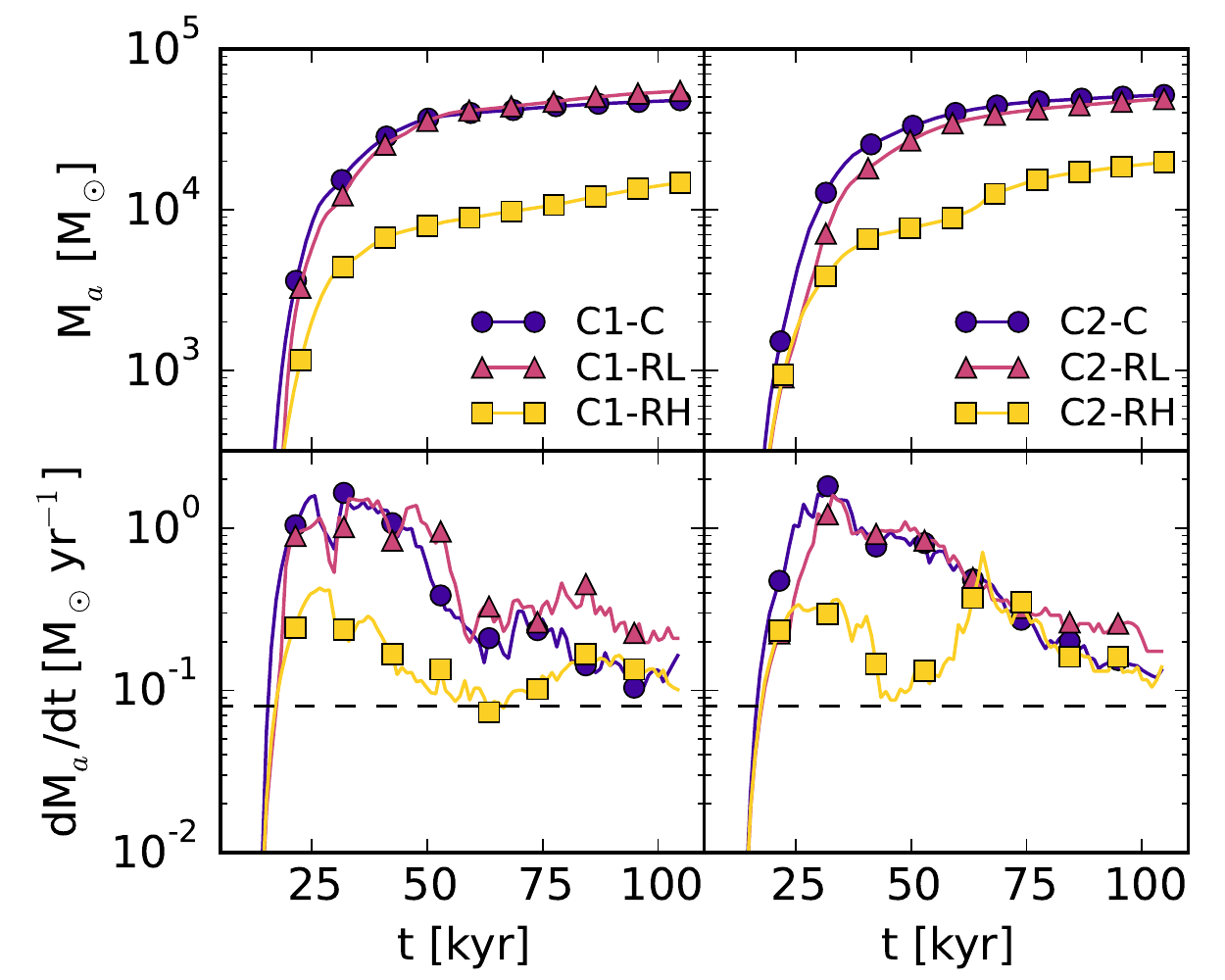} 
    \caption{Total mass accreted (top) and mass accretion rate (bottom) versus time for models with the C1 (left) and C2 (right) inflow conditions. The horizontal dashed line in the bottom panels is set at 0.08 M$_\odot$ yr$^{-1}$ corresponding to the mass accretion rate required for the Eddington luminosity assuming a 10\% radiative efficiency.}
    \label{fig:massacc_all}
\end{figure}

As mass passes through the inner boundary of the computational domain, mass and momentum are slowly removed from the simulation as detailed in \S~\ref{sec:accretionboundary}. In Figure~\ref{fig:massacc_all} we show the total mass accreted over time for all models. The figure also shows the mass accretion rate which is averaged in 1~kyr intervals. The general trend is the same for both inflow conditions. The mass accreted in C1-C, C2-C, C1-RL, and C2-RL grows throughout the simulation approaching an accreted mass of $\approx 5\times10^4~\text{M}_\odot$, corresponding to 30--50\% of the inflow mass. For both C1-RH and C2-RH, the total accreted mass is diminished with respect to the other models, yet it exceeds 10\% of the inflow mass in both cases. These values are in agreement with mass accretion reported for high mass misaligned-streamer models shown in \citet{2013MNRAS.433..353L}.

For all models, the mass accretion rate exceeds the Eddington rate of 0.08 $\text{M}_\odot \ \text{yr}^{-1}$ starting at t = 20 kyr, well before a dense central disc has formed. Peak accretion rates are over an order of magnitude greater than the Eddington rate, which is consistent with the maximum accretion rates expected for inflow driven by cloud-cloud collisions \citep{2009MNRAS.394..191H}. For low radiation fields, high accretion rates are reasonable given that disc formation is uninhibited. In simulations where disc formation is delayed, it follows that mass accretion would also be inhibited as mass is repelled from the origin.
\label{sec:massacc}
 
We use a constant radiation field, not accounting for the variability in radiative output expected to be driven by the accretion rate. This is not unlike previous work on the subject \citep{2010A&A...522A..24H,2011A&A...536A..41H,2014MNRAS.443.2018N,2011MNRAS.415..741S} in which the radiation field is assumed to be constant. Yet, in these models, gas clouds are not tracked to the point of the formation of an accretion disc, and in the most extreme case, the radiation source is removed by 50 pc. Our assumption of a constant radiation field may not be crippling for the following reasons. First, the accretion rates rapidly rise above the Eddington limit at $t=20$~kyr. At this time, gas is just beginning to collide with the SMBH. Stream collisions provide the necessary angular momentum loss for the onset of disc formation, though a disc is not detected until after the accretion rates have already risen above the Eddington rate. Second, as the disc is rather thin, shielding of the remaining gas inflow is nearly negligible.
                                                                              
\subsubsection{Viscous Accretion Timescales}
\label{subsec:viscous}
The accretion boundary of our simulations is set at $\approx~40~\text{mpc}$, roughly twice that used in smoothed particle hydrodynamic (SPH) models \citep{2008Sci...321.1060B}. Material which enters this boundary should have sufficient angular momentum to form an inner accretion disc which then feeds the central SMBH. The accretion rates measured in our models are therefore instantaneous values and should be considered as generous upper limits for the actual accretion rate onto the SMBH. Moreover, numerical accretion rates increase with the size of the accretion radius \citep{2011MNRAS.413.2633H}, leading to a further overestimate of our accretion rates.

To provide a more realistic estimate of the accretion rate, and resulting radiation strength, we consider the viscous timescale on which accretion is expected to occur:
\begin{equation}
    t_\text{visc}  = \alpha^{-1} \left( \frac{H}{R}\right)^{-2} \frac{1}{\sqrt{G M_\text{BH}}} R^{3/2} \ ,
\end{equation}
 where $H$ is the disc height, $R$ is the disc radius, and $\alpha$ is the viscosity parameter \citep{1973A&A....24..337S}. The value of $H/R$ can be approximated by rearranging the tidal stability criterion (Eq.~\ref{eq:toomreq}) to find $H/R \approx M_\text{disc}/M_\text{BH}$  \citep{2001ApJ...553..174G}. In our models, a disc mass of  $\text{few}\times 10^4 \text{M}_\odot$ yields $H/R \approx 0.01$. The value of $\alpha$ is less certain, but it is often assumed to be in the range of $0.01-1$. Again, our line of reasoning will lead to an overestimate of the accretion rate, since it assumes stability of the (subgrid) inner accretion disk. Yet, beyond a critical radius of $0.01$~pc for typical parameters, the disk is expected to become gravitationally unstable, leading to star formation rather than to
 increased angular momentum transport \citep{1989ApJ...341..685S,1999A&A...344..433C} and thus affecting the angular momentum transport either by removing gas mass into stars, or by expelling the gas via stellar feedback \citep{2016MNRAS.456L.109K}. 
 
Taking an inner disc radius of a few x 1000 AU, the viscous timescale for $\alpha = 0.01$ is $\approx 10^7~\text{yr}$. In comparison, the age of the central stellar disc in the GC is only a fraction of this value \citep{2006ApJ...643.1011P}, thus we would expect to see evidence of accretion resulting from this process if accretion occurs on this timescale. For a more liberal estimate of accretion with $\alpha=0.1$, the accretion timescale is $\approx 10^6~\text{yr}$. The accreted mass in our models ranges from $10^4 - 5\times10^4~\text{M}_\odot$. Spreading this accretion over a period of $10^6-10^7~\text{yr}$ yields accretion rates of 0.001 - 0.05~$\text{M}_\odot~\text{yr}^{-1}$, which is approximately 1\%--60\% of the Eddington limit. This agrees with the findings of both \citet{2008Sci...321.1060B} and \citet{2009MNRAS.394..191H}. Given these estimates, the range of radiation strengths used in our models is justified.

\subsection{Disc Substructure}
\label{subsec:discsubstructure}
\begin{figure}
    \includegraphics[width=\linewidth]{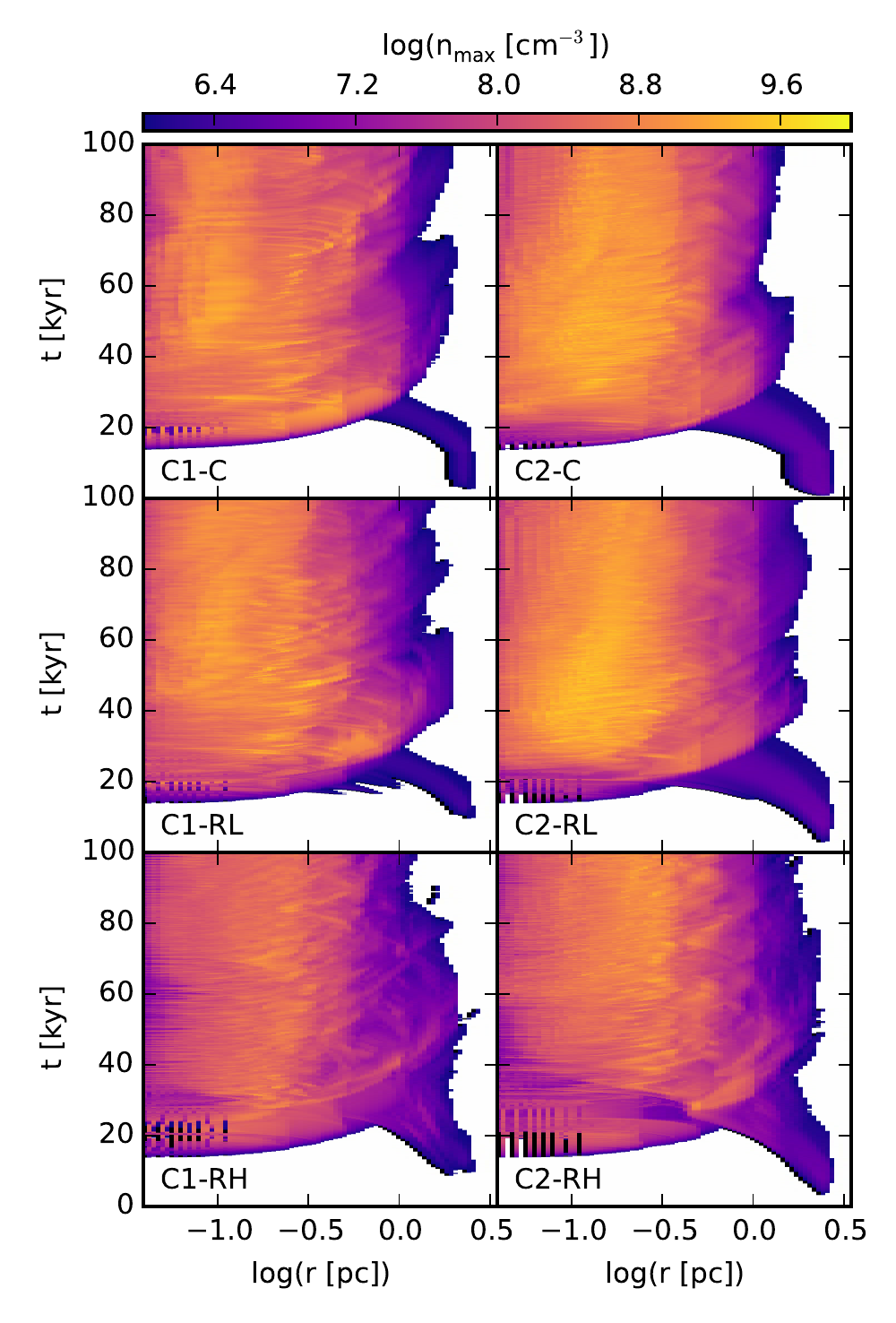} 
    \caption{Peak density distribution versus time and radius. The left boundary of each figure represents the accretion boundary, and the right edge corresponds to the maximum radius of the computational domain. Vertical lines in the image are a result of refinement boundaries. Vertical lines at log(r)~<~-0.75 and t~<~30 kyr are a result of the sparsely populated radial bins as a consequence of the logarithmic bin spacing.}
    \label{fig:multi_dens}
\end{figure}

For the duration of each simulation, we calculate peak densities in 80 logarithmically spaced radial bins extending from the accretion boundary to the corner of the simulation domain at 10 yr intervals. The results of this process are shown in Figure~\ref{fig:multi_dens}. Due to the logarithmic spacing, the number of grid cells available to the inner-most bins is low which causes streaks to appear for $r < 0.1$~pc at early times. Furthermore, due to the use of grid refinement, vertical lines can be seen in various locations where gas transitions from low to high grid refinement where compression and cooling of the gas is better resolved.  

For C1-C, the inflowing gas stream appears at $\log(r/\text{pc}) > 0$ for $t < 20$~kyr. The gas disc appears as a region with peak densities in excess of $10^8~\text{cm}^{-3}$ extending from the accretion boundary to $ \text{log}(r/\text{pc}) \approx -0.25$ for $t>20~$kyr. Peak densities are not contiguous due to strong shearing (Figure~\ref{fig:dynamics1}). The results of C1-RL are similar to C1-C, though peak densities are sustained at late times, consistent with the uniform surface density disc that manifests in this model. For C1-RH, peak densities are nearly an order of magnitude lower than in the previous cases, though they are also long-lived.

A disc is also present in C2-C, with the densest features between $\text{log}(r/\text{pc}) = -1$ and $\text{log}(r/\text{pc})=-0.5$ at late times. As the radiation field increases to 10\%  of the Eddington limit (C2-RL), peak densities increase in this same region. For C2-RH, the disc densities are lower than in the previous case but increase over time at $\text{log}(r/\text{pc}) \approx -0.5$ due to ongoing accretion. 

\subsubsection{Tidal Limit}     
\label{subsec:tidal_limit}
Though we do not include self-gravity and are unable to monitor the fragmentation of gas to the point of star formation, we can consider the tidal stability of dense gas structures approximately. From the tidal stability condition,
\begin{equation}
    \frac{G m_c}{R_c^2}  = \frac{2 G M_\text{BH} R_c}{r^3} \ ,    
\end{equation}  
we can determine the gas density required for a gas clump of mass $m_c$ and radius $R_c$ to remain self-gravitationally bound in the presence SMBH at a distance $r$:
\begin{equation}
\rho_\text{tidal} = 1.29 \times 10^{-16} \ \text{g} \ \text{cm}^{-3} \left(\frac{M_\text{BH}}{4 \times 10^6 \ \text{M}_\odot} \right) \left( \frac{r}{1 \text{pc}} \right)^{-3} 
\end{equation}
The corresponding number density, assuming a pure hydrogen gas, is 
\begin{equation}                            
    \label{eq:ntidal}
n_\text{tidal} = 7.73 \times 10^7 \  \text{cm}^{-3} \left(\frac{M_\text{BH}}{4 \times 10^6 \ \text{M}_\odot} \right) \left( \frac{r}{1 \text{pc}} \right)^{-3}   \ .
\end{equation}

\begin{figure}
    \includegraphics[width=\linewidth]{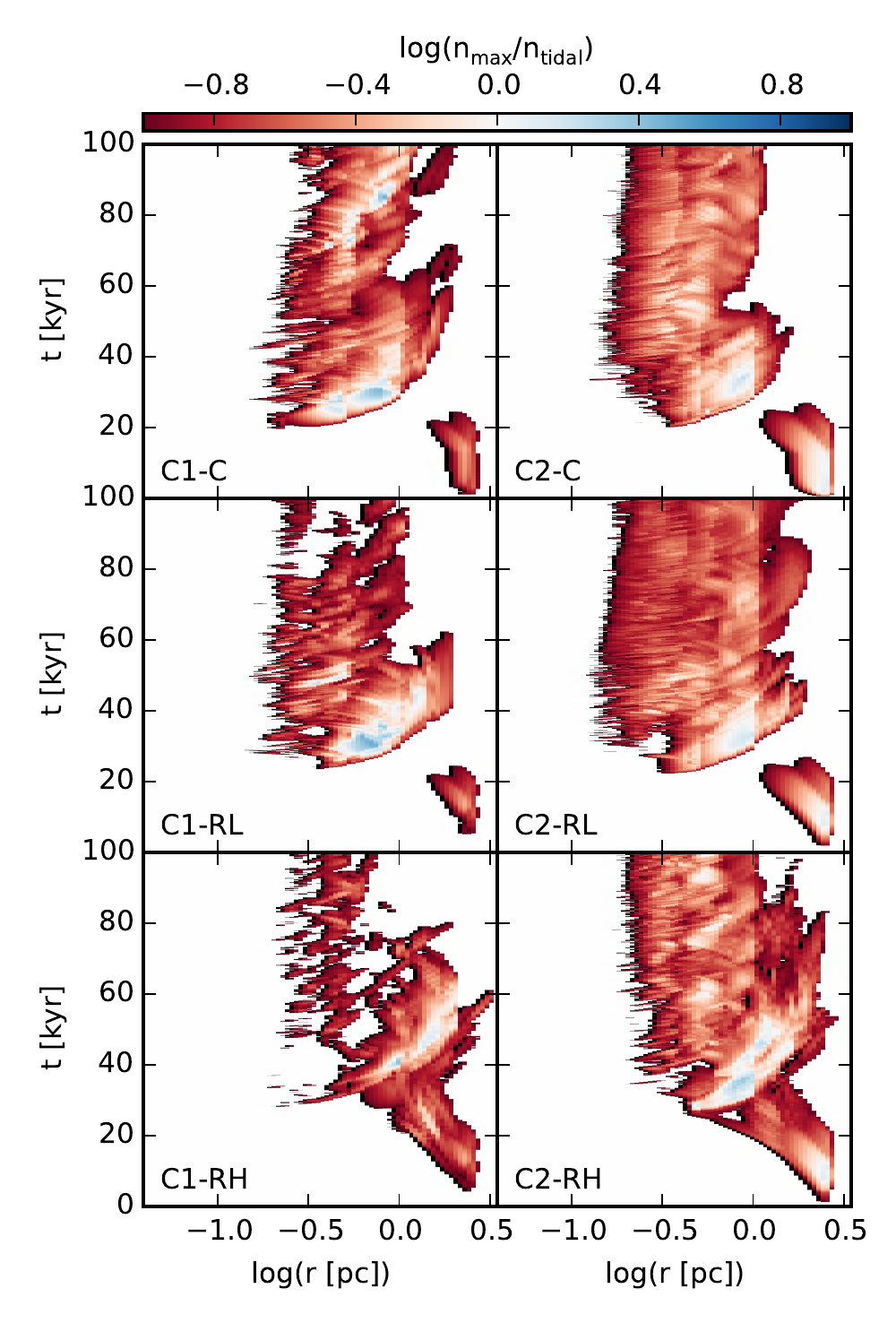} 
    \caption{Peak density distribution versus time and radius scaled to the radially dependant tidal density for both C1 and C2 models. The left boundary of each figure represents the accretion boundary, and the right boundary marks the maximum radius from the origin on the computational domain. Radiation strength increase from top to bottom, beginning with the control models. Vertical lines appearing in the image are caused by refinement zone boundaries.}
    \label{fig:multi_tidal}
\end{figure}

To compare the peak gas densities in our models to the radially dependent tidal limit, we divide the maximum densities seen in Figure~\ref{fig:multi_dens} by the corresponding tidal density using equation~\ref{eq:ntidal}. In Figure~\ref{fig:multi_tidal} we show the ratio of the peak densities to the tidal density for all models. We exclude bins with $n_\text{max}/n_\text{tidal} < 0.1$. The inflowing stream again appears in the lower right corner of each plot.  As gas first collides with the SMBH and stream collisions occur, strong compression and cooling drives the gas density above the tidal limit at $\text{log}(r/\text{pc}) \lesssim 0$ at $\approx 30$~kyr in all models. Densities above the tidal limit are not found for $\log(r/\text{pc})<-0.5$ in any model. Therefore the high densities seen in Figure~\ref{fig:multi_dens} in this region are well below the tidal limit.  

For C1-C, disc densities above the tidal limit are found at $\approx 80$~kyr. For C1-RL, a stream of super-tidal densities is seen around $\text{log}(r/\text{pc}) = -0.5$ at 50 kyr. As the radiation field increases to C1-RH, the gas disc is completely sub-tidal. Both C2-C and C2-RL show sub-tidal discs. As the radiation field increases, peak values migrate to larger radii. In C2-RH, periodic super-tidal values are seen at log(r/pc) $\approx$ -0.25 for $t>60$~kyr for C2-RH.

\begin{figure}
    \includegraphics[width=\linewidth]{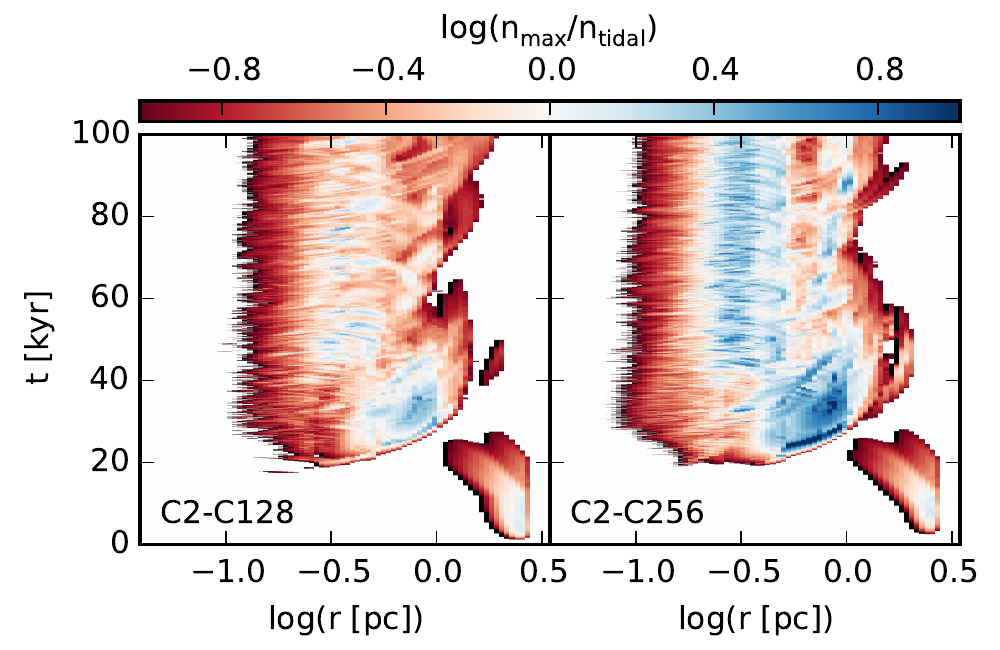} 
    \caption{Same as figure~\ref{fig:multi_tidal} for models C2-128 and C2-256.}
    \label{fig:multi_tidal_res}
\end{figure}

The lack of prevalent super-tidal densities in our models is a result of resolution limitations imposed by the exhaustive computational cost of radiative transfer. Because of this we are unable to resolve the cooling length at which dense gas cores are expected to form ($\lambda_\text{cool} =c_s \tau_\text{cool}$, \citet{2009A&A...508..725I}). The volume averaged densities in our models can therefore be treated as lower limits. To demonstrate this effect, we re-ran C2-C at base resolutions of $128^3$ and $256^3$ (i.e. two times and four times above our fiducial models). We show the peak densities relative to the tidal density for these test models in Figure~\ref{fig:multi_tidal_res}. As the resolution increases, the peak densities resulting from both stream collisions and disc formation increase. At a resolution of $256^3$, the central disc has super-tidal structures that are not present at the base resolution of $64^3$. However, these results are far from converged. For the average disc density of $10^8 \text{cm}^{-3}$, the cooling length is $\approx 20~\mu\text{pc}$, well below our resolution limit of $2.0$~mpc. The super-tidal densities seen at a resolution of $256^3$ are consistent with $Q_\text{T} < 1$ in these models. We cannot perform the same experiment for models with radiation because of current computational limitations. We expect that increasing the resolution will lead to higher densities as the cooling length is better resolved. This also increases shielding against destructive radiative effects leading to more pronounced differences in the evolution of diffuse (radiation-dominated) and dense (shielded) gas. The progression of fragmentation could be furthered by the presence of radiative losses \citep{2006ApJ...648.1052H}.

\subsection{Star Formation Rates} 
\label{subsec:SFRs}
The models presented in this work provide evidence that accretion-driven radiation at or below the Eddington limit does not quench the formation of dense gas discs in the immediate vicinity of an SMBH. Yet, the lack of self gravity in our models prohibits us from making definitive claims on any star formation occurring in such scenarios. In \S\ref{subsec:discform}, we show that the measured Toomre Q parameters for these discs are on the order of one, indicating that they may be gravitationally unstable. Here, we augment that analysis by considering approximate measures of star formation rates. 

Using the process outlined in Appendix~\ref{app:discfind}, we isolate formed discs in our simulations throughout time. For each extracted disc, we initially assume a constant star formation efficiency from which to compute the star formation rate density for each computational cell:

\begin{equation}
    \rho_\text{SFR} = e_\text{SFR} \frac{\rho}{t_\text{ff}} 
\end{equation}
where we have used the free-fall time of the gas,
\begin{equation}
    t_\text{ff} = \sqrt{\frac{3 \pi}{32 G \rho}}       \ .
\end{equation}
We integrate through the disc to compute the star formation rate surface density,
\begin{equation}
    \Sigma_\text{SFR} = \int_{z_\text{min}}^{z_\text{max}} \rho_\text{SFR} \ dz \ .
\end{equation}

\begin{figure}
    \includegraphics[width=\linewidth]{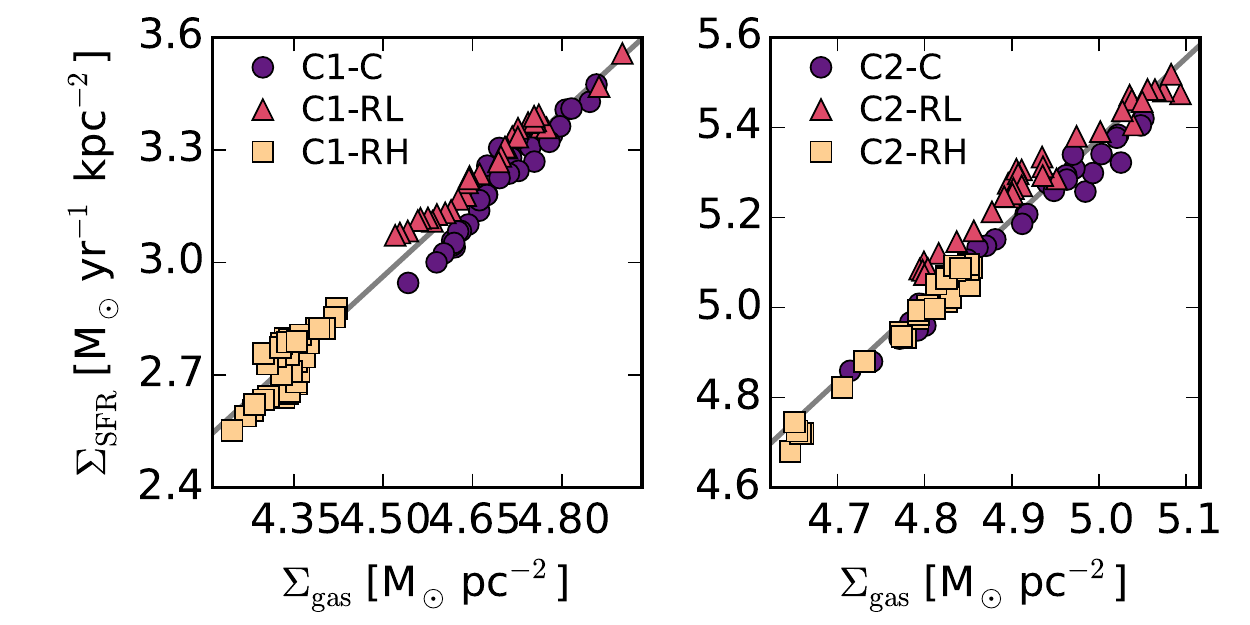}
    \caption{The mean star formation rate surface density vs. the mean gas surface density within extracted disks. The mean SFR surface density is calculated in a manner consistent with the gas surface density by averaging over all lines of sight perpendicular to the disk. The dashed lines assume a proportionality consistent with the Kennicutt-Schmidt (KS) law. For the low cloud mass (C1) models, the power-law index $n=1.45$. In the higher mass case (C2), the power law tends to $n=1.79$, breaking from observed values of the KS law ($n\approx1.5$). The implied star formation efficiencies are $0.02$\% for the C1 models and $1.0$\% for the C2 models.}
    \label{fig:SFR-relation}
\end{figure}

With the surface density of the disc and an estimate for the star formation rate surface density, we use the empirical KS relation \citep{1998ApJ...498..541K} to determine the star formation efficiency parameter, assuming that the proportionality constant for this relation is $2.5 \times 10^{-4}$. We determine the star formation efficiency by computing a linear fit in logarithmic space (see Figure~\ref{fig:SFR-relation}). Each fit yields an estimate for the star formation efficiency parameter and an exponent for the KS relation. For the C1 models, we find that the exponent is $n = 1.45$ and that $e_\text{SFR}=2\times 10^{-4}$. For the C2 models, we get the best fit values of $n=1.79$ and $e_\text{SFR}=10^{-2}$. Although the lower mass models are characterized by a slope consistent with the empirical value of $n=1.5$, the star formation efficiencies inferred by this fit are considerably lower than the average values of 0.01-0.1 reported in literature \citep{2018MNRAS.479.5544M,2012ApJ...761..156F,2013ApJ...763...51F}. For the higher mass models, the efficiency parameter falls within the expected range, though the slope is not consistent with the KS relation.  

It should be noted that the star formation efficiencies computed here are designed to force agreement with the KS relation; however, the scales considered in this study represent only the inner-most regions of the galaxies for which this relation is typically administered. Similarly, the ``low" values of the efficiency parameter do not indicate a lack of star formation potential. Conversely, the estimated star formation rate densities within the disc are sufficiently large to require a low efficiency if the KS relation is to be satisfied. Regardless, as the models here are subject to the same assumptions, a peer-to-peer comparison is appropriate. 

In Figure~\ref{fig:SFR-relation}, simulations with weak radiation fields are characterized by higher peak star formation rate densities and disc gas densities. This effect appears to be more pronounced for higher mass inflow. At the Eddington limit, however, the discs populate the lower end of the KS relation. The increase in peak surface density for weak radiation fields, coupled with the comparably lower values in models with strong radiation fields, suggests that there may be an ``ideal" radiation strength for fostering star formation in such a scenario. It is still a matter of investigation as to what degree feedback inhibits or promotes star formation near active black holes. \citet{2018MNRAS.479.5544M} recently explored the role of jet-driven feedback on gas discs surrounding AGN. They show that, depending on the strength of the feedback, that the effects can serve to enhance and quench star formation in this region. The same appears to be true in the case of radiative feedback, but we defer a more thorough parametric investigation to future studies.

\section{Conclusions}
\label{sec:conclusion}
Near radial gas flow towards nuclear SMBHs is thought to provide the most viable mechanism for the formation of a nuclear stellar disc on sub-parsec scales in the Galactic Centre \citep{2013MNRAS.433..353L,2008Sci...321.1060B, 2012ApJ...749..168M,2011MNRAS.412..469A}. Our results indicate that accretion onto the SMBH via this process can surge to large fractions of the Eddington limit \citep{2009MNRAS.394..191H,2008Sci...321.1060B}, though our accretion rates should be viewed as upper limits (Sec.~\ref{subsec:viscous}). Such high accretion rates suggest that the resulting radiative feedback could affect the disc formation and evolution. We present the first 3D RHD simulations following the infall of massive gas streams onto a $4\times10^6 \text{M}_\odot$ SMBH. The gas streams are exposed to a constant radiation field at 10\% or 100\% of the Eddington luminosity, approximating radiative feedback due to accretion. We consider inflow masses of $\approx 10^{5}~\text{M}_\odot$, and include the effects of ionization, photo-heating, and radiation pressure in our models. We find the following: 
\begin{enumerate}
    \item A  direct collision between a SMBH and a clumpy gas stream can produce gas discs that are characterized by $Q_T \lesssim 1$ (Figure~\ref{fig:discfindall}). This suggests that the discs may be gravitationally unstable, and that an episode of star formation may occur for such a scenario. 
    \item  At 10\% of the Eddington luminosity, radiation does not strongly influence the process of central disc formation (Figure~\ref{fig:dynamics1}). Photo-heating increases the average disc temperature, but an increase in surface density maintains conservative estimates of $Q_T \approx 1$ (Figure~\ref{fig:discfindall}), consistent with models that do not include radiation. 
        
    \item  At the Eddington luminosity, radiation pressure from UV photons delays the formation of a dense gas disc (Figure \ref{fig:dynamics1}). Radiative forces are not sufficient to drive gas from the vicinity of the SMBH, thus a disc builds gradually as mass accumulates along shielded lines of sight. Peak surface densities are below those seen for lower radiation fields (Figure~\ref{fig:discfindall}), thus the values of $Q_T$ are larger. Yet, for higher mass inflow, super-tidal densities manifest within the disc even at low resolution thus indicating that star formation may still be possible (Figure~\ref{fig:multi_tidal}). 
    
    \item The instantaneous accretion rates of all models are in excess of the Eddington luminosity (Figure~\ref{fig:massacc_all}), although viscous timescale constraints suggest that the resulting radiation strength may fall in the range of 1\% -- 60\% of the Eddington limit. These estimates do not account for on-going mass accretion beyond the simulation time. 

\item We lastly note that our results regarding gravitational instability are conservative due to resolution effects. We are unable to resolve the cooling lengths at which dense structures are expected to form. As a result, the densities observed in the disc are mostly sub-tidal (Figure~\ref{fig:multi_tidal}), but these values are not converged (Figure~\ref{fig:multi_tidal_res}). An increase in resolution is computationally prohibitive for models with radiation.   
\end{enumerate}

We conclude that star formation occurring via the collision of inflowing gas streams and the formation of a central gas disc may still occur despite radiative feedback from AGN activity.

\section*{Acknowledgements}
We thank the anonymous referee for the detailed report, 
helpful insights, and exhortations to expand our analysis.
The models presented in this publication were performed on
the Killdevil and Dogwood clusters at UNC-Chapel Hill. The authors gratefully
acknowledge support from the North Carolina Space Grant. FH would like to thank 
NASA ATP for initial support of this project through grant NNX10AC84G. CF would also like to thank both the Cato-SOAR Fellowship and the Paul Hardin 
Dissertation Completion Fellowship from the Royster Society at UNC-Chapel Hill for supporting this work.



\bibliographystyle{mnras}
\bibliography{main.bib} 



\appendix

\section{Modifications to Athena} \label{app:athenamod}

\subsection{Internal Energy Advection}
\label{app:duale}
In grid cells where the kinetic energy is a significant fraction of the total energy, it is possible that the resulting internal energy density ($ e = E - \frac{1}{2 \rho}  {\bf M} \cdot {\bf M}$) will be negative. For this reason we have implemented a procedure that is similar to that of \citet{2014ApJS..211...19B} and \citet{2016MNRAS.462.2777G} by additionally solving the internal energy equation: 
\begin{equation}\label{eq:eint}
    \fracd{e}{t} + \del \cdot ({\bf v} e)  = - P \del \cdot {\bf v}  \ . 
\end{equation}
The solution to this equation is split into two stages. First, thermal energy is transported proportionally to the density flux by solving the left hand side of Eq.~\ref{eq:eint} in conservative form: 
\begin{equation}
    \fracd{(C_e \rho)}{t} + \del \cdot ( C_e \rho {\bf v} ) = 0    \  ,
\end{equation}
where $C_e = \frac{e}{\rho}$ is the specific internal energy density. We calculate the interface states for the internal energy and calculate the flux in the same manner as a colour field. The resulting transport update in one dimension is then 
\begin{equation}
    e^{n+1} = e^n_{i} + \Delta t \left(F_{e,i-1/2} -  F_{e,i+1/2}\right)  \ ,
\end{equation}
where $F_{e}$ are the internal energy fluxes at each boundary of the cell. The right hand side of Eq.~\ref{eq:eint} is included as a source term in tandem with the source terms for the conserved variables. The resulting change in the internal energy is calculated via central difference (in one dimension):
 \begin{equation}
     \delta e = - P \Delta t \left(\frac{v_{x,i+1} - v_{x,i-1}}{2 \Delta x}\right)
 \end{equation}

Within the directionally un-split VL integrator, we calculate the source term at the half timestep using the pressure calculated from the initial condition. In the corrector step of the integrator, the pressure is calculated from the half timestep values. Our approach differs from previous implementations as we calculate the internal energy from the total energy at the beginning of every integration step. Pressure positivity is checked both at the predictor and corrector stages of the VL integrator. For cases where the pressure calculated from the conserved variables is negative, the internal energy solution is used as a replacement. The change in the total energy imposed by this process is typically $\delta E/ E < 10^{-5}$. 

\subsection{Coarse Grid Restriction}
The restriction scheme of the \textsc{athena} stock version uses conserved variables to synchronize the solutions of coarser grid levels with averaged values from finer levels. Under the physical conditions met in our simulations, this process led to unphysical feedback at the refinement boundaries. Unresolved motion from converging flows on the fine grid can be translated into thermal energy on the coarse grid. This results in an overpressure on the coarse grid that imprints the mesh structure onto the gas and dampens fluid flow onto higher refinement domains. We resolved this issue by restricting on the pressure rather than the total energy. By doing this, mass and momentum are perfectly conserved, but we sacrifice total energy conservation for improved continuity of fluid flow and increased stability. For the conditions in our models, this process leads to an average error of $ \delta E/E \lesssim \text{few} \times 10^{-3}$.

\subsection{Refinement Interpolation}
The directionally split boundary value interpolation method used for refinement synchronization in \textsc{athena} can result in negative pressures in regions with steep pressure, momentum, or density gradients. To avoid this issue, we have included the ``second-order A" interpolation scheme from the \textsc{enzo} code \citep{2014ApJS..211...19B}. This method uses a tri-linear interpolation to calculate the conserved variables at the corners of the coarse cell. Monotonized slopes are then calculated along each diagonal of the cell. An additional constraint is applied to the slopes to remove outliers. The resulting slopes are applied to calculate the refined zone boundary condition. We make two changes to this routine. First, we impose the condition that, for each conserved variable, the corner values must meet the condition that $ 0.1 q_c < q < 10 q_c$, where $q$ is the corner value and $q_c$ is the cell centered value. Additionally, rather than computing the interpolation on the total energy, we interpolate on the pressure as is done in the directionally split approach of \textsc{athena}.

\section{Radiative Transfer}   
\label{ap:radtrans}
The problem of radiative transfer is ubiquitous in astronomy and can be incorporated into numerical simulations in a variety of ways (see \citet{2009MNRAS.400.1283I} and references therein). A previous implementation of radiative transfer for \textsc{athena} is shown in \citet{2007ApJ...671..518K}, but this feature is not included in the public release of the code. Furthermore, it was not designed to be used with the static mesh refinement native to \textsc{athena} and does not include all the physics components required for this work. We roughly follow the structure of this previous implementation, but include features and simplifications from \citet{2011MNRAS.414.3458W} that were necessary for making our models computationally feasible. Here, we provide a brief overview of our radiative transfer module and defer the reader to \citet{2007ApJ...671..518K} and \citet{2011MNRAS.414.3458W} for more detail.

\subsection{Adaptive Ray Tracing}
\label{sec:raytrace}
Radiation in our models is treated using long ray characteristics. To prevent both oversampling near the source and minimize computational cost we use an adaptive ray tree based on the nested geometry of the Hierarchical Equal Area isoLatitude Pixelation (HEALPix, \citet{2005ApJ...622..759G}). We use this geometry specifically because it can be used to construct a quad-tree of rays as first demonstrated by \citet{2002MNRAS.330L..53A}. In short, we cast the $12$ rays from the radiation source using the angles supplied by the coarsest discretization of HEALPix. The rays extend radially outward until the ray density is too low to uniformly sample the grid. The radius at which this occurs is
\begin{equation}  
    \label{eq:Rmax_rad}
    R_\text{max} = \sqrt{\frac{N_\text{rays} \Delta x^2 }{f 4 \pi}} \ ,
\end{equation}
where $f$ is approximately the minimum number of rays that must trace each cell. We set $f=3$ for all the work presented in this paper. $N_\text{rays}$ is the number of rays at a given HEALpix resolution level ($l$) such that $N_\text{rays} = 12 \times 4^l$. Each terminating ray splits into four child rays. Each child continues outwards along the directions prescribed by HEALpix until they also terminate and spawn four new rays. This process continues until the mesh is completely sampled by the ray tree. 

At the beginning of every simulation, we calculate the geometry of the ray tree according to the geometric restrictions of the computational mesh. For each ray, we store the initial and final positions, the propagation angle prescribed by HEALpix, the resolution level, and the ray index. Similar to \citet{2011MNRAS.414.3458W}, we organize rays into a doubly-linked list such that each ray stores the address for both the parent ray and child rays. \textsc{athena} is parallelized with distributed memory, thus we only store rays which are either completely or partially contained within a processor's local boundary. 

Interfacing a spherical ray tree with a Cartesian grid can lead to geometric artifacts in the radiation profile and asymmetry in the absorption of energy and momentum within the gas. To mitigate this effect, \citet{2007ApJ...671..518K} introduce periodic rotations of the ray tree. Rotating by random angles at a preset interval suppresses numerical artifacts arising from ray tracing, particularly in spherically symmetric problems (Fig.11, \citet{2007ApJ...671..518K}). A keystone difference in this previous work and that which is showcased here is the requirement of SMR which complicates the ray-tracing geometry and increases the computational load of our simulations. As such, periodic rotations are not currently computationally feasible for the simulations presented in this work. As an alternative, we have found that a single, randomly oriented rotation was sufficient to eliminate strong grid effects along the principle axes. 

\subsection{Photon Absorption}
As rays pass through grid cells, ionization and heating takes place as a consequence of photon deposition. Assuming a zero emissivity gas and a constant absorption cross section, the radiative transfer equation can be solved to calculate photon absorption along rays such that \citep{1999ApJ...523...66A}:
\begin{equation}                                             
    \label{eq:rayatten}
    N_{\gamma,f} = N_{\gamma_i} e^{- \sigma_\text{pi} n_\text{H} (1-x) \Delta s}    ,
\end{equation}
where $N_{\gamma,i}$ and $N_{\gamma,f}$ are the initial and final photon numbers, $n_\text{H}$ is the neutral hydrogen density, $x$ is the ionization fraction, and $\Delta s$ is the path length through the cell. We take the photo-ionization cross section to be \citep{2011pok.....D}:
\begin{equation}
    \sigma_\text{pi} = 6.304 \times 10^{-18} \left(\frac{E_H}{h \nu}\right)^4  \frac{\exp\left(4\left(1-\text{arctan}(\eta)/\eta\right)\right)}{1 - \exp\left(-2\pi/\eta\right)}    \ \text{cm}^2
\end{equation}
with
\begin{equation}
     \eta = \sqrt{\frac{h\nu}{E_H} -1}             \ \  .
\end{equation}
where $E_\text{H} = 13.6 \ \text{eV}$ is the binding energy of atomic hydrogen and $\nu$ is the photon frequency. The photon deposition is calculated by taking the difference between the initial and final photon numbers:
\begin{equation}
    \delta N_\gamma =  N_{\gamma,i} \left( 1 -  e^{- \sigma_\text{pi} n_H (1-x) \Delta s}\right)  \ .
\end{equation}
Multiple rays will pass through each cell, thus, for every cell, we also assign a sub-volume to each ray so that photo-absorption does not lead to an overshoot in ionization or heating. For each ray segment, we calculate the weighting factor:
\begin{equation}
    W_\text{r} = \frac{\Delta s}{4^l}
\end{equation}
The sub-volume for each ray is then given by the fraction 
\begin{equation}
    V_\text{sub} = \frac{W_\text{r}}{\sum_i W_\text{r,i}} \ , 
\end{equation}
where the denominator is the sum of weights for all rays passing through the cell.

\subsubsection{Grid Walk}
At every radiation cycle, the ray tree must be walked through in sequence. We follow \citet{2007ApJ...671..518K} by storing a list of cells crossed and path lengths for each ray upon initialization. The initial grid walk follows the process outlined in \citet{1999ApJ...523...66A}. Emitted photons are equally split between the 12 initial rays. Each of these rays is then walked through its cell list, resulting in photon deposition. When the ray terminates, the remaining photon flux is split evenly between the four child rays. This process continues until either a grid (processor) boundary is hit, the ray reaches the edge of the computational mesh, or an optically thick region is encountered such that 99.99\% of the incident photon flux is absorbed in a single cell \citep{2011MNRAS.414.3458W}. 
                                                                                                                                     
For parallel jobs, rays may also terminate at grid boundaries, thus requiring parallel communication. One of two types of parallel communications will occur in these cases: (i) A ray will exit a grid and pass into another. (ii) A parent ray will terminate near a processor boundary, but one or more child rays will spawn on a neighboring processor. For each ray requiring parallel communication, we store the processor ID from which communications are expected and the processor ID to which any information will be sent. During the sequential grid walk, terminating rays post non-blocking messages containing the ray index, level, and photon number using the \textsc{mpi\_isend} function. For processors expecting messages, we sequentially post a blocking receive using \textsc{mpi\_recv} with the \textsc{mpi\_any\_source} flag so that messages can be received sequentially but asynchronously. Once a message is received, integration continues as usual before the next message is received. This process continues until all expected send and receive sequences have been completed.

\subsubsection{Static Mesh Refinement}
A major difference between the implementation of radiation into \textsc{athena} in previous work \citep{2007ApJ...671..518K} and that which is detailed here is that we require the use of grid refinement, thus we have designed our routine to be compatible with the SMR framework native to the \textsc{athena} code. In this case, The ray tree structure is complicated by non-uniform cell spacing on the computational mesh. Yet, the mesh geometry is completely determined and known to all processors for all time. The starting position of each ray is calculated using the restrictions that equation~\ref{eq:Rmax_rad} imposes on the ancestors of the ray. The termination point of each ray is then iteratively calculated using the same condition until the ray terminates within the bounds of a refinement domain. Cells which are overlapped by refinement zones are not included in the ray-trace, thus the effect of radiation only manifests through the refinement synchronization at the end of the last radiation sub-cycle.

\subsection{Time Integration}
\label{sec:raytime}
Radiation is included into our simulations in an operator split fashion. Because the radiation timescale is typically shorter than dynamical timescales, we allow the radiation module to sub-cycle during the hydrodynamic timestep but limit the number of sub-cycles to ten. In the event that the hydrodynamic timestep is less than the radiation timestep, the radiation field is calculated for the hydrodynamic timestep without sub-cycling. For each radiation cycle, we follow the structure seen in Figure 1 of \citet{2007ApJ...671..518K} with the exception that we calculate the timestep at the beginning of the cycle based on the conditions from the previous timestep. As a consequence, our scheme is more akin to \citet{2011MNRAS.414.3458W} in which we trace photon numbers and not rates. This helps to alleviate stringent restrictions imposed on the radiation timestep, helping to mitigate computational cost in our implementation of radiative transfer. We do not preserve the time derivative of the radiative transfer equation because the light transit times through our simulation domain are much shorter than typical dynamical timescales.

\subsubsection{Time Stepping}
For each radiation cycle, we calculate the timestep at each cell as
\begin{equation}
    \label{eq:dtrad}
    \Delta t_\gamma = \alpha \frac{x}{|dx/dt|} = \alpha \frac{n_H}{|I-R|}  \ ,
\end{equation}
where $\alpha$ is the maximum fractional change in the ionization fraction, x, allowed. We set $\alpha=0.1$ for our models. The radiation timestep is taken to be the minimum value across all cells in the simulation. To avoid prohibitively small times-eps-converted-to.pdf, we only include grid cells with $x>0.1$ in this timestep calculation and enforce a minimum timestep of $\Delta t_\gamma = \frac{\Delta x}{c}$, which is the light crossing time of a cell that depends on the cell size $\Delta x$. 

\subsubsection{Sub-cycling Thermal Physics}
To avoid over-cooling, ionization and thermal effects are included in a sub-cycle within each radiation step. After the heating rates are calculated, the sub-cycle timestep is calculated as:
\begin{equation}                           
    \label{eq:dtsub}
    \Delta t_\text{sub} =  \min \left(\beta \frac{e}{|\frac{de}{dt}|_\text{therm} + |\frac{de}{dt}|_\gamma}, \Delta t_\gamma\right)
\end{equation}
where $\beta$ is the fractional change allowed in the internal energy which is set to 0.1. Changes in internal energy due to the thermal physics prescription are calculated at each step in the sub-cycle. We assume that changes due to radiation are linear within this time interval. The internal energy is updated using a forward Euler step, 
\begin{equation}
    \label{eq:desub}
    e_f = e_i + \left(\left(\frac{de}{dt}\right)_\text{therm} + \left(\frac{de}{dt}\right)_\gamma\right) \Delta t_\text{sub} \ .
\end{equation}
The ionization fraction, which affects the cooling rate, is updated linearly in time so that
\begin{equation}
    \label{eq:dxsub}
    x_f = x_i + \frac{dx}{dt} \Delta t_\text{sub}
\end{equation}
The sub-cycle continues until the total time elapsed equals the radiation timestep. \\

\subsection{Accuracy Tests}           
\label{sec:raytests}
To demonstrate the fidelity of our radiative transfer method, we begin with three tests originally used for the first implementation of ray-tracing in the \textsc{athena} code. Our test cases differ for two reasons: (1) The radiation module used for this work is compatible with the refinement framework of \textsc{athena}, thus it is important to demonstrate agreement across grids of varying resolution. (2) The time-step restrictions imposed by the radiation routine have been designed to maximize the radiation timestep, thus lowering the computational cost. In addition, we present a new test for our implementation of radiation pressure, and demonstrate the effect of including secondary ionizations from X-rays.

For all tests, we use the mesh and refinement geometry shown in Table~\ref{tab:SMR2}. The radiation source in all cases is placed at the origin of a box which extends 10~pc in each direction. Two elongated SMR regions enclose the radiation source, and extend in the $+\hat{x}$ direction to the edge of the computational mesh.

\begin{table}
    \centering
    \caption{Ionization Tests Refinement Geometry}
    \begin{tabular}{cccccc}
        \hline
        Level & Dimensions & x$_0$ & y$_0$ & z$_0$  & Resolution   \\
              & [pc]       &  [pc]& [pc] &[pc]       &  \\
        \hline
        1 & 20$\times$20$\times$20     & -10  &-10 &-10 & 64$\times$64$\times$64   \\
        2 & 11.25$\times$10$\times$10  & -1.25&-5  &-5  & 72$\times$64$\times$64   \\
        3 & 10$\times$5$\times$5       & 0    &-2.5&-2.5& 128$\times$64$\times$64 \\
        \hline
    \end{tabular}
    \label{tab:SMR2}
\end{table}

\subsubsection{R-Type Ionization Front}
\begin{figure}
    \includegraphics[width=\linewidth]{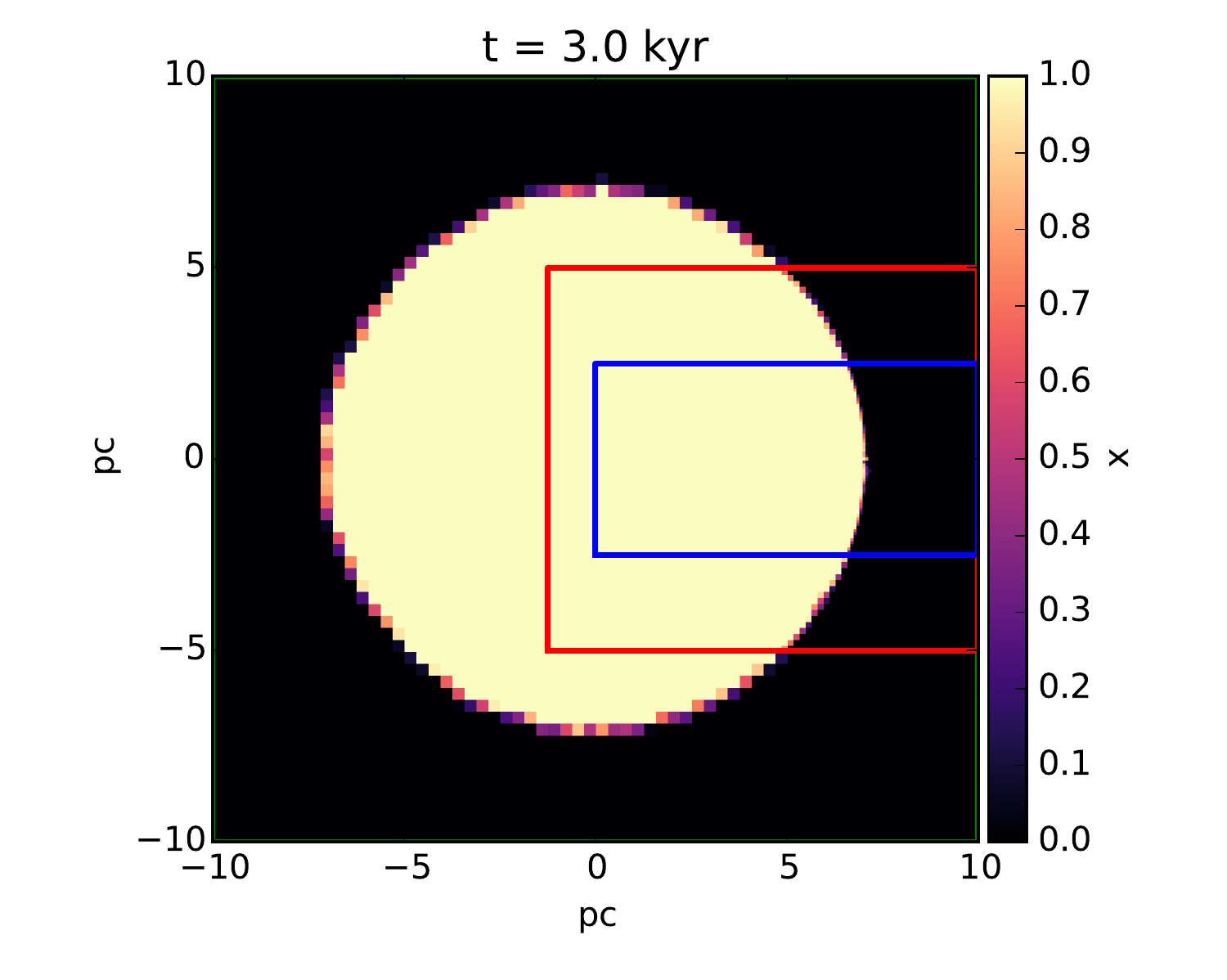} 
    \caption{Midplane slice of the ionization fraction in our simulation of an R-type I-front without recombination. The rectangular boxes outline the refinement domains used in this model.}
    \label{fig:rtype_norec_x}
\end{figure}

\begin{figure}
    \includegraphics[width=\linewidth]{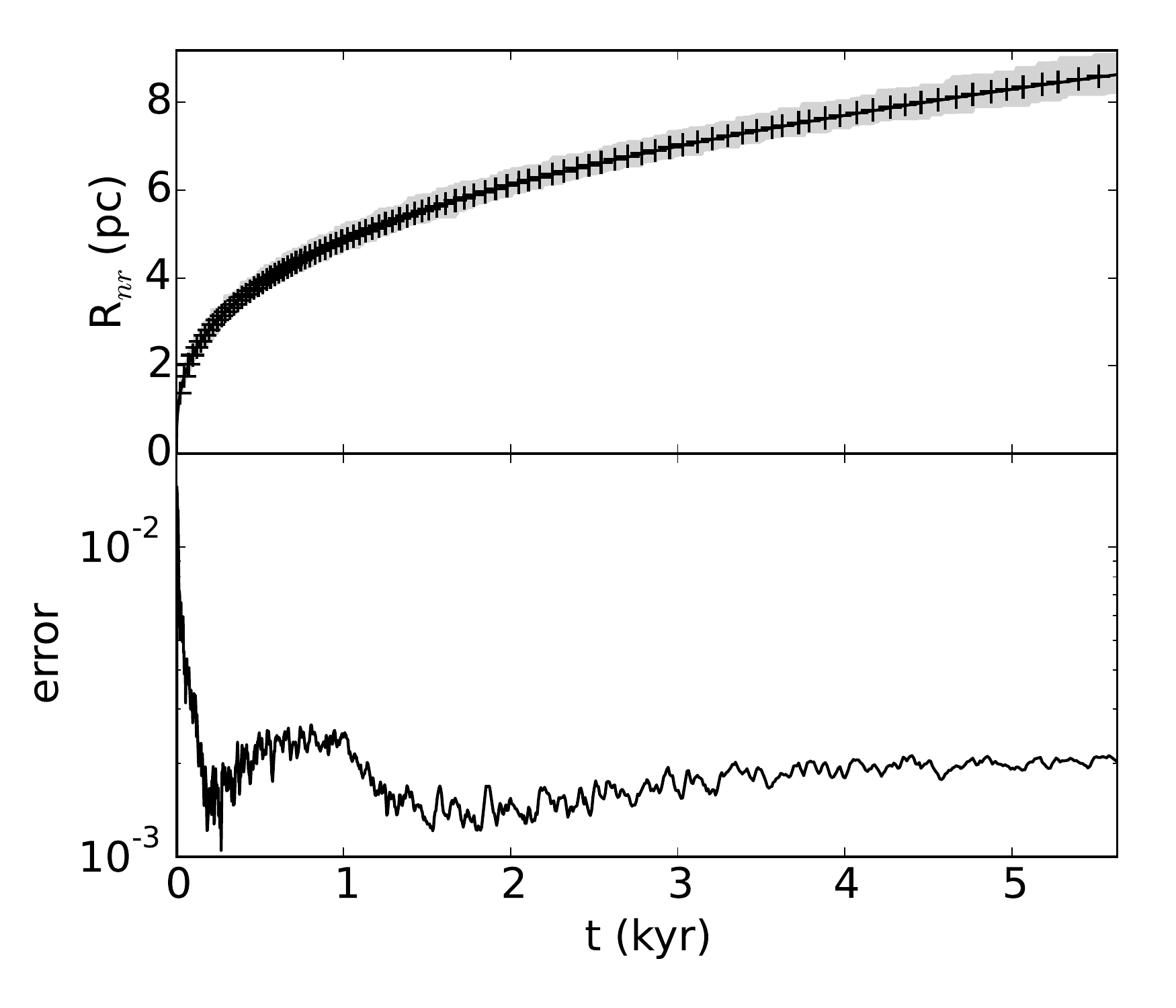} 
    \caption{Radius of the R-type ionization front without recombinations $R_{nr}$ vs time t (top), and the error relative to the analytic solution (bottom). The top figure shows the analytical solution (solid line) and average radius of the I-front in our simulation (plus signs). The gray shaded region shows the range of all radii included in this average. The bottom panel shows the relative error of the radius in our simulations with respect to the analytic solution.}
    \label{fig:rtype_norec}
\end{figure}

As a first test, we follow the evolution of an R-type ionization front (I-front) as detailed in \citet{2007ApJ...671..518K}. To disable photo-heating, we set $E_\gamma =13.6 \text{ eV}$, equal to the binding energy of hydrogen. We also set the recombination coefficient to $\alpha_B=0$ and disable radiation pressure. This test serves as an excellent probe of photon conservation as the number of ionizations that occur must exactly match the number of photons emitted. By equating these two quantities, an analytic expression for the radius of the expanding I-front is found to be
\begin{equation}
    R_{nr}(t) = \left(\frac{3 Q_\star t}{4 \pi n_H}\right)^{1/3}  \ ,
\end{equation}
where $Q_\star$ is the rate of photon emission, $n_H$ is the density of the ambient hydrogen gas, and $t$ is the elapsed time. We adopt parameters identical to those in \S~4.1 of \citet{2007ApJ...671..518K} so that $n_H = 100 \text{ cm}^{-3}$ and $Q_\star = 4.0 \times 10^{49} \text{ s}^{-1}$. The gas within the box is initially neutral and is in thermal equilibrium. In Figure~\ref{fig:rtype_norec_x}, we show a midplane slice of the ionization fraction for our model of the R-type I-front. We emphasize that good agreement is found across refinement levels. To calculate the numerical value of $R_{nr}$, we compute the average radius to cells with ionization fractions between 1\% and 99\%. In Figure~\ref{fig:rtype_norec}, we show a comparison of the numerically calculated I-front radius with the analytic solution. When the radius of the I-front is sufficiently large ($R_{nr} > \text{few} \times \Delta x$), the error between the numerical result and analytic result is on the order of $ .1\%$, which corresponds to an error less than that of a cell size on the highest refinement level (i.e. $\Delta x / R_{nr} \approx 1\%$). We conclude that our radiative transfer method is sufficiently photon conservative.


\subsubsection{R-Type Ionization with Recombination}
\begin{figure}
    \includegraphics[width=\linewidth]{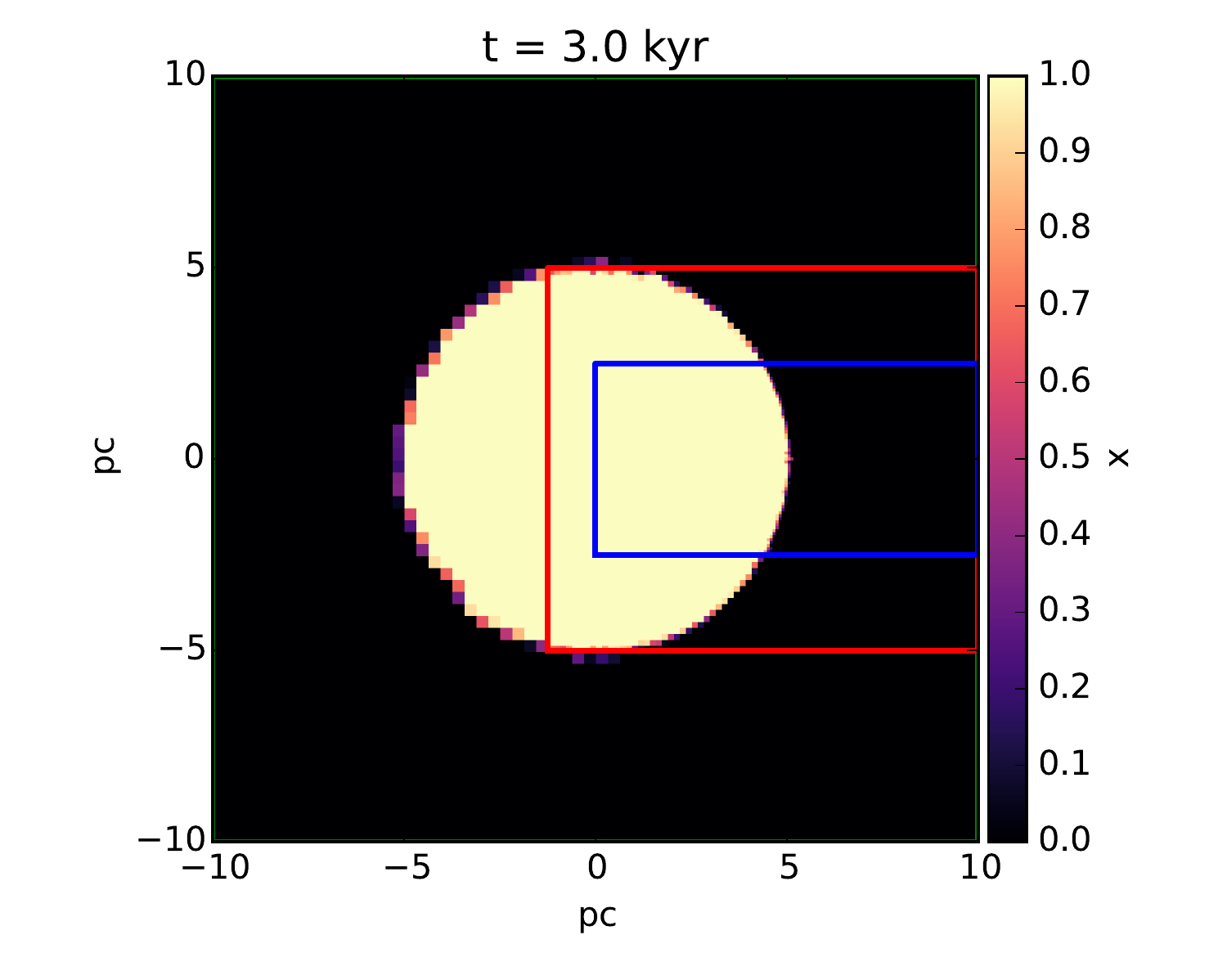} 
    \caption{Midplane slice of the ionization fraction in our simulation of an R-type I-front with recombinations. The rectangular boxes outline the refinement domains used in this model}
    \label{fig:rtype_x}
\end{figure}

\begin{figure}
    \includegraphics[width=\linewidth]{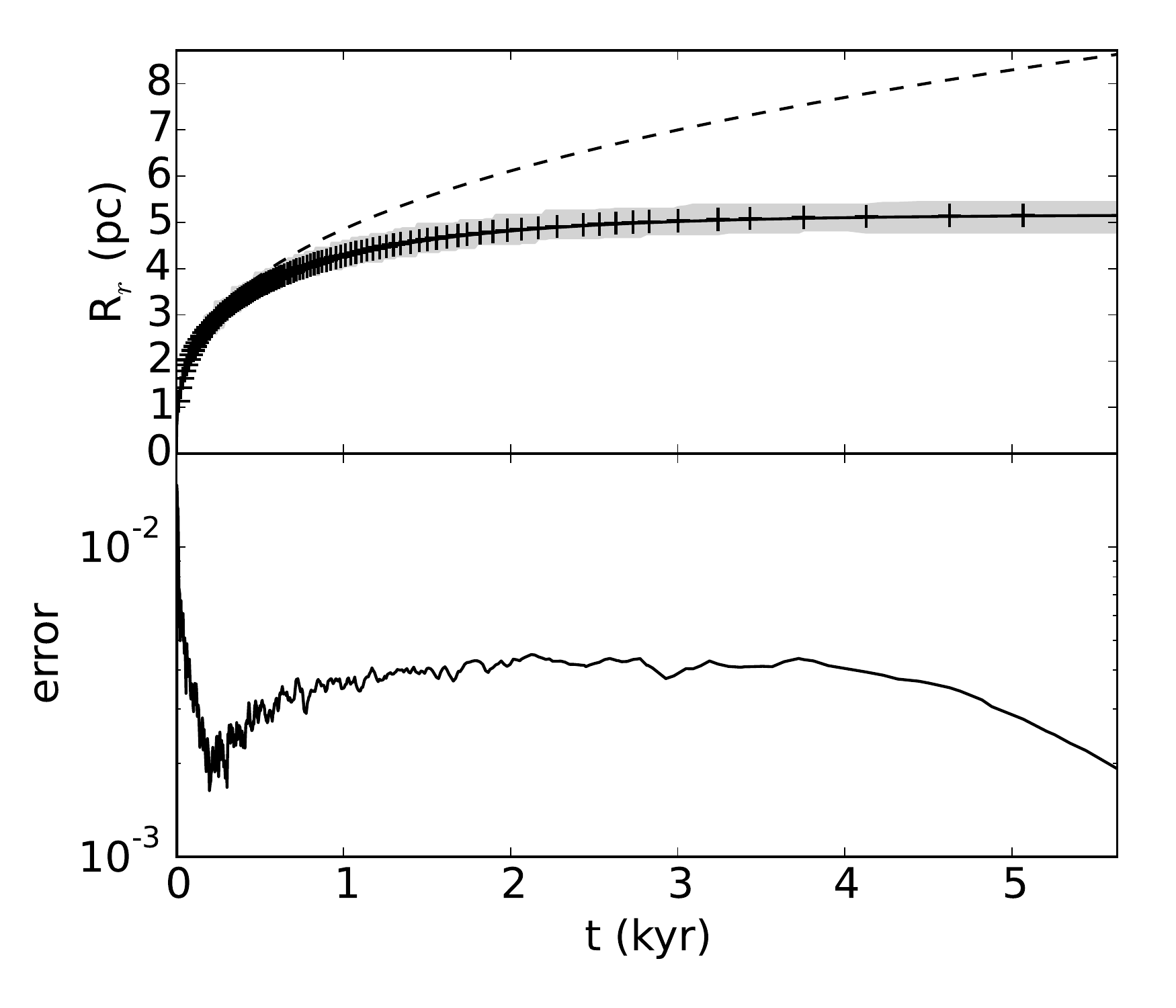} 
    \caption{Radius of the R-type ionization front with recombinations $R_r$ vs time t (top), and the error relative to the analytic solution (bottom). The top figure shows the analytical solution (solid line) and the numerical approximation to the radius of the I-front (plus signs). For reference, the dashed line shows the radius of the I-front without recombination. The gray shaded region shows the range of all radii included in this average. The bottom panel shows the relative error of the radius in our simulations with respect to the analytic solution.}
    \label{fig:rtype}
\end{figure}

Following \citet{2007ApJ...671..518K}, we repeat the previous test, but enable recombinations by setting $\alpha_B = 2.59 \times 10^{13}~\text{cm}^3~\text{ s}^{-1}$. In this case, an equilibrium radius exists at which the rate of photon emission perfectly balances the net recombination rate of gas interior to the I-front. By equating these two terms, the I-front radius has the analytic solution, 
\begin{equation}
    R_s = \left( \frac{3 Q_\star}{4 \pi \alpha_B n_H^2} \right)^{1/3}  \ ,
\end{equation}
which is the well known Str\"{o}mgren radius \citep{1939ApJ....89..526S}. By assuming a constant recombination coefficient, the time dependence of the ionization front radius as it approaches the Str{\"o}mgren radius also has the analytic solution
\begin{equation}
    R_r(t) = R_s \left(1 - e^{-t/\tau_r}\right)^{1/3} \ ,
\end{equation}
where $\tau_r = (n_H \alpha_B)^{-1}$ is the recombination timescale. For our initial conditions, the Str{\"o}mgren radius is $\approx5~\text{pc}$. In Figure \ref{fig:rtype_x} we show the ionization fraction in a midplane slice through the computational domain, which again demonstrates good agreement across refinement levels. The I-front radius is computed numerically in the same manner as the previous test. The result and the comparison to the analytic solution can be seen in Figure~\ref{fig:rtype}. The error in the result obtained from our radiative transfer method is again on the order of .1\%, demonstrating that our treatment of radiation in the absence of hydrodynamic response is theoretically sound.

\subsubsection{D-Type Ionization Front}
\begin{figure}
    \includegraphics[width=\linewidth]{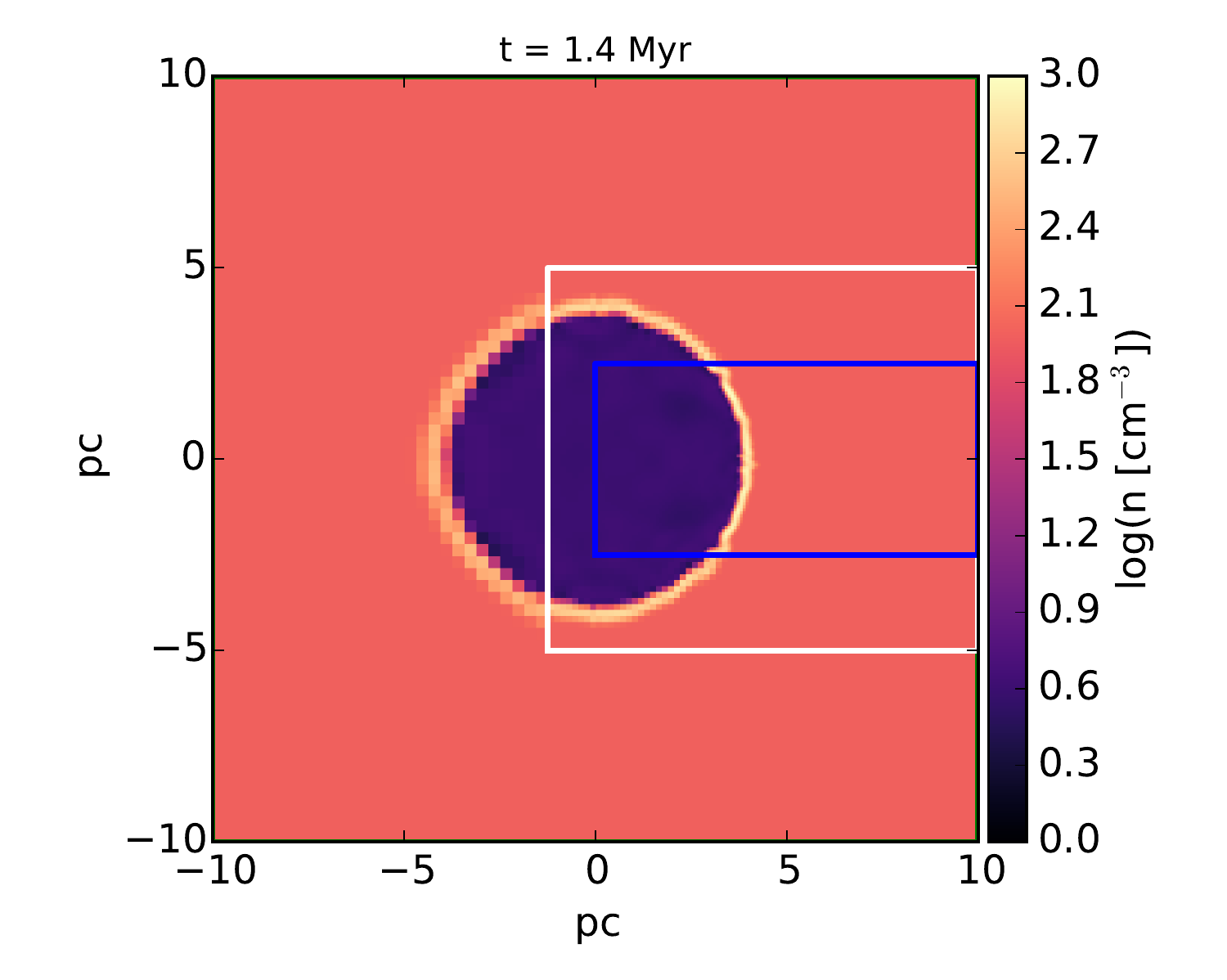} 
    \caption{Midplane slice of the density in our simulation of an D-type I-front. The rectangular boxes outline the refinement domains used in this model.}
    \label{fig:dtype_n}
\end{figure}
\begin{figure}
    \includegraphics[width=\linewidth]{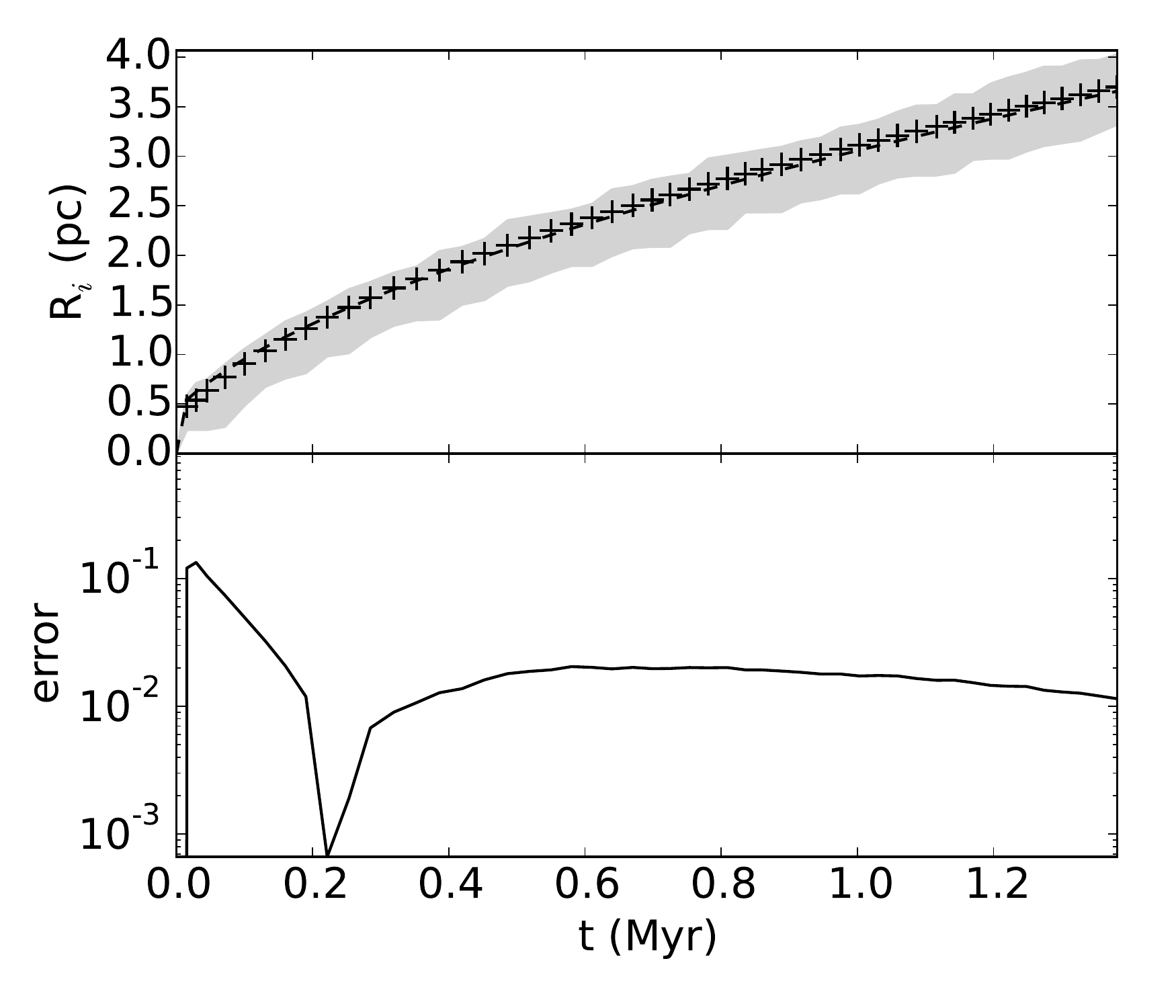} 
    \caption{Radius of the D-type ionization front $R_i$ vs time t (top), and the error relative to the analytic solution (bottom). The top figure shows the analytical solution (solid line) and the numerical approximation to the radius of the I-front (plus signs). The gray shaded region shows the range of all radii included in this average. The bottom panel shows the relative error of the radius in our simulations with respect to the analytic solution.}
    \label{fig:dtype}
\end{figure}
\label{sec:dtypeion}
To demonstrate that our code correctly models radiative-hydrodynamic coupling, we present an idealized model of a D-type I-front as seen in \citet{2007ApJ...671..518K}. Here, gas surrounding the radiation source will be ionized and rapidly heated. Shortly after, the over-pressured gas will expand into the surrounding medium, sweeping up a dense gas shell with a radius approximated by the analytic solution \citep{2011pok.....D}:  
\begin{equation}
    R_i(t) = R_D \left(1 + \frac{7}{4} \frac{c_s (t-t_D)}{R_D} \right)^{4/7}   \ ,
\end{equation}
where $R_D$ and $t_D$ are the radius and time at which the I-front transitions from R-type to D-type, and $c_s$ is the isothermal sound speed of the ionized gas interior to the dense shell. On the timescales considered in this test problem, we can safely take $t_D \approx 0$ and set $R_D \approx R_s$. For our model of the D-type ionization front, we use an identical simulation set-up as in the previous test cases. We lower the photon emission rate to $4\times 10^{46}~\text{s}^{-1}$, corresponding to a Str\"{o}mgren radius of $R_s = 0.5 \text{ pc}$. We set the photon energy to be 16 eV to include photo-heating and enable both the temperature dependent recombination rate in Eq.~\ref{eq:alphaB} and our full thermal physics prescription (\ref{sec:thermphys}). The equilibrium temperature of ionized gas in this model is 5803 K.

In Figure~\ref{fig:dtype_n}, we show a midplane slice of the density in our simulation of the D-type I-front, and in Figure~\ref{fig:dtype}, we show the comparison between I-front radius computed in our simulation with the analytical value. To calculate the numerical value of $R_i$, we find the radius to all cells that meet two criteria. First, the cell must have $\rho < \rho_\text{ambient}$. Second, the nearest neighbor in the the $+\hat{r}$ direction must have $\rho > 1.1 \rho_\text{ambient}$. These two conditions mark the inner edge of the dense gas shell and the extent of the I-front. We take an average of the radii of cells in this sample. The error with respect to the analytical solution is  $\approx1\%$, corresponding roughly to a cell size on the highest refinement level, indicating good radiative-hydrodynamic coupling.

\subsection{Additional Physics}           
\subsubsection{Radiation Pressure}
\label{sec:radprestest}
Here, we demonstrate the effect of radiation pressure through a simple test of a radiation source embedded in a uniform medium. We choose a monochromatic spectrum with $E_\gamma$ = 13.6 eV to disable photo-heating and choose an emission rate of $4\times10^{48}$ s$^{-1}$. The ambient medium is initialized with a density of $5$ cm$^{-3}$, a temperature of T=100 K, and is assumed to be fully ionized. We disable additional thermal physics for both the neutral and ionized gas and use the constant recombination coefficient of $\alpha_B = 2.59\times10^{-13} \text{ cm}^3 \text{ s}^{-1}$. The Str\"{o}mgren radius for this configuration is approximately 30~pc, which extends beyond the computational domain, thus the gas will remain completely ionized for the duration of the simulation. This is vital to the test as the photon absorption rate is only dependent on the recombination rate of the gas.

\begin{figure}
    \includegraphics[width=\linewidth]{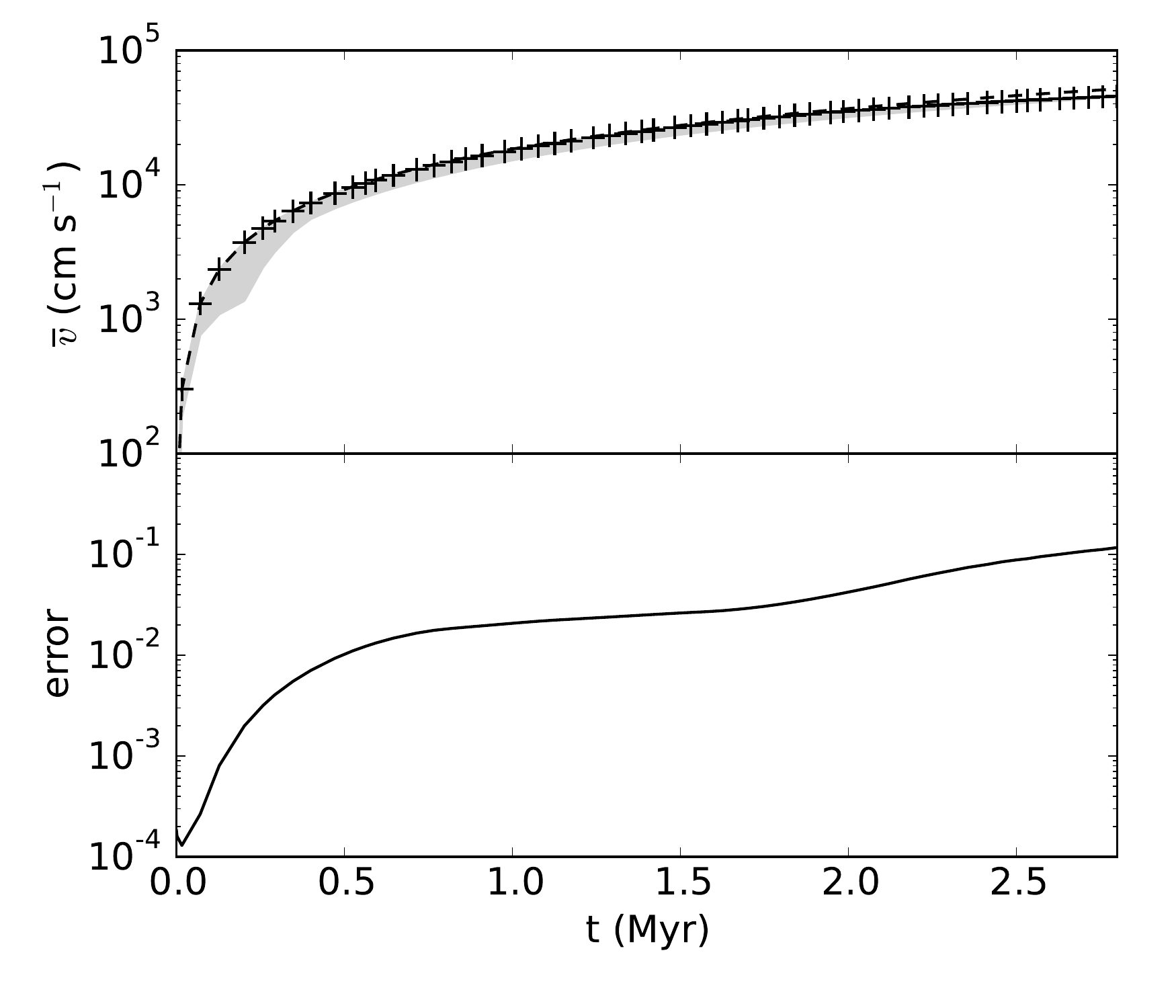} 
    \caption{Average velocity of an ambient ionized gas exposed to the radiation pressure of a central source vs. time (top) and relative error with the analytic velocity vs. time (bottom). The top panel shows the average velocity of all cells with $\rho > 0.95 \rho_\text{amb}$ in our model (plus signs) as well as the analytic velocity expected due to a constant acceleration. The gray shaded region shows the range of values for all cells that were used in the average. The bottom panel shows the relative error of the numerical result with the analytic result.}
    \label{fig:radpres}
\end{figure}

The analytic expression for the acceleration due to radiation pressure is given by:
\begin{equation}
    a_\text{HI}(r) = n_\text{HI} \frac{1}{ c\rho } \int\limits_{\nu_L}^{\infty} \frac{L_\nu e^{-\tau_\nu}}{4 \pi r^2} \sigma_{pi}(\nu) d\nu               
\end{equation}
In the case of our monochromatic spectrum, we may write the luminosity in terms of the ionizing photon emission rate:
\begin{equation}
L_\nu = Q E_\gamma \delta(\nu_\gamma)
\end{equation}
which simplifies the expression to 
\begin{equation}
    a_\text{HI}(r) = \frac{n_\text{HI} Q_\star e^{-\tau} \sigma_\text{pi}}{4 \pi r^2} \frac{E_\gamma}{c \rho} \ . 
\end{equation}
In our test case, the rate of photo-absorption is equal to the recombination rate of the ambient gas. Therefore, the expected acceleration of the ambient gas within the Str\"{o}mgren radius is
\begin{equation}
    a_\text{HI}(r) = n_\text{H}^2 \alpha_B \frac{E_\gamma}{c \rho}   \  , 
\end{equation}
which is independent of geometric and attenuation terms. We compare the average velocity of all cells with $\rho > 0.95 \rho_\text{amb}$ with the expected linear acceleration in Figure~\ref{fig:radpres}. Prior to the onset of hydrodynamic effects, the gas accelerates as expected with a relative error $< 1\%$. This figure also shows that the range of velocities calculated for all cells is exceptionally narrow which is in agreement with acceleration that depends exclusively on the local recombination rate.

\subsubsection{Secondary Ionizations}
\label{app:secondion}
\begin{figure}
    \includegraphics[width=\linewidth]{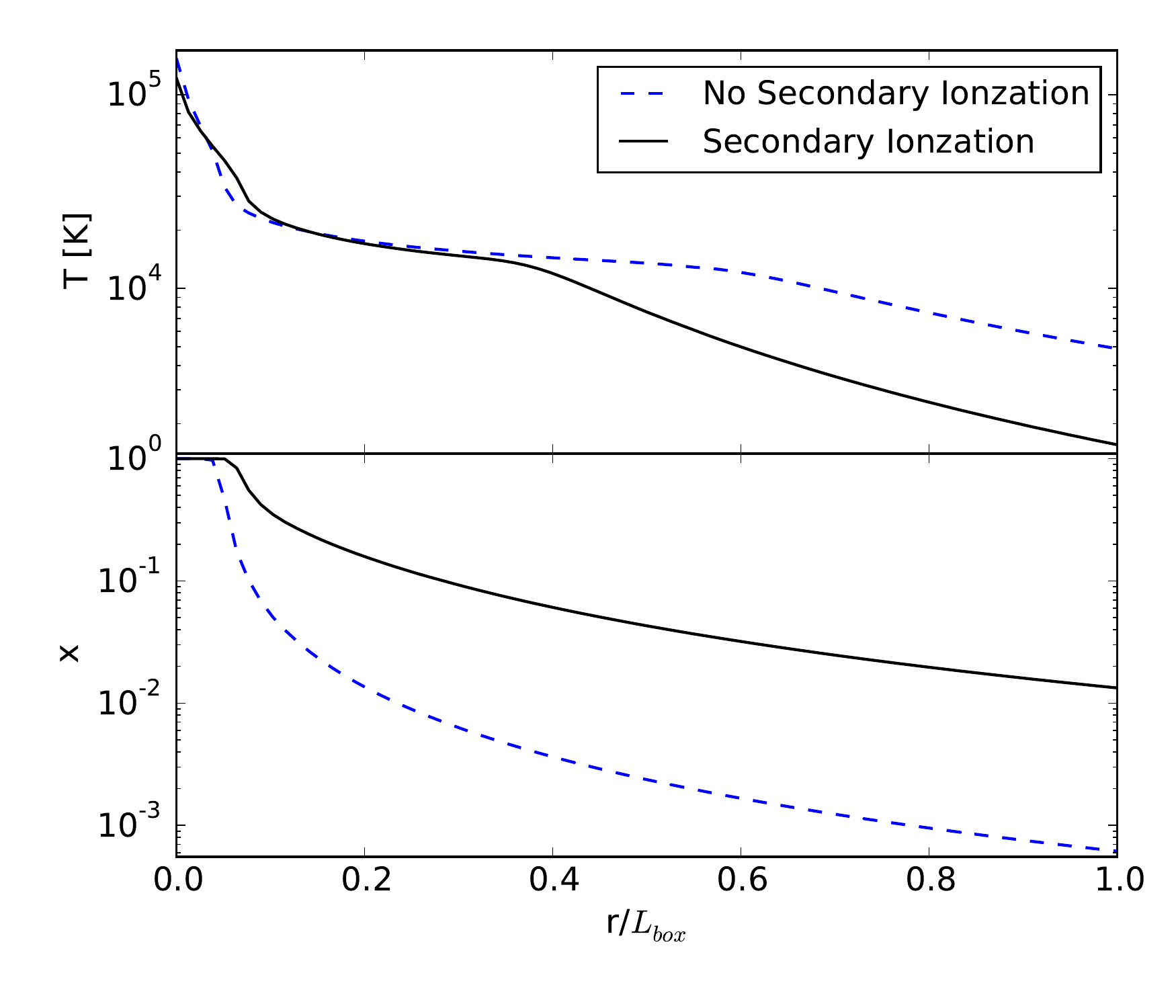} 
    \caption{Average radial profiles of temperature (top) and ionization fraction (bottom) vs. time for a monochromatic 1keV source with (black) and without (blue, dashed) secondary ionization.}
    \label{fig:secondion}
\end{figure}

Lastly, we replicate a test from \citet{2011MNRAS.414.3458W}, albeit on a smaller scale, to demonstrate the effect of including secondary ionizations. We use a monochromatic radiation source of 1~keV photons and an emission rate of Q=$10^{49}$~s$^{-1}$ placed at the center of the box. The computational domain is initialized with a uniform gas of $n_\text{H} = 1~\text{cm}^{-3}$ and temperature of 100 K. Our full thermal physics prescription, including temperature dependent recombinations, is included. Average ionization and temperature profiles are shown in Figure~\ref{fig:secondion}. In general, the inclusion of secondary ionizations extends the ionization profile while simultaneously reducing photo-heating, in agreement with \citet{2011MNRAS.414.3458W}.  

\section{Collisionally Excited Radiative Cooling}
\label{app:CLE}
In our models, we include the effect of optically thin line radiation resulting from collisional excitation of OII, OIII, NII, NeII, and NeIII. We assume abundances of $X_\text{Ne} = 7 \times 10^{-4}$, $X_\text{N} = 9 \times 10^{-5}$, and $X_\text{O} = 7 \times 10^{-4}$ and that 80\% of each species is in the singly ionized state and 20\% is in the doubly ionized state. For each species, we solve the collisional equilibrium equation: 
\begin{equation}
    \label{eq:rateeq}
    \sum_{j\ne i} n_j n_e q_{ji} + \sum_{j > i} n_j A_{ji} = \sum_{j\ne i} n_i n_e q_{ij} + \sum_{j<i} n_i A_{ij}
\end{equation}
where $A_{ij}$ is the spontaneous transition rate from level $i$ to level $j$, and $q_{ij}$ is the collisional excitation rate from level $i$ to level $j$. The spontaneous transition rates and collision strengths, which are used in the calculation for the collision rate, are taken from \citet{1989agna.book.....O}. We assume that the hydrogen gas is fully ionized so that $n_e = n_\text{H}$. This proves to be a fair assumption given that the resulting cooling rate peaks in a temperature regime where the gas is expected to be fully ionized.  

\begin{figure}
    \includegraphics[width=\linewidth]{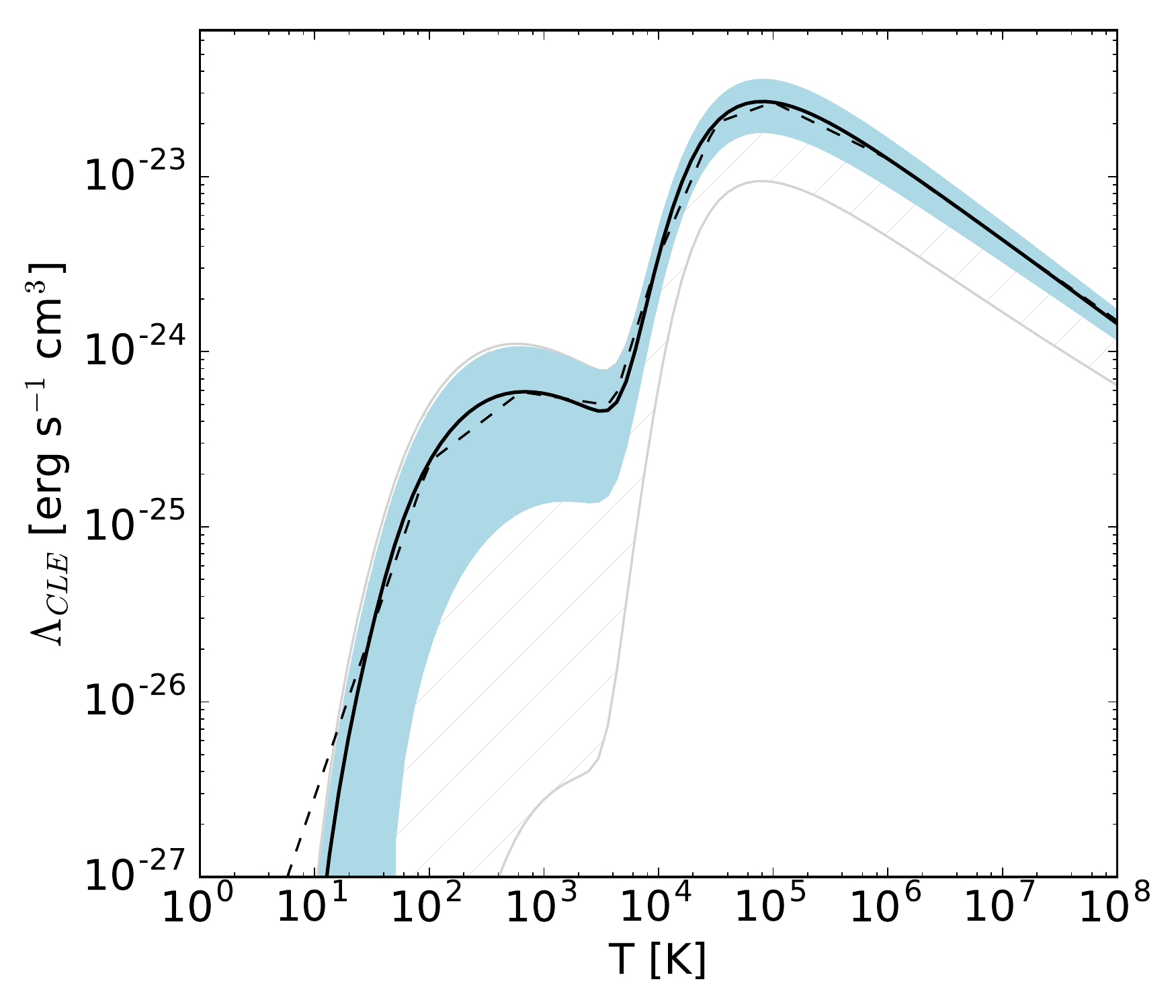}
    \caption{Average volumetric cooling rate (solid) and piecewise approximation (dashed) for collisionally excited line radiation. The shaded region indicates the 1$\sigma$ error region, and the hatched region highlights the range of values averaged over to obtain the approximate cooling function used.}
    \label{fig:CLEfit}
\end{figure}

To calculate an approximate cooling function, we solve Eq.~\ref{eq:rateeq} for a densities in the range $10^{-2}$ -- $10^{6}~\text{cm}^{-3}$ and in the temperature range of $1$ -- $10^8~\text{K}$. For each species, cooling rates are calculated at collisional excitation equilibrium. The net cooling is the sum of these terms:
\begin{equation}
 \Lambda_\text{CLE} = \sum_{s} \sum_i n_{i,s} \sum_{j<i} A_{ij} h \nu_{ij}
\end{equation}
where the first sum is over species, $n_{i,s}$ is the number density of an excited state for a given species, and $\nu_{ij}$ is the frequency of the photon emitted by a transition from level $i$ to level $j$. The two dimensional table of cooling rates is averaged along the density axis to get an average cooling rate as a function of temperature. The average cooling function is show in Figure~\ref{fig:CLEfit} alongside the piecewise fit used in our models (Eq.~\ref{eq:CLE}).

\section{Disc Finding Algorithm} \label{app:discfind}
We have developed an efficient parallelized disc finding routine that allows us to track the formation and evolution of formed discs in our models irrespective of scale or orientation. We implement this routine directly into \textsc{athena} for two reasons. First, this tool provides the advantage of ``on-the-fly" analysis which yields much greater time accuracy than is reasonably managed through post-processing. Second, the method is designed to take advantage of the parallelized structure of \textsc{athena}, which dramatically reduces the amount of time required to execute the necessary operations.

The algorithm takes the following s-eps-converted-to.pdf: (1) On each processor, or local grid, the total mass and angular momentum with respect to the origin are calculated within a search radius, $R_{\text{search}}$. For simulations with refinement, overlapped cells are excluded from this total. The net angular momentum ($\bf{L}_{\text{net}}$) and mass ($M_{\text{net}}$) are then calculated across all processors. (2) For each cell, the deviation angle is calculated as
\begin{equation}
    \theta_{\text{dev}} = \left|\cos^{-1}\left(\frac{{\bf L}_\text{net} \cdot \bf{L}_\text{cell}}{|\bf{L}_\text{net}||\bf{L}_\text{cell}|}\right) \right|   
\end{equation}
where $\bf{L}_\text{cell}$ is the angular momentum vector of the cell. (3) The total mass for cells with  $ \theta_\text{dev} <\theta_\text{thresh}$ is then calculated. For our simulations, we found that setting $\theta_\text{thresh} = 30^\circ$ was a sufficiently strict condition for capturing the formation of a disc. (4) A mass fraction is then calculated as
\begin{equation}
f_{M} = \frac{M(\theta_\text{dev} < \theta_\text{thresh})}{M_\text{net}}     \ .
\end{equation}
For our models, we consider $f_{M}$ > 0.75 to be indicative of a potential disc. (6) For disc candidates, we continue to a three-dimensional rotation of the mesh from the simulation reference frame into a reference frame in which the $\hat{z}'$ axis is parallel to $\bf{L}_{net}$, which we call the disc frame. We construct a new mesh in the disc frame with identical dimensions and hierarchical structure to the simulation frame of reference. We loop over all cells on the disc frame mesh and calculate the corresponding position in the simulation frame:
\begin{equation} 
    \label{eq:discrot}
    {\bf{x}} = 
    \begin{pmatrix}
        \cos \phi_L & - \sin \phi_L & 0   \\
        \sin \phi_L & \cos\phi_L & 0  \\
        0 & 0 & 1  
    \end{pmatrix} \
    \begin{pmatrix}
        \cos \theta_L & 0 & \sin \theta_L   \\
        0 & 1 & 0  \\
        -sin\theta_L & 0 & \cos\theta_L  
    \end{pmatrix}  \ 
    {\bf{x}}'      
\end{equation}
With $\bf{x}$ and $\bf{x}'$ representing simulation frame and disc frame coordinates, respectively. This rotation uses the angles of the net angular momentum vector calculated with respect to the simulation frame:  
\begin{align}
    \theta_L &= \tan^{-1} \left( \frac{\sqrt{L_\text{net,x}^2 + L_\text{net,y}^2}}{L_\text{net,z}}\right)
    \\    \phi_L &= \tan^{-1} \left( \frac{L_\text{net,y}}{L_\text{net,x}}\right)
\end{align}
(7) The eight cells surrounding ${\bf x}$ in the simulation frame are used to tri-linearly interpolate conserved variables in the disc frame. The interpolation is only performed if at least one of the surrounding cells in the simulation frame has $\theta_\text{dev}$ < $\theta_\text{thresh}$. (8) We use the disc frame mesh to calculate the average orbital frequency, sound speed, and column density of the disc.

\bsp	
\label{lastpage}
\end{document}